\PassOptionsToPackage{table}{xcolor}

\documentclass[acmsmall,screen=true]{acmart} 

\usepackage{tabularx}
\usepackage{dcolumn} 
\newcolumntype{d}[1]{D{.}{.}{#1}}

\usepackage{pifont}
\newcommand{\yes}{\ding{51}}
\usepackage{multirow}
\makeatletter
\@ifundefined{c@rownum}{%
  \let\c@rownum\rownum
}{}
\@ifundefined{therownum}{%
  \def\therownum{\@arabic\rownum}%
}{}
\makeatother

\AtBeginDocument{%
  \providecommand\BibTeX{{%
    \normalfont B\kern-0.5em{\scshape i\kern-0.25em b}\kern-0.8em\TeX}}}

\setcopyright{acmlicensed}
\acmPrice{15.00}
\acmDOI{10.1145/3567719}
\acmYear{2022}
\copyrightyear{2022}
\acmSubmissionID{iss22main-id8087-p}
\acmJournal{PACMHCI}
\acmVolume{6}
\acmNumber{ISS}
\acmArticle{566}
\acmMonth{12}

\begin{document}

\title[A Survey of Augmented Piano Prototypes]{A Survey of Augmented Piano Prototypes: Has Augmentation Improved Learning Experiences?}

\author{Jordan Aiko Deja}
\orcid{0001-9341-6088}
\affiliation{%
  \institution{University of Primorska}
  \city{Koper}
  \country{Slovenia}
 \postcode{6000}}
 \affiliation{%
  \institution{De La Salle University}
  \city{Manila}
  \country{Philippines}}
\email{jrdn.deja@gmail.com}

\author{Sven Mayer}
\orcid{0001-5462-8782}
\affiliation{%
  \institution{LMU Munich}
 \city{Munich}
  \country{Germany}}
\email{info@sven-mayer.com}

\author{Klen Čopič Pucihar}
\orcid{0002-7784-1356}
\affiliation{%
  \institution{University of Primorska}
  \city{Koper}
  \country{Slovenia}
  \postcode{6000}}
\email{klen.copic@famnit.upr.si}

\author{Matjaž Kljun}
\orcid{0002-6988-3046}
\affiliation{%
  \institution{University of Primorska}
  \city{Koper}
  \country{Slovenia}
  \postcode{6000}}
\email{matjaz.kljun@famnit.upr.si}


\begin{abstract}
Humans have been developing and playing musical instruments for millennia. With technological advancements, instruments were becoming ever more sophisticated. In recent decades computer-supported innovations have also been introduced in hardware design, usability, and aesthetics. One of the most commonly digitally augmented instruments is the piano. Besides electronic keyboards, several prototypes augmenting pianos with different projections providing various levels of interactivity on and around the keyboard have been implemented in order to support piano players. However, it is still unclear whether these solutions support the learning process. In this paper, we present a systematic review of augmented piano prototypes focusing on instrument learning based on the four themes derived from interviews with piano experts to understand better the problems of teaching the piano. These themes are (i) synchronised movement and body posture, (ii) sight-reading, (iii) ensuring motivation, and (iv) encouraging improvisation. We found that prototypes are saturated on the synchronisation themes, and there are opportunities for sight-reading, motivation, and improvisation themes. We conclude by presenting recommendations on augmenting piano systems towards enriching the piano learning experience as well as on possible directions to expand knowledge in the area.
\end{abstract}

\ccsdesc[500]{Human-centered computing~Human computer interaction (HCI)}
\ccsdesc[500]{Human-centered computing~Interaction devices}
\ccsdesc[300]{Applied computing~Sound and music computing}
\ccsdesc[300]{Applied computing~Interactive learning environments}
\keywords{augmented piano, music learning, systematic review, survey, piano}
\begin{teaserfigure}
\centering
  \includegraphics[width=\linewidth]{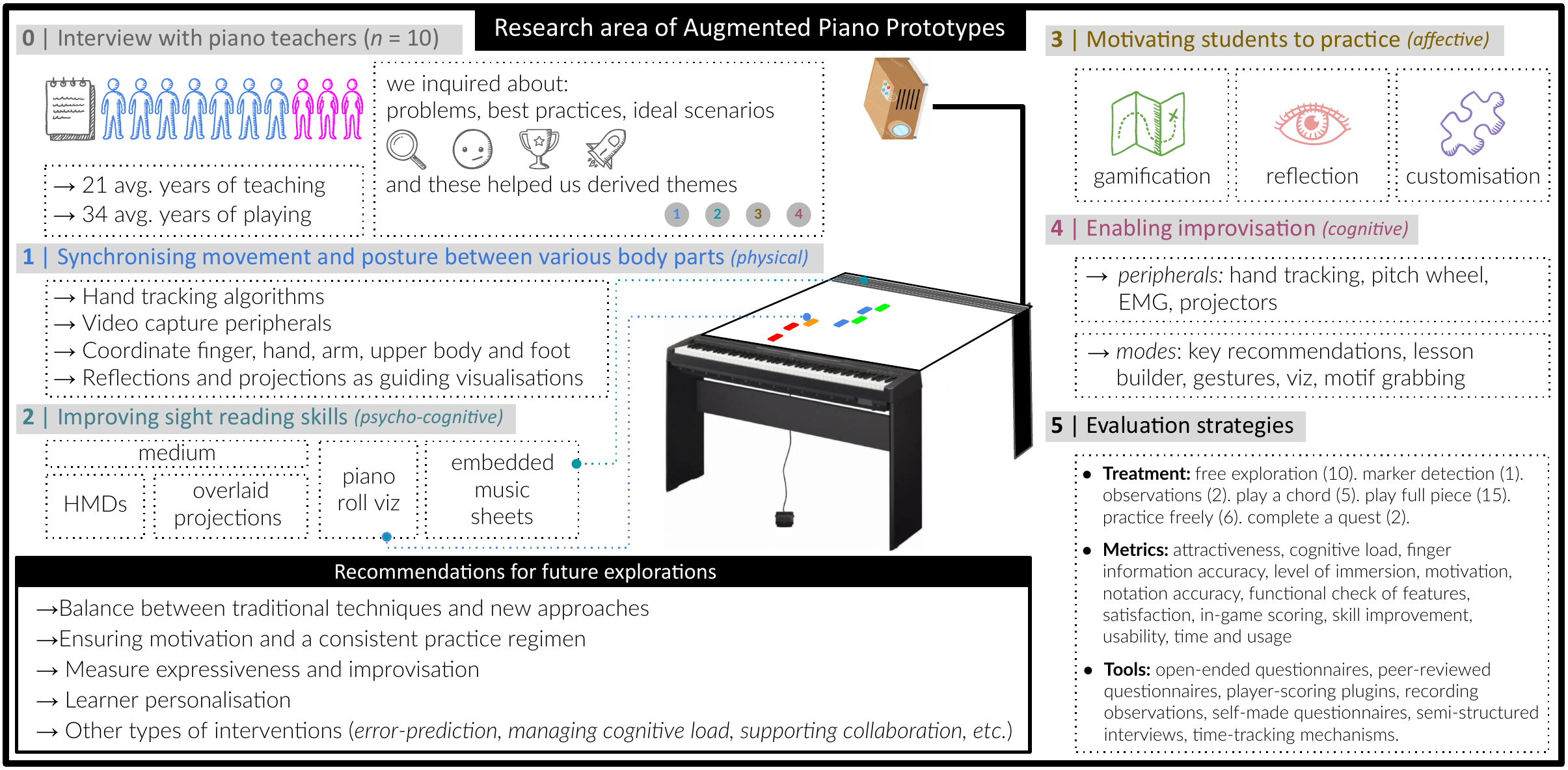}
  \caption{Visual narrative of our survey on augmented piano prototypes in the context of learning. We highlight the key steps and findings from our interviews (0), its themes (1-4), evaluation strategies (5) and recommendations (in highlight).}
  \Description{A photo of an augmented piano with annotations and labels of the important theme-related features.}
  \label{fig:teaser}
\end{teaserfigure}
\maketitle

\section{Introduction}
When learning an instrument, several factors and problems need to be considered. These can either be (i)~\textit{physical} (maintaining a proper posture, assuming a height level required for a particular instrument, length of fingers, arms and their flexibility, one-size-fits-all designs of instruments), (ii)~\textit{cognitive} (acquiring, understanding, retaining and applying music knowledge)~\cite{hanna2007new}, (iii)~\textit{psycho-cognitive} (visual - reading the notes and synchronising them with various body parts, auditory - hearing the sound of notes being played, tactile perception and strength of one's press)~\cite{conklin1987piano, simpson1966classification} or (iv)~\textit{affective} (willingness to learn, motivation to practice)~\cite{mccarthy2002music, burns2020using}. 

Using the piano as an example, playing it involves a proper sitting posture and distance from the instrument to ensure optimised movement of the arms and fingers when pressing keys. Playing also requires proper timing and coordination between hands, eyes, and feet (on the pedals). For most novice students, getting used to these motor skills along with the cognitive task of reading music sheet notation can be overwhelming~\cite{highben2004effects}. In addition, the physical characteristics of the instrument (e.g., grand piano vs. electronic keyboard) may introduce extra challenges in terms of accessibility, and comfort with prolonged usage, thereby affecting the student's practice. Practising and ensuring motivation can also be challenging. This is usually addressed by having regular practice sessions with an experienced teacher, while constant validation and progress monitoring help students learn faster, given their perseverance and consistent preparation. 

Despite these issues, the piano remains a popular choice among novice players~\cite{oneillboysgirlsinstrument}. As such, it also attracts  researchers~\cite{delzell1992gender} pushing for computer-supported contributions in the form of augmented interactive surfaces on and around the piano, providing supporting information to the learner. 
Some prototypes~\cite{zaqout2015augmented, goodwin2013key} were used to explore mobile augmented reality (AR), focusing also on hand-tracking by optimising detection and finger recognition~\cite{huang2011piano, liang2016barehanded}. Other works~\cite{xiao2011duet, xiao2014andante} explored varying forms of visualisations to enrich viewing and listening experiences. Given these contributions, it is yet to be investigated how these prototypes support the piano learning process through various types of augmentations. Thus, this paper aims to understand the general space, challenges, and opportunities of piano augmentations aimed at learning. 

We focus our study on two research questions: (i)~What has been augmented in pianos? and (ii) Do these prototypes address issues exposed by piano teachers? To this end, we conducted interviews with ten piano teachers and teachers of piano didactics to investigate the main themes around piano teaching and learning and did a literature review of augmented piano prototypes consisting of~56 papers. We synthesised our qualitative reviews and the insights we learned from our expert interviews into what we present as piano learner-based themes. Next, we explored how~56 prototypes address these themes and the focus of the technology shift throughout the years. We also propose possible future directions in the form of recommendations. Finally, we envision some and challenge our ideas towards an ideal augmented piano prototype that supports these learner-based themes and alleviates the factors and problems mentioned earlier.

\section{Background}
\label{background}
In the following, we give an overview of two key areas. First, we give an overview of the theoretical background of learning an instrument. Second, we introduce the concepts of instrument augmentations.

\subsection{Instrument Learning Methods, Theories, and Approaches}
\label{sec:background-methods}
The usual approach to music teaching is that an experienced musician, the teacher, passes the knowledge to a novice student. These sessions are complemented by personal practice to acquire the skills needed for the next session with a teacher. There are four primary methods that teachers can integrate when teaching music and instruments: the Kodály, Orff Schulwerk, Dalcroze, and the Suzuki method. The Kodály method~\cite{choksy1974kodaly} uses hand signals, shorthand notation, and rhythm verbalisation to prepare students to have a solid grasp of music theory and musical notation in both verbal and written forms. The Orff Schulwerk~\cite{shamrock1997orff} approach introduces pupils to the rudimentary forms of music at an early stage. It combines instruments, singing, movement, and speech to develop children’s innate musical abilities. As such, it fosters self-discovery and improvisation, which moves away from repetitive mechanical drills. The Dalcroze method~\cite{mead1994dalcroze} is considered the rhythm gymnastics approach to music learning. Students are instructed to emphasise physical awareness and engage with music involving all their senses and kinesthetic skills. Finally, the Suzuki method~\cite{peak1998suzuki} draws inspiration from music learning, similar to the approaches to learning one's native language. It describes an ideal environment that considers high-quality music samples, rote (mechanical) training, and repetition. Apart from these four internationally-renown methods, several others have been influential to music learning, such as Gordon, Reggio Emilia~\cite{gordon2003music, edwards1998introduction, carabo1969sensory, thomas1970manhattanville}, Russian Piano method~\cite{chang1994russian} and others. 

All these methods 
consider psycho-motor, cognitive, and affective domains. The psycho-motor domain in music education focuses on the development of movements and responses that the body performs based on visual, auditory, and tactile stimuli~\cite{simpson1966classification}. The cognitive domain describes the process of how a student acquires, retains, and applies knowledge of essential concepts and foundations in music, which leads to more effective music learning experience~\cite{hanna2007new}. This domain supports various music-making phases, such as performing, improvising, composing, arranging, and conducting. Having a concrete foundation in this domain ensures encompassing the development of the student~\cite{westerlund2003reconsidering}. While the affective domain covers students' willingness to receive, reflect and share what they have acquired during the music learning process, as well as music appreciation and sensitivity as a response to the emergence of music education as an aesthetic learning process~\cite{mccarthy2002music}.
Music teachers are using a selected method (or methods) to teach music and musical instruments, but all need to take into consideration all domains to deliver their instructions effectively~\cite{burns2020using}. 

Beyond the scope of music learning methods, other general learning frameworks have also helped to describe and understand the music learning process. Among these are Social Learning Theory~(SLT)~\cite{waldron2009exploring, gordon2011roots, maisto1999social}, Experiential Learning~(ExL)~\cite{webster2011construction, russell2013mission, kolb2014experiential}, and Active Learning~(AL)~\cite{scott2011contemplating, michael2003active}. Since learning a musical instrument such as the piano requires a tactile perception, familiarising with the equipment is also important. As some experts claim, learning a musical instrument is like learning how to ride a bicycle or play tennis~\cite{kastner2014exploring}. Students improve their skills by actually \textit{doing} them repeatedly until they master them~\cite{stryker1997content}. 

In recent years, gamification has been introduced as a strategic attempt to improve existing systems, thereby motivating and engaging users~\cite{blohm2013gamification}. Game-design elements and principles can also be used in non-game contexts~\cite{deterding2011game, robson2015all} such as learning and playing the piano. Furthermore, gamification's scope also includes making existing tasks feel more like a game. It has been observed that introducing game-based elements helps students learn, which in turn improves the flow, engagement, and immersion~\cite{hamari2016challenging}. Similarly, the incorporation of game-based elements has effects on the cognitive and psycho-cognitive domains of learning, as seen in other experiments~\cite{yang2012building}. Gamification focuses on maximising engagement and capturing the student's interest, thereby contributing to the affective domain in learning as well. Specifically, learning is facilitated by incorporating some (but not limited to the following) elements: (i) progress mechanics, (ii) narrative and characters, (iii) player control, (iv) immediate feedback, (v) learning by scaffolding and many others~\cite{toda2019analysing, toda2019gamify}.  

\subsection{Instrument Augmentations}

In the last couple of decades, digital technology interventions have been introduced to musical instruments to improve one or several of their features and properties~\cite{magnusson2007acoustic} such as portability, automatic tuning, immunity to harmful conditions like humidity, recording capabilities, volume control, and logging of student input, to name a few. For sure, this depends on the instrument itself. Nevertheless, all classes of instruments can be equipped with auxiliary hardware, peripherals, and sensors to improve the sound quality~\cite{mcpherson2011multidimensional} or to track user motion while playing them~\cite{hadjakos2012pianist}. We have seen digital augmentations of string~\cite{overholt2005overtone}, wind~\cite{silva2008interaction}, and other instruments~\cite{turchet2018some, newton2011examining} as well as the piano. 

As electronic pianos are commercially available and given their popularity as a learning instrument of choice~\cite{sloboda1992transitions}, augmentations in their design have been of interest to both the academic community and the industry. Augmentations in the piano have either been done to emulate the grand piano as an instrument or to introduce newer experiences in either learning~\cite{weing2013piano}, playing~\cite{xiao2011duet}, improvising~\cite{karolus2020hit} and performing~\cite{xiao2013mirrorfugue} and in general supporting various of above presented learning theories, methods and approaches with the aim to support students during the learning process (e.g., preview the playback, get feedback based on a recording, gamifying the playing experience, etc.) 

Even though several augmented piano prototypes have been designed with implicit or explicitly-defined gamified, game-based and other learning elements, we argue that their effects have not been well-understood. Some prototypes have applied SLT or ExL as the guiding principle in their instrument design. What has yet to be explored is whether these prototypes align with the piano teachers' requirements during their teaching process and if they can help learners or improve the learning process. 

\section{Interviews with Experts}
To better understand the teaching and learning piano process as well as problems, issues, and needs related to these processes, we first conducted interviews with piano teachers. The goal of the interviews was to learn more about the piano learning process and difficulties experts encounter in teaching, as well as their insights about musical instrument augmentation.

\subsection{Recruitment and Interview Protocol}
\label{interviews}
In total, we recruited ten piano teachers from various music schools, different stages of their careers, and different music teaching systems (private or public schools) from the Philippines, Slovenia, Switzerland, and United States, see \autoref{tab:expertinfo}.
 
In the semi-structured interviews, we inquired about the following questions and asked follow-up questions if needed:
\begin{enumerate}
\item What specific method, framework, or approach do experts use when teaching? 
\item What are the problems and good experiences they faced or their pupils face when learning the piano?
\item What are the problems and good experiences they face when teaching the piano?
\item What specific physical factors or nuances do they focus on to lessen or mitigate errors? 
\item What concepts/topics in the music curriculum do they focus on more and/or which concepts/topics do they believe need more focus? 
\item What is their level of acceptance/openness to the use of digital technology and augmentations when teaching the piano?
\end{enumerate}

Additionally, we presented the experts with video demonstrations of digitally augmented piano prototypes (specifically the works of ~\cite{weing2013piano, takegawa2012piano, xiao2013mirrorfugue}). Based on that, we solicited their opinions on the value of various features introduced by these augmentations. The interviews lasted from 30 to 45 minutes. 
All interviews were conducted in English, recorded, and transcribed. Next, we reviewed the transcripts and conducted a thematic analysis as in~\cite{guest2011applied, braun2012thematic}. We looked for common problems, wishful thoughts, methods employed, and piano teachers' ideas about digital augmentations. 

\begin{table}[t]
\centering
\rowcolors{2}{gray!15}{white}
\caption{Piano Expert Demographics and Overview. *E3 and E4 shared that they use their own inspiration from previous existing methods learned.}
\label{tab:expertinfo}
\begin{tabular}{llccl} \toprule
\rowcolor{white}
\textbf{Expert \#}  & \textbf{Sex}       & \textbf{Years Playing} & \textbf{Years Teaching}   & \textbf{Method(s)}  \\ \midrule
E1      & M     & 32            & 17              & Russian piano method   \\
E2      & M     & 36            & 17              & Russian piano method   \\
E3      & M     & 31            & 18              & personal piano method*   \\
E4      & M     & 19            & 7               & personal piano method* \\
E5      & F     & 21            & 21              & Russian piano method    \\
E6      & F     & 41            & 30              & Eclectic, Kodaly method   \\
E7      & F     & 49            & 38              & Eclectic, Dalcroze method  \\
E8      & M     & 17            & 7               & Eclectic, classical method   \\
E9      & M     & 56            & 40              & Suzuki and Kodaly method  \\
E10     & M     & 25            & 19               & Eclectic method \\
\toprule
\rowcolor{white}
\begin{math}n=10\end{math}   & \textbf{Mean}          & \textbf{SD}          & \textbf{Range}  & \textbf{n [\%]}\\ \midrule
      Years Playing          & 32.7         & 13.0      & 17-56  & \\ 
      Years Teaching         & 21.4         & 11.4      & 7-40   & \\ 
     Male-Female ratio       &              &          &        & 7:3 [70\%:30\%] \\
     \bottomrule
\end{tabular}
\end{table}

\subsection{Insights and Feedback from Experts}

\begin{figure}
\centering
  \includegraphics[width=\linewidth]{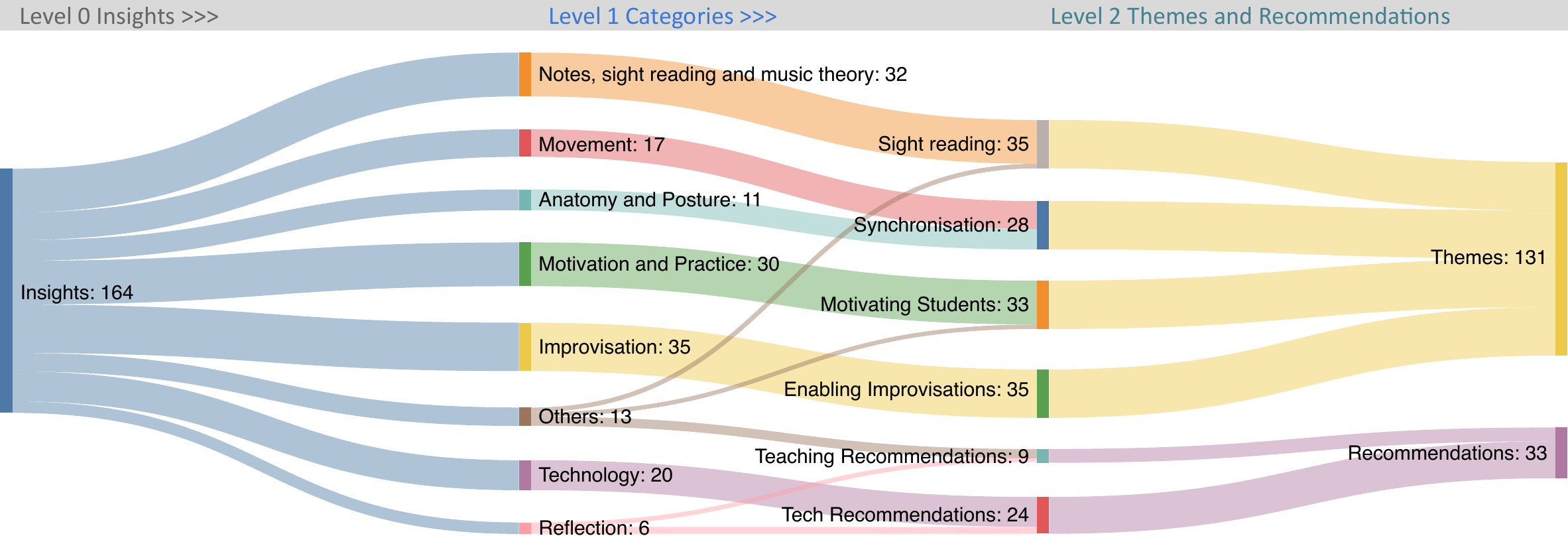}
  \caption{Sankey diagram depicting the coding process. We show Level 0 (Insights) to Level 1 (Categories), Level 2 (Subcategories), and Final Level (themes and recommendations). The main branches in the final level are used to guide the categorisation of the papers in the corpus. The recommendations were also synthesised with the data analysed from the papers to develop future explorations on piano augmentation.}
  \Description{A Sankey diagram showing the coding process and its categories. There are four levels.}
  \label{fig:sankey-theme}
\end{figure}

The overview of transcripts' analysis and coding is illustrated in \autoref{fig:sankey-theme}. In the first stage (level 0), the transcripts were coded into 164 insights. Using the affinity diagram technique, we came up with 8 initial Categories (level 1) based on how related they are to each other. These categories are (1-A) Notes, sight-reading and music theory, (1-B) Movement, (1-C) Anatomy and posture, (1-D) Motivation and practice, (1-E) Improvisation, (1-F) Technology, (1-G) Reflection, and (1-H) Others. Next, we draw inspiration from the domains (psycho-cognitive, cognitive, physical, and affective) described in \autoref{sec:background-methods} and conceptualised four themes (2A--2D, level 2): (2-A) Sight-reading (psycho-cognitive), (2-B) Synchronisation (physical), (2-C) Motivating students (affective), (2-D) Enabling Improvisations (cognitive). The other categories that we thought did not belong to any of these four themes were then categorised as recommendations (2E--2F): (2-E) Teaching and (2-F) Tech recommendations. The complete affinity diagram with the individual insights from the experts is included in the supplementary material.

\subsubsection{On Teaching and Music Curricula}
All our experts (E1--E10) use sight-reading as an essential skill when learning the piano. E4 thinks that younger people learn music theory and sight-reading faster than their older counterparts. E10 emphasises this by saying \textit{``there is no basis of comparison; you need to read the notes if you want to learn.''} They argue that learning sight-reading along with the appropriate music theory (E2) and historical interpretation of the musical piece (E3, E6, E9) helps the students appreciate the piece, which in turn may improve the learning experience. Based on their experience (E1, E3, E6), using visual aids and animations (e.g., animated videos of frogs jumping on a lotus leaf to denote beat and timing) also helped their students learn other concepts such as time signature and rhythm. 

E7 and E9 mentioned that there are usually two types of learners - those who are aurally gifted (those who can learn and play by ear) and those who are not. Students who cannot learn or play by ear need sight-reading, and those who can play by ear may rely more on auditory feedback. E8 notes that students learn music theory first so they can later on \textit{``break these rules''}, especially when learning improvisation. When it comes to teaching sight-reading, our experts vary in some of their approaches. E2, E3, and E10 teach the basics of note reading first as fast as possible, contrary to how E9 focuses on the problematic parts of sight-reading (measures, chord combinations) and lets the student learn the easy parts on their own. E2 and E6 believe that the current system needs a better approach to teaching sight-reading since many still struggle in this area.

Along with sight-reading and music theory, we note that experts highly regard physical anatomy concerning proper movement (E1, E2, E3, and E8). \textit{``There are statistics on people with posture problems and injuries, so this area needs focus also''} as shared by E1. \textit{``[Body] technique is important in the long run, but this depends on the focus of the teacher and goal of the student''} E5 says. To address this, they sometimes borrow methods from dance following Dalcroze principles (E9) or using regular exercise routines (E3). This potentially helps the body, especially the fingers and arms, warm up in preparation for more extended practice periods. When some students struggle because of these anatomical features, E6 prepares specialised routines to help them while E9 focuses on different body parts (right hand first, then left hand next ...). In addition, E6 mentions the VARK framework in music pedagogy, which stands for Visual, Aural, Reading, and Kinesthetic approaches to learning the piano. E3 and E6 also incorporate a learner-centred approach to customise and adjust their methods based on where the students strive or struggle. E1, E2, and E3 record their sessions and use them as a reflective approach to improve their subsequent sessions. E6 and E7 use a journaling/annotative technique alternatively. E5 uses a piano-teaching chart to monitor the practice sessions of their students.

All of our respondents have emphasised that practice is crucial for learning the piano. While younger [learners] quickly learn music theory, they struggle more in the practice department (E7). \textit{``Talent is not enough, you need to practice and work hard''} added by E9. While dedicated practice time is incorporated into the delivery of their lessons, some experts also noted the importance of immediate feedback and positive reinforcement during their sessions (E4, E6, E7, E8, and E10). Introducing variety in their assigned music pieces to learn has also helped the students' experience (E2 and E4), while E3 and E5 think repetition is key to mastering a specific piece. E2, E4, and E10 also believe that other approaches (such as game-based) might be able to motivate students and help them in their struggle to practice. 

\subsubsection{On Improvisation}
We specifically inquired about expert methods, frameworks, pains, and experiences during our interviews when teaching the piano. Interestingly, during our interviews, improvisation emerged as a notable topic for discussion (only E4 did not mention or share insights about it). To quote E3, \textit{``there are [always] 50 [and] more rules in music to follow, and I'm trying to find a hack''} as they referred to improvisation. E2, E3, E9, and E10 believe that improvisation leans on more creative aspects of music as it encourages \textit{``playing by heart''} (E2) or gives focus on having \textit{``your own interpretation of the song''} (E3) as it \textit{``adds artistry to one's talent''} (E9). The notion of improvisation involves a \textit{``listen, imitate then try''} approach to music (E1), where even though you need an understanding of the music rules, the exercise gives you a certain level of freedom in terms of interpretation (E5).

Despite all of this, our experts believe that improvisation needs extra focus in the piano learning process. E1 thinks that improvisation is no longer common in Western systems, while E2 believes that improvisation is challenging to assess and measure, underrated and hard to teach (E8). E6 believes that improvisation requires advanced aural ability and feedback, while E9 mentioned that students struggle in learning and are confused when they try to improvise. The experts also believe that teaching improvisation may benefit from approaches such as the use of cadenzas\footnote{In music, a cadenza is, generically, an improvised or written-out ornamental passage played or sung by a soloist or soloists, usually in a ``free''   rhythmic style, and often allowing virtuosic display. Source: \url{https://en.wikipedia.org/wiki/Cadenza}} for references (E8), giving immediate feedback in terms of sound quality (E9) and possibly with the use of technology-aided visualisations.

\subsubsection{On Technology and Digital Augmentation}
Our experts have been teaching piano for an average of 21.4 years. This means that they have spent a significant amount of time teaching the piano before the covid-19 era, and at present, they are forced to teach remotely using technology. In our interviews, we explored their openness to technology, their tools, and their wishful thoughts on the technology that may help teach piano.

All experts argue that teachers should still be the primary providers in the piano learning process instead of being replaced by digital apps and tools. E1 thinks that while technology is their weakest point, they have expressed their openness to using technology when teaching. Other experts provided several points on how technology can aid them in piano teaching. E1--E3, and E7 record their students' sessions using video and multimedia capture devices. \textit{``I wish I could annotate my comments on their performance''} as shared by E7. E5 believes that the best way to learn the piano using tech should be [almost] similar to how it is taught traditionally. A metronome-related app could be an ideal concept, as shared by E4.
Similarly, E5 resonates with this idea but adding \textit{``I should know the tempo element in these apps''} possibly referring to the visuals commonly used in piano roll apps. Concepts in dynamics (such as crescendo and descrendo) are not typical and evident in these apps (E6). Beyond apps and visualisations, software features should be able to encourage and motivate students to practice outside teaching sessions and foster collaboration. E3 and E6 use available tools to share notes, and \textit{``musical repositories''} consist of different materials and lessons.

During the interviews, we showed our experts videos and previews of augmented piano apps (such as the works of~\cite{rogers2014piano, takegawa2012piano}) and solicited their opinion about these prototypes. \textit{``Combine traditional notes and falling bars and you get the best of both worlds''} stated by E3. E7--E10 resonates with this sentiment by emphasising including or mapping musical notation somewhere, so students do not neglect sight-reading. They think they can remain fluent with notes if they have prior training when using these augmented pianos. From a different angle, E5 believes that when you teach students using piano rolls alone, you only teach them how to play a song. They add \textit{``if you already know how to play [by sight-reading], using the piano roll might be hard for you since you need to unlearn the notes you have learned prior.''} E3 believes that learning the piano roll alone would be \textit{``static.''} E6 thinks that some students will benefit from the piano roll if they have the excellent aural ability. As these opinions may vary, it is clear that our experts would recommend that piano roll visualisations should be mapped with an additional level akin to that of classical musical notation.

\subsubsection{Other Insights on the Piano Learning Process}
Our experts also shared some other insights into the context of learning the piano. E10 believes that tactile feedback is also essential when learning the piano. Similarly, E6 and E8 think the same, which is why they also consider the type of interface (classical piano, organ, or a Clavinova) their students use when learning. E8 shares that students also lack training with sound feedback - they need to know what a note should sound. This helps significantly in helping students learn improvisation. E9 supports this claim from experience. They have observed that students grasp a piece listening by listening to it by ear and then integrating the Suzuki method to integrate sound, movement, and music notation easily. Some experts (E7, E10) also believe that the human touch should be incorporated when teaching the piano since some students learn by heart (E3). In the experience of E10, they customise their lessons based on their students' personalities. We believe these comments and insights warrant further investigation and may be worth exploring in the future. 

\subsection{Synthesis of Findings and Deriving of Themes}
\begin{figure*}[t]
    \centering
    \includegraphics[width=\textwidth]{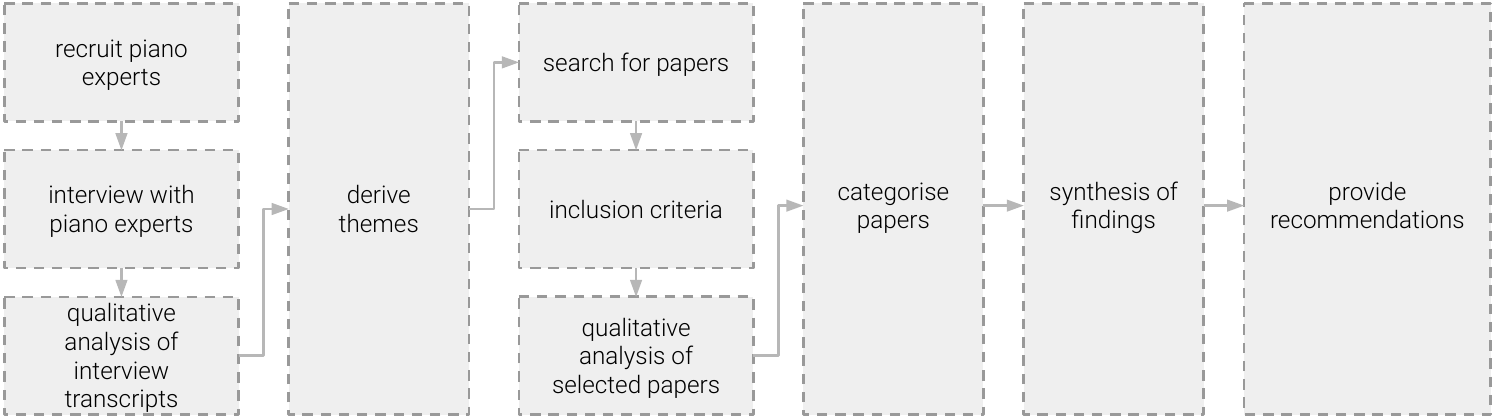}
    \caption{Overview of steps in this paper. PRISMA principles were incorporated in the qualitative review. Thematic analyses were done in the interviews and synthesis section.}
    \label{fig:method}
\end{figure*}

The coding resulted in four (4) recurring learner-based themes concerning piano teaching and learning: (i)~synchronising movement and posture between various body parts, (ii)~improving sight-reading skills, (iii)~motivating students to practice, and (iv)~encouraging improvisation. These will be discussed further in \autoref{sec:themes}. These four themes were then used to categorise the prototypes found in literature as explained in the next section. \autoref{fig:method} shows different steps involved in this research. It includes two branches whose results were then synthesised to come up with recommendations for our research question. 

\section{Systematic Review of Prototypes}
\begin{figure*}[t]
    \centering
    \includegraphics[width=\textwidth]{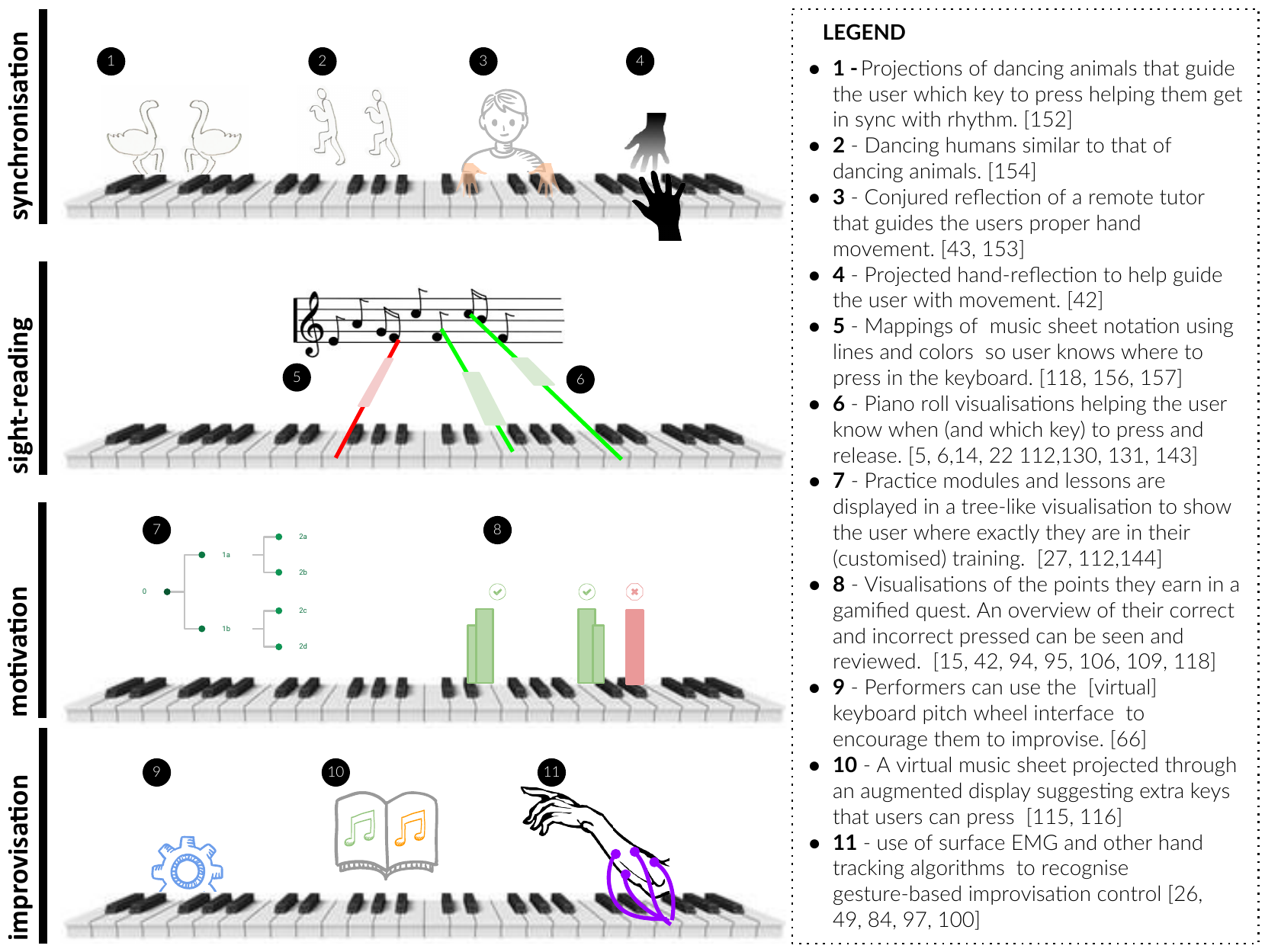}
    \caption{Approaches to Piano Augmentation based on themes.}
    \label{fig:augmentations}
\end{figure*}

To understand the space of augmented piano prototypes and their support for learning experiences, we followed the \textit{Preferred Reporting Items for Systematic Reviews and Meta-Analyses} (\textbf{PRISMA}) technique~\cite{moher2009preferred}. Approaches from similar reviews on related topics such as augmented reality in education~\cite{santos2013augmented, blattgerste2019augmented}, mobile augmented reality applications~\cite{schneegass2016mobile}, mixed reality~\cite{speicher2019what}, learning and information management~\cite{kljun2015transference}, as well as reviews in music systems and prototypes~\cite{mcpherson2015buttons, delgado2011state} were also followed. A visual abstract of these augmentations is displayed in \autoref{fig:augmentations}. 

\begin{table*}[t]
\caption{The corpus of the papers on digitally augmented piano prototypes and the learner-based themes they cover, sorted by the year of publication. Legend: \textit{\#} = number of citations; \textit{Synch} = Synchronising movement and posture between various body parts; \textit{Sight} = Improving sight reading skills; \textit{Motiv} = Ensuring motivation of students; \textit{Impro} = Encouraging improvisation.}
\label{tab:piano_themes}
\rowcolors{2}{gray!15}{white}
\resizebox{\textwidth}{!}{%
\begin{tabular}{llrr*{4}{c@{\hskip4pt}}l} \toprule
\rowcolor{white}
& \textbf{Author(s)}                    & \textbf{Year} & \textbf{\#}    & \textbf{Synch} & \textbf{Sight} & \textbf{Motiv} & \textbf{Impro} & \textbf{Additional info} \\ \midrule
P01            &~\citet{barakonyi2005augmented}        & 2005          & 58             & \yes &           &           & \yes & suggests chords and harmonising melodies\\
P02            &~\citet{schmalstieg2007experiences}    & 2007          & 298            &           & \yes & \yes &           & rewards users with artefacts   \\
P03            &~\citet{correa2009computer}            & 2009          & 75             & \yes &           &           &           & considers hand, arm, leg mobility\\
P04            &~\citet{mcpherson2010toward}           & 2010          & 3              &           &           &           & \yes & gestures for improvising \\
P05            &~\citet{mcpherson2010augmenting}       & 2010          & 55             &           &           &           & \yes & cont. of P04   \\
P06            &~\citet{zhang2010affordable}           & 2010          & 25             & \yes &           &           &           & repetitive motion for use of glove;  \\
P07            &~\citet{mcpherson2011multidimensional} & 2011          & 31             & \yes &           &           &           & identifiers improper actions by user \\
P08            &~\citet{huang2011piano}                & 2011          & 60             & \yes &           &           &           & detailed tracking of fingers\\
P09            &~\citet{xiao2010mirrorfugue}           & 2011          & 42             & \yes &           & \yes &           & piano collaboration space\\
P10            &~\citet{xiao2011duet}                  & 2011          & 8              & \yes &           & \yes &           & watch recording for self reflection \\
P11            &~\citet{hadjakos2012pianist}           & 2012          & 43             & \yes &           &           &           & head, shoulder, arm detection, depth sensing \\
P12            &~\citet{nicolls2012gesturally}         & 2012          & 12             &           &           &           & \yes & gesture controlled live impro, motif grabbing \\
P13            &~\citet{yang2012augmented}             & 2012          & 23             &           & \yes &           & \yes & guided note viz, gesture to adjust quality \\
P14            &~\citet{p2012problem}                  & 2012          & 47             & \yes &           & \yes &           & detects shallow or deep presses \\
P15            &~\citet{takegawa2012piano}             & 2012          & 33             & \yes & \yes &           &           & notation is mapped to viz \\
P16            &~\citet{mcpherson2013space}            & 2013          & 23             & \yes &           &           &           & tracks intentional and unintentional presses \\
P17            &~\citet{yang2013visual}                & 2013          & 9              &           & \yes &           & \yes & same as P13 \\
P18            &~\citet{mcpherson2013portable}         & 2013          & 26             & \yes &           &           &           & captures continuous key motion \\
P19            &~\citet{chow2013music}                 & 2013          & 64             &           & \yes & \yes &           & piano roll and notation mapping \\
P20            &~\citet{weing2013piano}                & 2013          & 37             & \yes &           & \yes &           & specific finger mapping, has practice mode \\
P21            &~\citet{chouvatut2013virtual}          & 2013          & 8              &           & \yes &           &           & notes are accompanied by sound feedback\\
P22            &~\citet{oka2013marker}                 & 2013          & 30             & \yes &           &           &           & recognises multiple fingering \\
P23            &~\citet{xiao2013mirrorfugue}           & 2013          & 17             & \yes &           &           & \yes & conjured projection to review performance\\
P24            &~\citet{goodwin2013key}                & 2013          & 15             & \yes &           &           &           & observing of hands from a monitor\\
P25            &~\citet{zandt2014piaf}                 & 2014          & 12              & \yes &           &           &           & monitors synchronisation of presses\\
P26            &~\citet{xiao2014andante}               & 2014          & 36             & \yes &           &           &           & uses body rhythm to guide movement\\
P27            &~\citet{raymaekers2014game}            & 2014          & 26             & \yes &           & \yes &           & game based incentives for practise\\
P28            &~\citet{de2014infrared}                & 2014          & 9              & \yes &           &           &           & teaches proper finger movement \\
P29            &~\citet{chiang2015oncall}              & 2015          & 3              &           & \yes &           &           & teaches which key to press based on sound \\
P30            &~\citet{dahlstedt2015mapping}          & 2015          & 4              &           &           &           & \yes & focused on gesture impro, harmonics\\
P31            &~\citet{zaqout2015augmented}           & 2015          & 1              & \yes &           &           &           & mobile based keystroke gesture detection\\
P32            &~\citet{fernandez2016piano}            & 2016          & 9              &           &           & \yes &           & fun way of practising, agent points out errors\\
P33            &~\citet{liang2016barehanded}           & 2016          & 27             & \yes &           &           &           & detects soothing rhythm movements \\
P34            &~\citet{ogata2017keyboard}             & 2017          & 1              &           &           &           & \yes & allows users to use other hand for gesture improv\\
P35            &~\citet{liang2017piano}                & 2017          & 9              & \yes &           &           &           & tracks foot pedalling \\
P36            &~\citet{hackl2017holokeys}             & 2017          & 13              & \yes &           &           &           & HMD viz to guide key pressing \\
P37            &~\citet{das2017music}                  & 2017          & 12              & \yes &           & \yes & \yes & has jazz, blues rock; has lesson builder\\
P38            &~\citet{rogers2014piano}               & 2017          & 50             & \yes & \yes & \yes &           & colours to support correct fingering; practice modes\\
P39            &~\citet{birhanu2017keynvision}         & 2017          & 5              & \yes & \yes &           &           & Hololens; maps keys to notes; detects posture \\
P40            &~\citet{trujano2018arpiano}            & 2018          & 8              & \yes &           &           &           & teaches length of notes when pressing \\
P41            &~\citet{sun2018mr}                     & 2018          & 4              & \yes & \yes & \yes &           & practise mode \\
P42            &~\citet{pan2018pilot}                  & 2018          & 2              & \yes &           & \yes &           & paired user play; shows note information\\
P43            &~\citet{granieri2019reach}             & 2019          & 3              &           &           &           & \yes & hand tracking of gesture impro \\
P44            &~\citet{zeng2019funpianoar}            & 2019          & 9              & \yes  &          &           &            & pairwise collaboration and key finger harmony\\
P45            &~\citet{molloy2019mixed}               & 2019          & 8              & \yes  &          & \yes &           & measures cognitive load, gamification of notation\\
P46            &~\citet{cai2019designa}                & 2019          & 3              & \yes  &          & \yes &           & competition as a motivating component\\
P47            &~\citet{gerry2019adept}                & 2019          & 3              & \yes  &          &           &           & teaches improved motor sensing in performances\\
P48            &~\citet{cai2019designb}                & 2019          & 2              & \yes  & \yes&           &       & key presses with note mapping \\
P49            &~\citet{sandnes2019enhanced}           & 2019          & 3              & \yes  & \yes &          & \yes & jazz chords; chord info mapped on keypress\\
P50            &~\citet{santiniaugmented}              & 2020          & 2              &           & \yes &           & \yes & virtual note sheet that moves with keypress\\
P51            &~\citet{karolus2020hit}                & 2020          & 10              &           &           &           & \yes & EMG to measure user flow in impro\\
P52            &~\citet{moro2020performer}             & 2020          & 0              & \yes &           &           &\yes & continuous key sensing and gestural technique of pianist; \\
P53            &~\citet{molero2021novel}                & 2021          & 7              &           &          & \yes &           & visualise music concepts (using metaphors), gamified\\
P54            &~\citet{stanbury2021holokeys}           & 2021          & 0              & \yes &          &           &           & instructors remotely teach students, live demo of keyboard view\\
P55            &~\citet{guo2021hand}                    & 2021          & 3              & \yes & \yes &\yes &            & 3D animation of natural hand motion, piano roll hint\\
P56            &~\citet{kilian2021em}                   & 2021          & 1              &           &          &           & \yes & interface to replace piano pitch wheel for improv\\ \midrule
\rowcolor{white}
               &                  &               &                & 39 (80\%) & 14 (27\%) & 16 (31\%) & 16 (31\%) & \\ \bottomrule
\end{tabular}%
}
\end{table*}

\subsection{Search for Prototypes}
We conducted a literature search in Google Scholar, ACM Digital Library, and other digital libraries. Scientific articles ranging from January 2005 until December 2021 were considered in the search. We used all combination of the following keywords \{\texttt{``augmented reality'', ``AR'', ``augmented''}\} with these \{\texttt{``piano'', ``keyboard'', ``guitar'', ``drum'', ``violin'', ``flute''}\}, resulting in search terms such as \{\texttt{``augmented reality piano''}\}. We included the words \{\texttt{``guitar'', ``drum'', ``violin''} and \texttt{``flute''}\} to understand how many contributions have been published for a particular instrument and to confirm if the piano is the most augmented one. We considered only scientific papers written in English and ended up with a total of 1,307 search results. Among found, 635 papers focused on piano and keyboard (48.6\% of the results), 260 on violin (19.9\% of the results), 206 on guitar (15.8\% of the results), 174 on the drum (13.3\% of the results), and 32 on flute (2.4\% of the results). Based on these numbers alone, we can observe that the piano is a popular choice for augmentation. 

\subsection{Inclusion Criteria}

We defined selection criteria to identify a subset of papers that fit our research questions' context. We included papers about: 
\begin{enumerate}
    \item a physical prototype involving a piano or keyboard, which has been augmented or equipped with auxiliary hardware (e.g., sensors, cameras, projectors) and software (e.g., features, lighting, modules) OR
    \item a virtual piano prototype that has been implemented in a mixed/virtual/augmented reality environment;
    \item a usable piano prototype, and not an abstract or a schematic concept only;
    \item the augmentation is intended towards solving a specific piano user problem; and
    \item an augmentation that uses digital technology beyond what is already commercially available (e.g., an electronic keyboard is technically an augmented classical piano, but since these are commercially available, we look at augmentations beyond their features). 
\end{enumerate}
Following the said criteria, we narrowed down the set of 595 piano and keyboard papers to 56 papers. 

\subsection{Qualitative Analysis of Prototypes} 
\label{qualitative-analysis}
Following the PRISMA approach, we collected qualitative data on the included papers. We extracted categorical information from the papers, such as year of publication, the number of citations, the technology used, type of augmentation, the number of participants in a user study, metrics or constructs measured (e.g., satisfaction, user experience, immersion, etc.), experiment treatment, tools, and other descriptive information. This information was tabulated and stored for the succeeding steps. It is important to note that the aim of this review is \textit{not} to define a digitally augmented piano but rather to collect as many examples of pianos that have been digitally augmented.  

We then coded the extracted information from each of the 56 papers selected in the process (as in~\cite{williams2019art}) to see whether there were patterns or similarities in how the prototypes presented were designed. Initially, the four~(4) domains presented in \autoref{background} (physical, cognitive, psycho-cognitive, and affective) guided us in this conceptualising the themes, which in turn was used as the basis in organising our corpus of papers. We categorise the prototypes based on learner-based themes for the piano (see \autoref{tab:piano_themes}). 

\section{Learner-based Themes}
\label{sec:themes}
This section discusses the four (4) learner-based themes we have derived from the interviews. In addition, we present 56 augmented piano prototypes and categorised them on how they subscribed to these learner-based themes. 

\subsection{Synchronising Movement and Posture Between Various Body Parts}
\label{learnin_themes_posture}
Playing the piano involves several motor skills, such as maintaining proper posture, using both hands and all fingers to press keys, and coordinating everything with the foot movement (pedalling). These skills fall under a the physical domain used when learning a particular instrument~\cite{goebl2009synchronization}. In retrospect, most traditional music teaching frameworks (see \autoref{background}) introduced various methods (in the form of mechanically repetitive exercises or kinesthetic activities) that help novice students to develop the required skills to play an instrument. However, the difficulty in synchronising movement goes beyond motor skills and is also related to understanding basic music rudiments and theories. Synchronising movement and posture and an almost natural flow to the piece's rhythm is essential in ensuring sound quality. Here, we highlight augmented piano prototypes that captured, tracked body movement (key presses, hand movement, body posture) and provided feedback to the user.

Earlier prototypes began with the goal of detecting key presses and giving immediate or post-performance feedback to the user~\cite{barakonyi2005augmented, mcpherson2011multidimensional, huang2011piano, sun2018mr} (displaying errors in timing, colouring which press was correct). Some prototypes focused on synchronising finger-hand-arm movement with the ultimate goal of improving the moving flow or enabling faster movement among users~\cite{correa2009computer, zhang2010affordable} - a type of feedback that learners will usually receive when being observed by a piano teacher. With the help of these augmentations, piano teaching systems can show the learner if there are areas for improvement in their movement or posture almost immediately. Some augmented piano prototypes distinguished between deep and shallow presses~\cite{p2012problem}, or intentional and unintentional presses~\cite{mcpherson2013space}. These prototypes were meant to develop systems to help recognise ``natural'' flow of movement of the body parts involved.  

There were 39 papers in total that discussed the synchronisation of body parts as (one of) their focus (foci). Out of these, only one covered foot pedalling~\cite{liang2017piano} while the rest focused on monitoring and giving feedback on finger presses alone~\cite{oka2013marker, weing2013piano, mcpherson2013portable, raymaekers2014game, molloy2019mixed, de2014infrared, hackl2017holokeys, rogers2014piano, trujano2018arpiano, pan2018pilot, sandnes2019enhanced}, finger-and-hand movement~\cite{zaqout2015augmented, liang2016barehanded, cai2019designb}, finger-hand-and-arm~\cite{correa2009computer} and finger-hand-arm-and-shoulder~\cite{hadjakos2012pianist} movement as well as whole body movement~\cite{birhanu2017keynvision}. 

The MirrorFugue series~\cite{xiao2010mirrorfugue, xiao2011duet, xiao2013mirrorfugue, xiao2014andante} along with some other prototypes~\cite{goodwin2013key, zandt2014piaf, stanbury2021holokeys} tackled the problem of synchronisation of body movement with the use of self-reflection. In this technique, the researchers used various visualisations (showing a remote tutor or players' body reflection) projected and seen by the user, thereby allowing them to watch, review, and reflect on their own movement. The technique acts as indirect feedback that enables users to learn from their mistakes and improve on their movements. Besides self-reflection, some prototypes utilised real-time visualisations to help learners improve their movement and synchronisation. For example. the Augmented Design to Embody a Piano Teacher (ADEPT)~\cite{gerry2019adept} prototype provided a virtual teacher (or a tutor in the case of the prototypes by~\cite{cai2019designa, zeng2019funpianoar, moro2020performer, guo2021hand}) visible through an AR head-mounted display in front of the student. At the same time, the virtual hands of the teacher were visible on top of the physical keyboard, together with blue highlights on the currently pressed keys. \citet{fernandez2016piano} also tracked the keys on the keyboard, highlighted the keys needed to be pressed, and entertained players with \textit{anime}-inspired agents to teach piano. The prototype also featured a piano roll visualisation to guide the student. 

Sitting posture has not been explored yet in the context of digitally augmented piano prototypes. However, various technology-supported monitoring of a proper sitting posture exist and have used computer vision techniques with camera(s) either in front (e.g.,~\citet{mu2010sitting}) or on the side of users (e.g.,~\citet{zorvc2019preprivcljive}), as well as sensors either on the body (e.g.,~\citet{dunne2007system}), chair (e.g.,~\citet{tan2001sensing}), or clothes (e.g.,~\citet{mattmann2006design}). Foot dynamics while playing the piano is usually not in the users' sight and is, together with monitoring sitting posture, least explored in the context of augmented piano prototypes. 

\subsection{Improving Sight Reading Skills}

A prior understanding of music notation is necessary for students to play an instrument effectively. Ideally, a user can sit down in front of a piano, read the music sheet notation, and play the piano following the said notation. This skill is referred to as \textit{sight-reading}, which includes reading the notes and synchronising them with various parts of the body. While the previous theme focuses on the movement aspect (thus the physical domain in piano learning), sight-reading takes the skill to a different level, which also uses the sense of vision and cognitive processes. Therefore, sight-reading skill falls under the domain of a psycho-cognitive task~\cite{wristen2005cognition}. This skill is also challenging and not quickly learned by piano learners. 

Several papers addressed the difficulty in sight-reading by incorporating visualisations. These prototypes omitted the music sheet notation entirely, intending to lessen the learning curve in sight-reading~\cite{rogers2014piano}. They do this by overlaying moving piano roll visualisations, making them more straightforward~\cite{walder2016modelling}, thereby teaching players which key to press, when to press and when to release/hold it. The authors of these prototypes believe that the complexity of the music sheet notation (especially in more advanced musical pieces) adds to the cognitive load that overwhelms piano learners. In these visualisations, information usually denoted in music sheets is abstracted in coloured bars that may be translated into finer details such as the length of a beat, trill, staccato, and many others. In effect, these prototypes may teach users to play a piece or a song \textit{but not necessarily to play the piano} - with sight-reading as a skill. The interviewed teachers often do not agree with such an approach since they believe this does not use all the skills that piano learners acquire when they traditionally learn the instrument. According to our interview findings, an ideal augmented prototype must represent sheet notation into a compelling visualisation that helps the learner process this information - not remove them entirely. In some instances (especially for complex chords and progressions), audio feedback is also a desirable augmentation. We used these findings to categorise which piano prototypes had sight-reading as a theme of focus. Based on our analysis (\autoref{tab:piano_themes}), we found that 14 (out of 56, roughly 27\%) of the papers included in this review complied with this theme based on this criteria. 

To make piano roll visualisations more effective, we argue these visualisations should be mapped with the music sheet notation to teach or help in sight-reading. The papers seen in \autoref{tab:piano_themes} marked under \textit{Sight} have been categorised to satisfy this criteria. Their piano roll visualisations came with an extra layer of information that allowed the user to review and map these visualisations to their representation in a musical sheet~\cite{schmalstieg2007experiences, yang2012augmented, yang2013visual}. The most common approach involved a projector that overlaid visualisations of moving piano bars~\cite{takegawa2012piano, rogers2014piano, birhanu2017keynvision, sun2018mr, cai2019designb, weing2013piano, rogers2014piano}. Head-mounted displays (HMDs) were also used to show these piano visualisations~\cite{chow2013music, birhanu2017interactive}. While the majority of such prototypes featured classical music in piano roll visualisations, other music genres were also explored, such as jazz~\cite{sandnes2019enhanced}. Some included additional visualisation guiding user gestures~\cite{santiniaugmented}. Other prototypes expanded the scope of piano roll visualisations on other devices (e.g.,~\citet{chow2013music}) or other use cases such as gestures~\cite{santiniaugmented, yang2012augmented, yang2013visual}, game-based learning (quizzes and hints)~\cite{schmalstieg2007experiences, guo2021hand}. Lastly, some prototypes augmented visualisations with sound feedback~\cite{chouvatut2013virtual, chiang2015oncall, sun2018mr} to teach learners how to map music notes, piano rolls and corresponding sounds. 

The papers categorised in this theme focused on whether learners understand music sheet notation and know which keys and when to press them. Thus, a construct measured in these studies is key-press accuracy. While this can be a good metric of performance (number of correct presses vs. number of total presses; the number of correctly-released presses vs number of total presses), we argue that measuring actual sight-reading is more challenging to assess. A psycho-cognitive task would require more factors or parameters to be measured more accurately (such as knowledge of music theory and, in some instances, sound quality). Playing a musical instrument is similar to speaking a second language, but their differences end with the tactile part of properly pressing the right keys. As emphasised in our interviews, a solid understanding of musical notation is necessary for learners to play the piano in the long term. Ideally, as being able to press the right keys at the right time needs to be synchronised with it, the language represented by a (complex) notation of notes, rests, staffs, and other musical notations are considered simultaneously. We provide more details on this in our recommendations; see \autoref{sec:recomm}.

\subsection{Motivating Students to Practice}
\label{learnin_themes_motivating}
Traditionally, piano students attend one-on-one classes run by a teacher. In these classes, students gain fundamental theory and practical lessons to play a particular piece of music. However, their skills can only be improved further by regular practice. Thus, the teacher often provides students with \textit{take-home activities}, while students are expected to practice and master the content in preparation for the next class. Yet, being self-regulated to practice continuously is a difficult task on top of practising the piano itself. Thus, several approaches have been introduced to encourage and sustain the motivation of piano students. This makes the theme of motivating students to practice effective since it considers the context of emotion and motivation. 

Monitoring students' progress, tracking their performance and accuracy, encouraging regular practising, and generally motivating students are critical elements in the learning process. Experienced teachers believe that novices and proficient students are set apart by the hard work dedicated to mastering the craft~\cite{bandura1977social}. In the context of piano learning, several strategies to motivate students have been introduced by digitally augmented piano prototypes. The most common approach is through the use of gamification, where the learner becomes motivated by the elements of game playing~\cite{schmalstieg2007experiences, molero2021novel} and other game-based incentives~\cite{raymaekers2014game, molloy2019mixed, fernandez2016piano} such as earning points by regularly playing the piano.
Another strategy is introducing competitions in collaborative piano environments~\cite{cai2019designa, pan2018pilot}. 

Self-regulation has also been observed among students as they are encouraged to maintain a regular practice schedule with the help of augmentations that introduce interpersonal and remote spaces (involving a tutor)~\cite{xiao2010mirrorfugue, xiao2011duet}. With this feature, students can watch their performance either in real-time or after performing a piece. This feature draws inspiration from the theory of~\citet{zimmerman2009self} on how self-reflection promotes self-regulated learning, which has been supported by numerous experiments (e.g.,~\citet{deja2016discovering, lyons2011monitoring}), showing that students reflecting on their performance feel motivated afterwards. Aside from self-regulation and self-reflection theory, social learning theory (SLT) has also been explored in the context of augmented piano prototypes. SLT emphasises four distinct steps in learning: attention, retention, reproduction, and motivation~\cite{bandura1977social}. In \textit{attention} phase, a piano student observes how a certain piece should be played by carefully watching a teacher perform. During the \textit{retention} phase, the student tries to remember what they have observed. The \textit{reproduction} step covers performing the activities by the student, which further supports retention~\cite{stryker1997content}. This learning process becomes sustainable in the med- to long-term by learner's \textit{motivation}, where reinforcement (either positive or negative) can also ensure continuous practice.

In the P.I.A.N.O prototype~\cite{weing2013piano, rogers2014piano}, authors employed SLT through the design of their learning modes. The \textit{listen (attention)} mode allows students to listen to a song and observe its visualisation. The \textit{practice (retention)} mode provides students with feedback by highlighting the correct and the wrongly-pressed keys. Lastly, the \textit{play (reproduction and motivation)} mode provides additional feedback on students' performance while playing. Besides the correctly or incorrectly pressed keys, details on expected notes (missed keys), irregular duration, and a summary of their performance through a progress bar were also displayed.

Several digitally augmented prototypes introduced extra modules in their software, such as practice mode~\cite{sun2018mr, weing2013piano, chow2013music, rogers2014piano, guo2021hand}, which gave learners additional content to practice on, or a lesson builder~\cite{das2017music}, which allowed learners to customise their lessons. 

\subsection{Enabling Improvisation}
\label{learnin_themes_improvisation}
For experienced and advanced piano players, being able to improvise is proof of a wide musical vocabulary. Also, according to our experts, talented people with ``perfect pitch'' can easily perform improvisation. The practice involves the application of current music knowledge skills and applying them in an impromptu (or any scenario without prior practice) to a performance which produces a musically-sound piece~\cite{beaty2013first}. This makes improvisation a task under the cognitive domain. Improvisation at all levels of music education ``creates an environment where children can express their creativity''~\cite{burnard2000children}. Teaching improvisation to novice and intermediate students can improve their rhythmic accuracy and note reading skills, concentration, self-actualisation, imagination, and nonetheless increases their confidence level~\cite{montano1984effect, chyu2004teaching}. Besides mastering proper body/hands/fingers posture, reading music sheet notation and motivation, confidence is also an essential aspect of playing an instrument for novice students as well as more experienced and even expert piano players~\cite{deja2021adaptive, allen2013free}. This is true especially as it has been shown that piano students can experience anxiety and being overwhelmed during their performances~\cite{allen2013free}. Teaching improvisation to novice and intermediate students has also been observed to help improve their rhythmic accuracy and note-reading skills~\cite{montano1984effect}. This could be one reason why roughly 27\% (14 out of 51) of augmented piano prototypes reviewed have considered improvisation. While this theme has appeared several times in our review, experts argue there is much room for development since the skill of improvisation is challenging to assess from the perspective of a piano teacher. 

Augmented piano prototypes have enabled improvisations in various forms and approaches. The most common approach is to incorporate gesture detection allowing musicians to use their hands freely during performances, thereby improving their musical vocabulary~\cite{schmalstieg2007experiences, mcpherson2010augmenting, mcpherson2010toward, sandnes2019enhanced}. Some prototypes enabled users to move freely and perform natural gestures, which allowed them to express themselves further~\cite{karolus2020hit, granieri2019reach, ogata2017keyboard}. These studies also investigated several constructs, such as levels of expressiveness and improvisation. Their findings show that their finger-based control of additional features does not interfere with the playing flow of a performer. Researchers proposed that improvisations improved user experience, especially in live performances, as observed inherently in the prototypes of~\cite{nicolls2012gesturally, xiao2013mirrorfugue, das2017music, santiniaugmented}. Furthermore, in some instances, improvisations were enabled to improve the sound quality~\cite{dahlstedt2015mapping, yang2012augmented, yang2013visual}, which in turn has effects on listeners besides performers.

\section{Trends in Piano Augmentation User Studies}
In this section, we analyse user studies in surveyed papers. This will help us to understand how the reviewed prototypes were studied and where the focus should shift in the future. 

\begin{table}[t]
\caption{List of studies with user evaluation. This table provides an overview of their treatment, metrics or constructs, and tools.}
\label{tab:us-all}
\rowcolors{2}{gray!15}{white}
\resizebox{\textwidth}{!}{%
\begin{tabular}{lr*{27}{c|}l} \toprule
 &  &   \multicolumn{1}{c}{} &  \multicolumn{7}{c}{\textbf{TREATMENT}}  & \multicolumn{12}{c}{\textbf{METRICS}}  & \multicolumn{7}{c}{\textbf{TOOLS}}  &   \setcounter{rownum}{0}\\
\cmidrule(r){4-10}\cmidrule(rl){11-22}\cmidrule(l){23-29}
\textbf{ID} & \textbf{Ref}  & \textit{\# of users} & \rotatebox[origin=l]{90}{ex - free exploration }  & \rotatebox[origin=l]{90}{md - marker detection} & \rotatebox[origin=l]{90}{ob - observation} & \rotatebox[origin=l]{90}{pc - play a chord} & \rotatebox[origin=l]{90}{pl - play a full piece} & \rotatebox[origin=l]{90}{pr - practice freely} & \rotatebox[origin=l]{90}{qu - complete a game quest} & \rotatebox[origin=l]{90}{At - attractiveness} & \rotatebox[origin=l]{90}{CL - cognitive load} & \rotatebox[origin=l]{90}{Fi - finger accuracy} & \rotatebox[origin=l]{90}{Im - level of immersion} & \rotatebox[origin=l]{90}{Mo - level of motivation} & \rotatebox[origin=l]{90}{No - notation accuracy} & \rotatebox[origin=l]{90}{Op - functional check of features} & \rotatebox[origin=l]{90}{Sa - satisfaction} & \rotatebox[origin=l]{90}{Sc - scoring (for gamified)} & \rotatebox[origin=l]{90}{Sk - skill improvement} & \rotatebox[origin=l]{90}{Us - usability} & \rotatebox[origin=l]{90}{Ti - time and usage} & \rotatebox[origin=l]{90}{OEQ - open ended questionnaires}  & \rotatebox[origin=l]{90}{QUE - peer-reviewed questionnaires}  & \rotatebox[origin=l]{90}{PSP - player scoring plugins}  & \rotatebox[origin=l]{90}{REC - observation of recordings}  & \rotatebox[origin=l]{90}{SMQ - self-made questionnaires}  & \rotatebox[origin=l]{90}{SSI - semi structed interviews}  & \rotatebox[origin=l]{90}{TTM - time tracking mechanisms}  &   \textbf{Notes} \\ \midrule

 P02         & \cite{schmalstieg2007experiences}    & 6                &      &      &      &      & \yes  &      & \yes &      &      &      &      &      &      &      & \yes &      &      &      & \yes &      &      & \yes  &      &      & \yes  & \yes  & \\
P03         & \cite{correa2009computer}            & 1                & \yes  &      &      &      &      &      & \yes  &      &      &      &      &      &      & \yes  &      &      &      & \yes  &      &      &      &      & \yes  &      &      & \yes  & patient motor effects \\
P07         & \cite{mcpherson2011multidimensional} & 30*              & \yes  &      &      &      & \yes  &      &      &      &      & \yes  &      &      & \yes  & \yes  &      &      &      & \yes  & \yes  &      &      &      & \yes  &      &      & \yes  & 3 tests with 10 n each                         \\
P09         & \cite{xiao2010mirrorfugue}           & 5                &      &      &      & \yes  & \yes  &      &      &      &      &      & \yes  &      &      & \yes  &      &      &      &      &      &      &      & \yes  & \yes  &      & \yes  & \yes  & improvise a piece         \\
P10         & \cite{xiao2011duet}                  & 3                & \yes  &      & \yes  &      &      &      &      &      &      &      & \yes  &      &      &      &      &      &      & \yes  &      &      &      &      &      &      & \yes  &      &  \\
P15         & \cite{takegawa2012piano}             & 9                &      &      &      &      & \yes  & \yes  &      &      &      & \yes  &      &      & \yes  &      &      & \yes  &      &      & \yes  &      &      & \yes  &      &      & \yes  & \yes  &  \\
P16         & \cite{mcpherson2013space}            & 8                &      & \yes  &      &      & \yes  &      &      &      &      & \yes  &      &      & \yes  & \yes  &      &      &      &      &      &      &      & \yes  &      &      &      & \yes  &  \\
P19         & \cite{chow2013music}                 & 7                &      &      &      &      & \yes  &      &      &      &      &      &      &      &      &      & \yes  &      &      & \yes  &      & \yes  &      &      &      &      &      &      &  \\
P20         & \cite{weing2013piano}                & 5                & \yes  &      &      &      &      & \yes  &      &      & \yes  & \yes  &      &      & \yes  &      & \yes  &      &      &      &      &      &      &      &      & \yes  &      &      &  \\
P23         & \cite{xiao2013mirrorfugue}           & 15               &      &      & \yes  &      &      &      &      &      &      &      & \yes  &      &      & \yes  &      &      &      & \yes  &      &      &      &      &      & \yes  & \yes  &      &  \\
P27         & \cite{raymaekers2014game}            & -*               & \yes  &      &      &      & \yes  & \yes  &      & \yes  &      &      &      &      &      &      & \yes  &      &      & \yes  &      & \yes  &      &      & \yes  &      &      &      & open demo UT              \\
P34         & \cite{ogata2017keyboard}             & 3                & \yes  &      &      &      & \yes  &      &      &      &      &      &      &      &      & \yes  & \yes  &      &      &      &      &      &      &      &      & \yes  &      &      &  \\
P38         & \cite{rogers2014piano}               & 74*              &      &      &      & \yes  & \yes  & \yes  &      & \yes  & \yes  &      &      &      &      &      & \yes  &      & \yes  & \yes  & \yes  &      & \yes  &      &      &      &      &      & *$n_{1}$=56, \begin{math}n_{2}\end{math}=18, $^\dagger$\cite{ekstrom1976manual, klepsch2012subjective, hassenzahl2003attrakdiff, wrigley2013ecological} \\
P41         & \cite{sun2018mr}                     & 20               & \yes  &      &      & \yes  & \yes  &      &      &      &      &      &      &      &      &      &      & \yes  & \yes  & \yes  & \yes  &      &      & \yes  &      &      &      & \yes  &  \\
P42         & \cite{pan2018pilot}                  & 13               &      &      &      &      & \yes  & \yes  &      &      &      &      &      &      &      &      &      & \yes  & \yes  &      &      & \yes  &      & \yes  &      & \yes  & \yes  &      &  \\
P43         & \cite{granieri2019reach}             & -*               & \yes  &      &      &      &      &      &      &      &      &      & \yes  &      &      &      &      &      &      &      &      &      &      &      & \yes  &      &      &      & \textit{*open demo UT, n not reported}         \\
P45         & \cite{molloy2019mixed}               & 23               &      &      &      &      & \yes  &      &      & \yes  &      &      & \yes  & \yes  &      &      &      &      &      & \yes  &      & \yes  & \yes  &      &      &      & \yes  &      & *SUS~\cite{lewis2009factor}                    \\
P51         & \cite{karolus2020hit}                & 12               & \yes  &      &      &      & \yes  &      &      & \yes  & \yes  &      &      &      &      & \yes  & \yes  &      &      &      & \yes  & \yes  & \yes  &      &      &      &      & \yes  & *TLX~\cite{hart1988development}, CSI~\cite{carroll2009creativity}, HEMA~\cite{huta2010pursuing}               \\
P52         & \cite{moro2020performer}             & 6                & \yes  &      &      & \yes  & \yes  &      &      &      &      &      &      &      & \yes  & \yes  &      &      &      &      &      & \yes  &      & \yes  &      & \yes  & \yes  & \yes  &  \\
P53         & \cite{molero2021novel}               & 13               &      &      &      &      & \yes  & \yes  &      &      &      &      &      &      &      & \yes  &      &      &      & \yes  &      &      & \yes  & \yes  &      &      &      &      & *5 students, 8 teachers, **TAM~\cite{davis1993user}                 \\
P55         & \cite{guo2021hand}                   & -                &      &      &      & \yes  &      &      &      &      & \yes  &      &      &      &      &      &      &      &      &      &      &      &      & \yes  & \yes  &      &      &      &  \\ \midrule
\rowcolor{white}
\multicolumn{3}{l|}{\# of dimensions (\textit{\={x}}=13)} & 10 & 1 & 2 & 5 & 15 & 6  & 2 & 4 & 4 & 4  & 5   & 1 & 5 & 9 & 7 & 3 & 3 & 10 & 6 & 6 & 4 & 9 & 6 & 5 & 8 & 9 & \\ 
\bottomrule \rowcolor{white}
\end{tabular}%
}
\end{table}

The results in \autoref{tab:us-all} show that the experiments in papers published before 2014 were designed with measuring general usability as a goal. The authors considered notation accuracy, general satisfaction, finger information, level of immersiveness, and in-game scoring as constructs measured (labelled as \textit{No, Sa, FI, Im, Sc} respectively in \autoref{tab:us-all}). These studies focused on whether the related prototypes \textit{technically} work and can emulate the piano in the closest way possible. They measured these constructs by either (1) letting the users play a specific piano chord or a (2) full piece, (3) generally practising or (4) exploring the augmented piano prototype or (5) completing a quest (if the prototype came with game-related features) (labelled as \textit{pc, pl, pr, ex, qu} respectively in \autoref{tab:us-all}). To gather data, the authors used several instruments and data sources such as player scoring plugins (\textit{PSP}), semi-structured interviews (\textit{SSI}), time-tracking mechanisms (\textit{TTM}), and manually observed recordings of users (\textit{REC}). The study size averaged around 5-6 participants for these papers. Analysing the trend during this period reveals that technology augmentation in pianos is relatively new and is focused on the success of the prototypes themselves: (\textit{Do the keys play a sound? Can users play a chord? Can they play a piece? Is there a delay between the sound produced, the graphics rendered and with how the player uses the prototype?}). Looking at the relation with the four learner themes introduced in this paper, there is a general lack of learner-centric considerations that have been measured other than the quantitative metrics (such as finger and notation accuracy) mentioned.

With technology-specific APIs and hardware improving as the years progressed, we can notice that there is also a shift from prototype performance to user performance in the way these papers have conducted user studies. Different metrics have been considered such as attractiveness (\textit{At}),  satisfaction (\textit{Sa}), ease of use and usability (\textit{Us}), cognitive load (\textit{CL}), skill improvement (\textit{Sk}), and motivation (\textit{Mo}). Similarly, the treatments used in more recent studies have not changed (they have also used \textit{pc, pl, pr, ex, qu}). However, even if they used the same treatment, having a different set of constructs would also mean having to use different instruments and data sources to measure these metrics accurately. These studies have developed their own questionnaires (\textit{SMQ}- which they built on top of a specific paradigm or framework) or used peer-reviewed questionnaires (\textit{QUE}). Some notable instruments used were the NASA-TLX~\cite{hart1988development}, Creativity Support Index~\cite{carroll2009creativity}, and the measurement of hedonic and eudaimonic attributes~\cite{huta2010pursuing}. The study size averaged 20-21 participants per study, a noticeable increase from the studies done before 2014. The trend in recent years has focused on the success of the piano users rather than the success of the prototypes (addressing questions such as \textit{Do users play properly with the piano? Do users learn faster with this augmentation? Does this feature motivate or encourage users [to improvise]? Does augmentation overwhelm the user?}). Despite having a broader user base in the later studies, there is a lack of longitudinal studies to better understand the long-term effects of augmented prototypes. In addition, the prototypes presented do not take into account the specific learner-based themes identified by interviewing expert teachers. In the next section, we provide recommendations for future development, implementations, and research of the augmented piano prototypes. 

\section{Recommendations for Future Explorations}
\label{sec:recomm}

The difficulties when learning the piano have pushed for several digitally augmented innovations within the last two decades. While we understand that learning the piano is a physical, cognitive, psycho-cognitive and affective task, we have reported that some of these augmentations may lack the essential learner-based focus that emerged during the interviews with piano experts (teachers). Moreover, hardware or software augmentations have shown how technology can also introduce newer problems brought about by affordances~\cite{dede1996evolution}. At the same time, we posit that most of these augmentations bring about short-term improvements to the student. As such, we present recommendations that provide improvements to the student in the long term. We base these recommendations on the core elements of gamification and game-based learning, social learning theory (SLT), experiential learning theory (ExL), and many others. We map these recommendations with our music experts' insights and verify them on classical music pedagogy. 

\paragraph{Balance Between Traditional Techniques and New Approaches.} Music experts claim that playing the piano is a centuries-old technique and is best experienced when learned traditionally. Piano roll visualisations and other gamified elements have been a popular choice to help teach a more ``\textit{natural}'' flow of movement of hands and posture when playing the piano. However, most visualisation approaches implemented missed or neglected the elements in a traditional music sheet (such as time signature, octave, which finger to use and many others). For example, based on our expert interviews, users get to learn how to press the right keys at the right time but do not necessarily acquire the skill to learn to read notes to play musical pieces independently. Thus, piano roll visualisations should visualise the abstraction of music notes that work in both ways (help the user learn sight reading and recognise heard audio with its equivalent notation, thus achieving balance on both traditional and newer techniques). Users should be able to match the moving rolls and their equivalent musical sheet counterparts (both in notation and in sound). While some prototypes implemented this, there is a lack of user studies on what is the best way to achieve this while not overwhelming users with extra information shown. Future user studies should explore various visualisations and combinations of, e.g., piano roll and music notation. It should also guide the user to learn from both aural (sound and tune) feedback to visual (notes) form and back.  This also satisfies the needed skill to hear and recognise music rudiments and be able to translate them back into sounds (thus being able to do it the other way around). As most systems have fed students with visualisations on which keys to press, when and for how long, it is also crucial that these systems would recognise and know how these sheet notations might sound (or might form a general harmonic melody in a greater sense). This would equip learners with longer-lasting skills that include knowing how to play the piano and integrating theory with practice. However, this should also be empirically tested in longitudinal studies. In this approach, we present students with both traditional approaches (sight reading, feedback, practising, music theory) and newer approaches (augmented visualisations, practice, and reflection modes) to learning the piano.

\paragraph{Ensuring Motivation and a Consistent Practice Regimen.} Motivation plays an essential role in helping piano students. There have been several studies that have explored motivation and various constructs to measure it: (i) motivating students to be competitive~\cite{vassileva2012motivating}, (ii) using gamification~\cite{rodrigues2021personalization}, (iii) via self regulation~\cite{dimitracopoulou2007computer} and others~\cite{orji2021modelling}. Allowing students to reflect on their learning and their progress, giving them a clearer view of the mistakes they may have committed in their recent performance, can help motivate dedicated learners. While these contributions have presented study designs that measured engagement, immersiveness, motivation, and cognitive load as constructs, these have yet to be fully explored in the context of music learning (such as in the piano). As these factors contribute to assessing motivation and how it affects learning, there is a need to conduct more studies involving a broader array of metrics or constructs that may lead to a consistent and sustained learning experience. This can be facilitated by borrowing established methods from other domains such as cognitive load, physiological signals (e.g., GSR, ECG), eye-tracking, and many others to make measuring engagement and motivation more accurate as possible. While these have been incorporated in short-term studies~\cite{yuksel2016learn}, it would be interesting to see whether there are greater benefits when observed in the long term.  
In addition, several prototypes mentioned had gamification elements embedded to keep users engaged. However, the long-term effects are not known as learning a piano is a years-long task. It should also be explored what game elements are suitable for a particular age since every age requires a different approach~\cite{brox2017user}. 

\paragraph{Measuring Expressiveness and Improvisation.} Being able to apply a creative approach in one's performance is a measure of how confident a person is when playing the piano~\cite{smith1983homer}. Unfortunately, most augmented piano prototypes have contributed to mechanical and intra-personal piano playing so far. Thus, we found that expressiveness and improvisation in piano learning are underrepresented~\cite{das2017music, xia2017improvised, donahue2019piano} but are equally important as well. As mentioned earlier, improvising on the piano demonstrates a higher level of skill that involves mastery of music theory and natural movement in the piano. Piano experts believe that expressiveness and improvisation are difficult-to-measure skills in the piano~\cite{laroche2014enacting, solis2009musical}, which is why it is usually skipped even though they are prescribed in the standardised music curriculum. Developing important key features that promote and encourage learners to improvise is desired despite being a relatively-new domain if we will look at it from the perspective of piano augmentation. We recommend that exploring user studies and technology augmentations that aid in measuring and enabling expressiveness following the initial work by~\citet{karolus2020hit} could open more opportunities for personalised and pleasant improvisations during learning. While improvisation is also effective in other related domains such as music therapy~\cite{wigram2004improvisation}, it would be interesting to explore whether these techniques can also help the general learner be more confident in their performance or help them have a broader musical vocabulary. 

\paragraph{Learner Personalisation.} User modelling has been a long-existing technique in the broader domain of technology-aided learning. However, work on personalised student experiences in the context of piano learning is a less-explored research area. There have been some studies on understanding the pace of the user and their preferences~\cite{goguey2021interaction}, personalised error interventions based on user input~\cite{lam2021effects} and how learners studied (for better or worse) with personalised visualisations and others~\cite{uddin2021image}. Yet, much of these studies have been explored in the context of using a general graphic user interface and not specifically for music learning (like in the piano). Their results show potential for personalised learner experiences enhanced by user modelling and are worth exploring specifically for this context. On the other hand, there are also several studies that attempted to understand user proficiency~\cite{karolus2021proficiency, karolus2018proficiency} in a specific context and the internal mechanisms~\cite{huang2019modeling, lee2016modelling, park2020intermittent} that go with it - but outside the domain of learning. Their results show that they can model user movement and predict user errors which later on they could use to design interventions to mitigate these errors. While their works have been used on general pointing tasks or in game-based environments, we argue that the elegant structure and form of music make it an interesting domain that goes well with personalised learning. Furthermore, as user modelling has been explored with various input modalities, these personalised experiences can supplement teacher-based piano sessions. Therefore, we suggest looking into movement patterns, music features, and even expert heuristics, among many others, when building general or user-specific models of piano students. Studies involving these variables have been proposed prior~\cite{deja2021adaptive,deja2021encouraging} but have yet to be explored further and deeper. 

\paragraph{Exploring Other Types of Interventions.} As several types of students and a broad spectrum of music-based pedagogies exist, we recommend future directions in digital technology augmentation based on the appropriate interventions that fit students' corresponding learning types~\cite{deja2022vision}. Apart from gamified approaches to piano learning, we noticed that most existing augmented piano prototypes have not used other forms of interventions such as (but not limited to): (i)~predicting user errors in key-pressing (e.g.,~\citet{mecke2019exploring, buschek2015improving}), (ii)~managing the cognitive load of students (e.g.,~\citet{kosch2018identifying}), (iii)~simplifying or streamlining the number of piano rolls (e.g.,~\citet{kosch2018look}), and (iv)~supporting collaborative or grouped learning (e.g.,~\citet{wozniak2016ramparts}). Different types of intervention can also be introduced at varying levels of interaction (from macro to micro e.g., improving synchronisation based on the ideal finger or hand-angle and positioning~\cite{mecke2019exploring, buschek2015improving}, analysing where the learners are looking/gazing to understand better cognitive load~\cite{kosch2018identifying, kosch2018look}, and understanding spatial elements around the user to explore better context involve the task~\cite{wozniak2016ramparts}). This way, augmented piano prototypes can understand the context and introduce interventions to the user at the specific pain point that needs improvement or adjusting. 

\section{Conclusion}
In this paper, we did a systematic review of digitally augmented piano prototypes based on the four learner-based themes that were conceptualised during the interviews with piano teachers: (1) synchronising movement and posture between various body parts, (2) improving sight-reading skills, (3) motivating students, and (4) enabling improvisation. Based on the data we collected from the papers, we saw how smaller (local) experiments succeeded with their objectives (e.g., several prototypes helped users become better in terms of movement and synchronisation, and other prototypes noticed a change in behaviour or motivation for students). We found that synchronisation and motivation studies worked, while authentic sight-reading and improvisation studies require further exploration. Combining these findings with the themes we derived, we call for more studies that involve larger samples done over a longer duration to properly-assess learning success. For such large-scale and long-term studies, we laid out a set of recommendations to guide researchers in designing their prototypes and experiments. 

\begin{acks}
The authors wish to thank the several experts who willingly offered their free time to share their experiences and opinions during our interviews. The authors also would like to thank Sarah Rüller for helping them generate the high-resolution affinity diagram that is part of this paper's supplementary material. This research was funded by the Slovenian Research Agency, grant number P1-0383, J1-9186, J1-1715, J5-1796, and J1-1692, and European Commission, InnoRenew CoE project (Grant Agreement 739574) under the Horizon2020 Widespread-Teaming.
\end{acks}

\newpage
\appendix
\section{Affinity Diagram}
\begin{figure}[h]
    \includegraphics[width=\linewidth,origin=c]{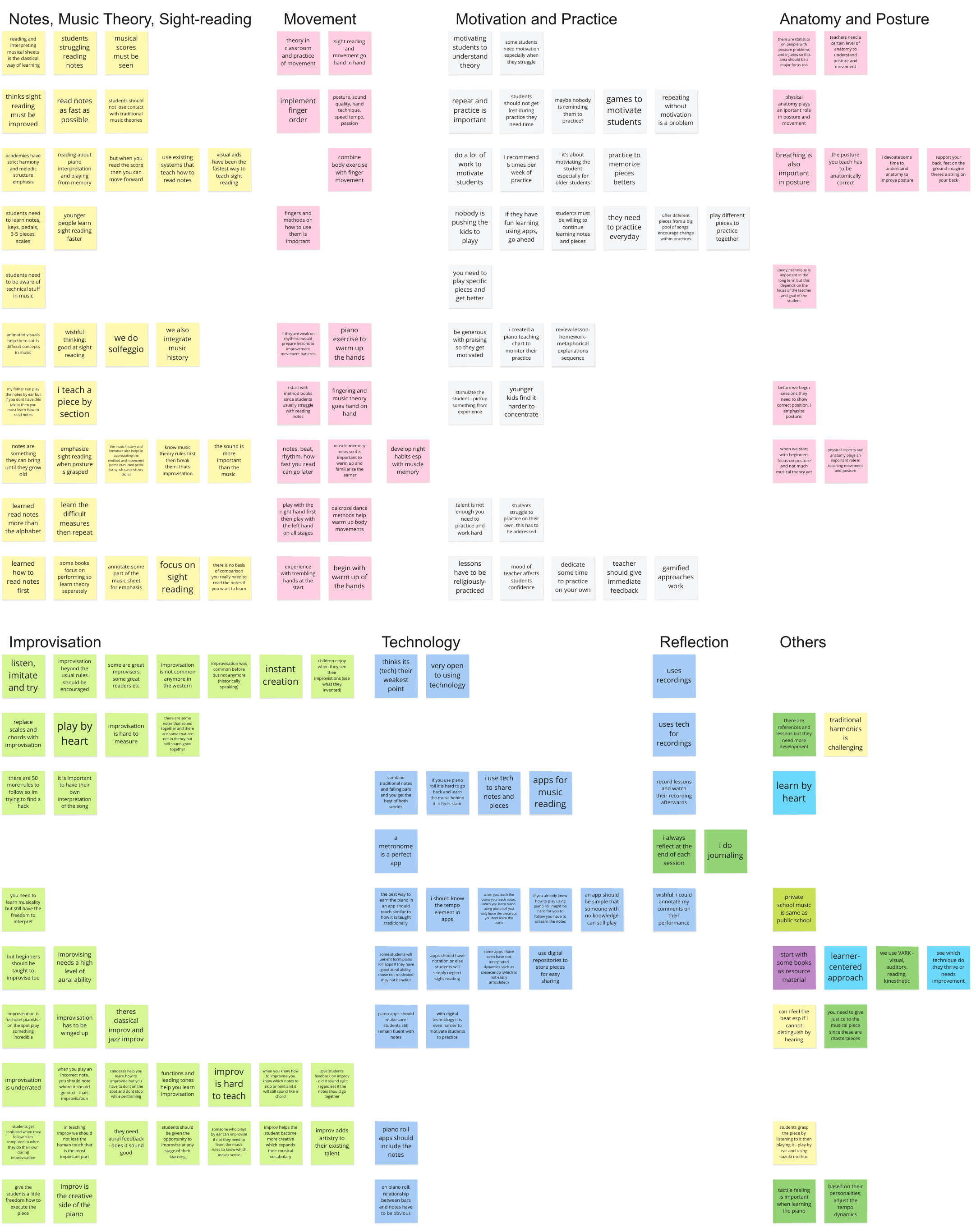}
    \caption{Affinity diagram containing insights that were coded and sorted. These were extracted from the transcripts of interviews with the experts. Each row represents insights from E1 down to E10.}
    \label{fig:affinity}
\end{figure}

\newpage
\bibliographystyle{ACM-Reference-Format}
\bibliography{main}


\begin{thebibliography}{164}


\ifx \showCODEN    \undefined \def \showCODEN     #1{\unskip}     \fi
\ifx \showDOI      \undefined \def \showDOI       #1{#1}\fi
\ifx \showISBNx    \undefined \def \showISBNx     #1{\unskip}     \fi
\ifx \showISBNxiii \undefined \def \showISBNxiii  #1{\unskip}     \fi
\ifx \showISSN     \undefined \def \showISSN      #1{\unskip}     \fi
\ifx \showLCCN     \undefined \def \showLCCN      #1{\unskip}     \fi
\ifx \shownote     \undefined \def \shownote      #1{#1}          \fi
\ifx \showarticletitle \undefined \def \showarticletitle #1{#1}   \fi
\ifx \showURL      \undefined \def \showURL       {\relax}        \fi
\providecommand\bibfield[2]{#2}
\providecommand\bibinfo[2]{#2}
\providecommand\natexlab[1]{#1}
\providecommand\showeprint[2][]{arXiv:#2}

\bibitem[\protect\citeauthoryear{Allen}{Allen}{2013}]%
        {allen2013free}
\bibfield{author}{\bibinfo{person}{Robert Allen}.}
  \bibinfo{year}{2013}\natexlab{}.
\newblock \showarticletitle{Free improvisation and performance anxiety among
  piano students}.
\newblock \bibinfo{journal}{\emph{Psychology of Music}} \bibinfo{volume}{41},
  \bibinfo{number}{1} (\bibinfo{year}{2013}), \bibinfo{pages}{75--88}.
\newblock
\urldef\tempurl%
\url{https://doi.org/10.1177/0305735611415750}
\showDOI{\tempurl}


\bibitem[\protect\citeauthoryear{Bandura and Walters}{Bandura and
  Walters}{1977}]%
        {bandura1977social}
\bibfield{author}{\bibinfo{person}{Albert Bandura} {and}
  \bibinfo{person}{Richard~H Walters}.} \bibinfo{year}{1977}\natexlab{}.
\newblock \bibinfo{booktitle}{\emph{Social learning theory}}.
  Vol.~\bibinfo{volume}{1}.
\newblock \bibinfo{publisher}{Prentice-hall Englewood Cliffs, NJ},
  \bibinfo{address}{New Jersey}.
\newblock


\bibitem[\protect\citeauthoryear{Barakonyi and Schmalstieg}{Barakonyi and
  Schmalstieg}{2005}]%
        {barakonyi2005augmented}
\bibfield{author}{\bibinfo{person}{Istv{\'a}n Barakonyi} {and}
  \bibinfo{person}{Dieter Schmalstieg}.} \bibinfo{year}{2005}\natexlab{}.
\newblock \showarticletitle{Augmented Reality Agents in the Development
  Pipeline of Computer Entertainment}. In
  \bibinfo{booktitle}{\emph{International Conference on Entertainment
  Computing}}, \bibfield{editor}{\bibinfo{person}{Fumio Kishino},
  \bibinfo{person}{Yoshifumi Kitamura}, \bibinfo{person}{Hirokazu Kato}, {and}
  \bibinfo{person}{Noriko Nagata}} (Eds.). \bibinfo{publisher}{Springer Berlin
  Heidelberg}, \bibinfo{address}{Berlin, Heidelberg},
  \bibinfo{pages}{345--356}.
\newblock
\showISBNx{978-3-540-32054-8}
\urldef\tempurl%
\url{https://doi.org/10.1007/11558651_34}
\showDOI{\tempurl}


\bibitem[\protect\citeauthoryear{Beaty, Smeekens, Silvia, Hodges, and
  Kane}{Beaty et~al\mbox{.}}{2013}]%
        {beaty2013first}
\bibfield{author}{\bibinfo{person}{Roger~E Beaty}, \bibinfo{person}{Bridget~A
  Smeekens}, \bibinfo{person}{Paul~J Silvia}, \bibinfo{person}{Donald~A
  Hodges}, {and} \bibinfo{person}{Michael~J Kane}.}
  \bibinfo{year}{2013}\natexlab{}.
\newblock \showarticletitle{A first look at the role of domain-general
  cognitive and creative abilities in jazz improvisation.}
\newblock \bibinfo{journal}{\emph{Psychomusicology: Music, Mind, and Brain}}
  \bibinfo{volume}{23}, \bibinfo{number}{4} (\bibinfo{year}{2013}),
  \bibinfo{pages}{262}.
\newblock


\bibitem[\protect\citeauthoryear{Birhanu}{Birhanu}{2017}]%
        {birhanu2017interactive}
\bibfield{author}{\bibinfo{person}{Amare Birhanu}.}
  \bibinfo{year}{2017}\natexlab{}.
\newblock \emph{\bibinfo{title}{Interactive AR Experiences as Training
  Applications: Guidelines and Requirements for Piano Pedagogy in Mixed
  Reality}}.
\newblock \bibinfo{thesistype}{Ph.D. Dissertation}. \bibinfo{school}{Drexel
  University}.
\newblock
\showISBNx{978-0-355-11725-7}


\bibitem[\protect\citeauthoryear{Birhanu and Rank}{Birhanu and Rank}{2017}]%
        {birhanu2017keynvision}
\bibfield{author}{\bibinfo{person}{Amare Birhanu} {and} \bibinfo{person}{Stefan
  Rank}.} \bibinfo{year}{2017}\natexlab{}.
\newblock \showarticletitle{KeynVision: Exploring Piano Pedagogy in Mixed
  Reality}. In \bibinfo{booktitle}{\emph{Extended Abstracts Publication of the
  Annual Symposium on Computer-Human Interaction in Play}} (Amsterdam, The
  Netherlands) \emph{(\bibinfo{series}{CHI PLAY '17 Extended Abstracts})}.
  \bibinfo{publisher}{Association for Computing Machinery},
  \bibinfo{address}{New York, NY, USA}, \bibinfo{pages}{299–304}.
\newblock
\showISBNx{9781450351119}
\urldef\tempurl%
\url{https://doi.org/10.1145/3130859.3131336}
\showDOI{\tempurl}


\bibitem[\protect\citeauthoryear{Blattgerste, Renner, and Pfeiffer}{Blattgerste
  et~al\mbox{.}}{2019}]%
        {blattgerste2019augmented}
\bibfield{author}{\bibinfo{person}{Jonas Blattgerste}, \bibinfo{person}{Patrick
  Renner}, {and} \bibinfo{person}{Thies Pfeiffer}.}
  \bibinfo{year}{2019}\natexlab{}.
\newblock \showarticletitle{Augmented Reality Action Assistance and Learning
  for Cognitively Impaired People: A Systematic Literature Review}. In
  \bibinfo{booktitle}{\emph{Proceedings of the 12th ACM International
  Conference on PErvasive Technologies Related to Assistive Environments}}
  (Rhodes, Greece) \emph{(\bibinfo{series}{PETRA '19})}.
  \bibinfo{publisher}{Association for Computing Machinery},
  \bibinfo{address}{New York, NY, USA}, \bibinfo{pages}{270–279}.
\newblock
\showISBNx{9781450362320}
\urldef\tempurl%
\url{https://doi.org/10.1145/3316782.3316789}
\showDOI{\tempurl}


\bibitem[\protect\citeauthoryear{Blohm and Leimeister}{Blohm and
  Leimeister}{2013}]%
        {blohm2013gamification}
\bibfield{author}{\bibinfo{person}{Ivo Blohm} {and} \bibinfo{person}{Jan~Marco
  Leimeister}.} \bibinfo{year}{2013}\natexlab{}.
\newblock \showarticletitle{Gamification}.
\newblock \bibinfo{journal}{\emph{Business \& information systems engineering}}
  \bibinfo{volume}{5}, \bibinfo{number}{4} (\bibinfo{year}{2013}),
  \bibinfo{pages}{275--278}.
\newblock


\bibitem[\protect\citeauthoryear{Braun and Clarke}{Braun and Clarke}{2012}]%
        {braun2012thematic}
\bibfield{author}{\bibinfo{person}{Virginia Braun} {and}
  \bibinfo{person}{Victoria Clarke}.} \bibinfo{year}{2012}\natexlab{}.
\newblock \bibinfo{booktitle}{\emph{Thematic analysis.}}
\newblock \bibinfo{publisher}{American Psychological Association},
  \bibinfo{address}{Washington, DC}.
\newblock
\urldef\tempurl%
\url{https://doi.org/10.1037/13620-004}
\showDOI{\tempurl}


\bibitem[\protect\citeauthoryear{Brox, Konstantinidis, Evertsen,
  et~al\mbox{.}}{Brox et~al\mbox{.}}{2017}]%
        {brox2017user}
\bibfield{author}{\bibinfo{person}{Ellen Brox}, \bibinfo{person}{Stathis~Th
  Konstantinidis}, \bibinfo{person}{Gunn Evertsen}, {et~al\mbox{.}}}
  \bibinfo{year}{2017}\natexlab{}.
\newblock \showarticletitle{User-centered design of serious games for older
  adults following 3 years of experience with exergames for seniors: a study
  design}.
\newblock \bibinfo{journal}{\emph{JMIR serious games}} \bibinfo{volume}{5},
  \bibinfo{number}{1} (\bibinfo{year}{2017}), \bibinfo{pages}{e6254}.
\newblock


\bibitem[\protect\citeauthoryear{Burnard}{Burnard}{2000}]%
        {burnard2000children}
\bibfield{author}{\bibinfo{person}{Pamela Burnard}.}
  \bibinfo{year}{2000}\natexlab{}.
\newblock \showarticletitle{How Children Ascribe Meaning to Improvisation and
  Composition: Rethinking pedagogy in music education}.
\newblock \bibinfo{journal}{\emph{Music Education Research}}
  \bibinfo{volume}{2}, \bibinfo{number}{1} (\bibinfo{year}{2000}),
  \bibinfo{pages}{7--23}.
\newblock
\urldef\tempurl%
\url{https://doi.org/10.1080/14613800050004404}
\showDOI{\tempurl}


\bibitem[\protect\citeauthoryear{Burns}{Burns}{2020}]%
        {burns2020using}
\bibfield{author}{\bibinfo{person}{A.M. Burns}.}
  \bibinfo{year}{2020}\natexlab{}.
\newblock \bibinfo{booktitle}{\emph{Using Technology with Elementary Music
  Approaches}}.
\newblock \bibinfo{publisher}{Oxford University Press, Incorporated},
  \bibinfo{address}{New York, NY, USA}.
\newblock
\showISBNx{9780190055646}
\showLCCN{2019059944}
\urldef\tempurl%
\url{https://books.google.si/books?id=pyT3DwAAQBAJ}
\showURL{%
\tempurl}


\bibitem[\protect\citeauthoryear{Buschek, De~Luca, and Alt}{Buschek
  et~al\mbox{.}}{2015}]%
        {buschek2015improving}
\bibfield{author}{\bibinfo{person}{Daniel Buschek}, \bibinfo{person}{Alexander
  De~Luca}, {and} \bibinfo{person}{Florian Alt}.}
  \bibinfo{year}{2015}\natexlab{}.
\newblock \showarticletitle{Improving Accuracy, Applicability and Usability of
  Keystroke Biometrics on Mobile Touchscreen Devices}. In
  \bibinfo{booktitle}{\emph{Proceedings of the 33rd Annual ACM Conference on
  Human Factors in Computing Systems}} (Seoul, Republic of Korea)
  \emph{(\bibinfo{series}{CHI '15})}. \bibinfo{publisher}{Association for
  Computing Machinery}, \bibinfo{address}{New York, NY, USA},
  \bibinfo{pages}{1393–1402}.
\newblock
\showISBNx{9781450331456}
\urldef\tempurl%
\url{https://doi.org/10.1145/2702123.2702252}
\showDOI{\tempurl}


\bibitem[\protect\citeauthoryear{Cai, Amrizal, Abe, and Suganuma}{Cai
  et~al\mbox{.}}{2019a}]%
        {cai2019designb}
\bibfield{author}{\bibinfo{person}{Minya Cai}, \bibinfo{person}{Muhammad~Alfian
  Amrizal}, \bibinfo{person}{Toru Abe}, {and} \bibinfo{person}{Takuo
  Suganuma}.} \bibinfo{year}{2019}\natexlab{a}.
\newblock \showarticletitle{Design and Implementation of AR-Supported System
  for Piano Learning}. In \bibinfo{booktitle}{\emph{IEEE 8th Global Conference
  on Consumer Electronics}} \emph{(\bibinfo{series}{GCCE '19})}.
  \bibinfo{publisher}{IEEE}, \bibinfo{address}{New York, NY, USA},
  \bibinfo{pages}{49--50}.
\newblock
\urldef\tempurl%
\url{https://doi.org/10.1109/GCCE46687.2019.9015530}
\showDOI{\tempurl}


\bibitem[\protect\citeauthoryear{Cai, Amrizal, Abe, and Suganuma}{Cai
  et~al\mbox{.}}{2019b}]%
        {cai2019designa}
\bibfield{author}{\bibinfo{person}{Minya Cai}, \bibinfo{person}{Muhammad~Alfian
  Amrizal}, \bibinfo{person}{Toru Abe}, {and} \bibinfo{person}{Takuo
  Suganuma}.} \bibinfo{year}{2019}\natexlab{b}.
\newblock \showarticletitle{Design of an AR-Based System for Group Piano
  Learning}. In \bibinfo{booktitle}{\emph{IEEE International Symposium on Mixed
  and Augmented Reality Adjunct}} \emph{(\bibinfo{series}{ISMAR-Adjunct '19})}.
  \bibinfo{publisher}{IEEE}, \bibinfo{address}{New York, NY, USA},
  \bibinfo{pages}{20--21}.
\newblock
\urldef\tempurl%
\url{https://doi.org/10.1109/ISMAR-Adjunct.2019.00020}
\showDOI{\tempurl}


\bibitem[\protect\citeauthoryear{Carabo-Cone}{Carabo-Cone}{1969}]%
        {carabo1969sensory}
\bibfield{author}{\bibinfo{person}{Madeleine Carabo-Cone}.}
  \bibinfo{year}{1969}\natexlab{}.
\newblock \bibinfo{booktitle}{\emph{A Sensory-Motor Approach to Music Learning.
  Book I-Primary Concepts.}}
\newblock \bibinfo{publisher}{ERIC}, \bibinfo{address}{New York, NY, USA}.
\newblock


\bibitem[\protect\citeauthoryear{Carroll and Latulipe}{Carroll and
  Latulipe}{2009}]%
        {carroll2009creativity}
\bibfield{author}{\bibinfo{person}{Erin~A. Carroll} {and}
  \bibinfo{person}{Celine Latulipe}.} \bibinfo{year}{2009}\natexlab{}.
\newblock \showarticletitle{The Creativity Support Index}. In
  \bibinfo{booktitle}{\emph{CHI '09 Extended Abstracts on Human Factors in
  Computing Systems}} (Boston, MA, USA) \emph{(\bibinfo{series}{CHI EA '09})}.
  \bibinfo{publisher}{Association for Computing Machinery},
  \bibinfo{address}{New York, NY, USA}, \bibinfo{pages}{4009–4014}.
\newblock
\showISBNx{9781605582474}
\urldef\tempurl%
\url{https://doi.org/10.1145/1520340.1520609}
\showDOI{\tempurl}


\bibitem[\protect\citeauthoryear{Chang}{Chang}{1994}]%
        {chang1994russian}
\bibfield{author}{\bibinfo{person}{Anita Lee-Ling Chang}.}
  \bibinfo{year}{1994}\natexlab{}.
\newblock \emph{\bibinfo{title}{The Russian School of advanced piano technique:
  Its history and development from the 19th to 20th century}}.
\newblock \bibinfo{thesistype}{Ph.D. Dissertation}. \bibinfo{school}{The
  University of Texas at Austin}.
\newblock


\bibitem[\protect\citeauthoryear{Chiang and Sun}{Chiang and Sun}{2015}]%
        {chiang2015oncall}
\bibfield{author}{\bibinfo{person}{Pei-Ying Chiang} {and}
  \bibinfo{person}{Chung-Hsuan Sun}.} \bibinfo{year}{2015}\natexlab{}.
\newblock \showarticletitle{Oncall Piano Sensei: Portable AR Piano Training
  System}. In \bibinfo{booktitle}{\emph{Proceedings of the 3rd ACM Symposium on
  Spatial User Interaction}} (Los Angeles, California, USA)
  \emph{(\bibinfo{series}{SUI '15})}. \bibinfo{publisher}{Association for
  Computing Machinery}, \bibinfo{address}{New York, NY, USA},
  \bibinfo{pages}{134}.
\newblock
\showISBNx{9781450337038}
\urldef\tempurl%
\url{https://doi.org/10.1145/2788940.2794353}
\showDOI{\tempurl}


\bibitem[\protect\citeauthoryear{Choksy}{Choksy}{1974}]%
        {choksy1974kodaly}
\bibfield{author}{\bibinfo{person}{Lois Choksy}.}
  \bibinfo{year}{1974}\natexlab{}.
\newblock \bibinfo{booktitle}{\emph{The Kod{\'a}ly method: Comprehensive music
  education from infant to adult}}.
\newblock \bibinfo{publisher}{Prentice-Hall Englewood Cliffs, NJ},
  \bibinfo{address}{New Jersey}.
\newblock


\bibitem[\protect\citeauthoryear{Chouvatut and Jindaluang}{Chouvatut and
  Jindaluang}{2013}]%
        {chouvatut2013virtual}
\bibfield{author}{\bibinfo{person}{Varin Chouvatut} {and}
  \bibinfo{person}{Wattana Jindaluang}.} \bibinfo{year}{2013}\natexlab{}.
\newblock \showarticletitle{Virtual piano with real-time interaction using
  automatic marker detection}. In \bibinfo{booktitle}{\emph{2013 International
  Computer Science and Engineering Conference}} \emph{(\bibinfo{series}{ICSEC
  '13})}. \bibinfo{publisher}{IEEE}, \bibinfo{address}{New York, NY, USA},
  \bibinfo{pages}{222--226}.
\newblock
\urldef\tempurl%
\url{https://doi.org/10.1109/ICSEC.2013.6694783}
\showDOI{\tempurl}


\bibitem[\protect\citeauthoryear{Chow, Feng, Amor, and W\"{u}nsche}{Chow
  et~al\mbox{.}}{2013}]%
        {chow2013music}
\bibfield{author}{\bibinfo{person}{Jonathan Chow}, \bibinfo{person}{Haoyang
  Feng}, \bibinfo{person}{Robert Amor}, {and} \bibinfo{person}{Burkhard~C.
  W\"{u}nsche}.} \bibinfo{year}{2013}\natexlab{}.
\newblock \showarticletitle{Music Education Using Augmented Reality with a Head
  Mounted Display}. In \bibinfo{booktitle}{\emph{Proceedings of the Fourteenth
  Australasian User Interface Conference - Volume 139}} (Melbourne, Australia)
  \emph{(\bibinfo{series}{AUIC '13})}. \bibinfo{publisher}{Australian Computer
  Society, Inc.}, \bibinfo{address}{AUS}, \bibinfo{pages}{73–79}.
\newblock
\showISBNx{9781921770241}


\bibitem[\protect\citeauthoryear{Chyu}{Chyu}{2004}]%
        {chyu2004teaching}
\bibfield{author}{\bibinfo{person}{Yawen~E. Chyu}.}
  \bibinfo{year}{2004}\natexlab{}.
\newblock \emph{\bibinfo{title}{Teaching improvisation to piano students of
  elementary to intermediate levels}}.
\newblock \bibinfo{thesistype}{Ph.D. Dissertation}. \bibinfo{school}{Ohio State
  University}.
\newblock
\showISBNx{978-0-496-94607-5}


\bibitem[\protect\citeauthoryear{Conklin~Jr}{Conklin~Jr}{1987}]%
        {conklin1987piano}
\bibfield{author}{\bibinfo{person}{Harold~A Conklin~Jr}.}
  \bibinfo{year}{1987}\natexlab{}.
\newblock \showarticletitle{Piano design factors—their influence on tone and
  acoustical performance}.
\newblock \bibinfo{journal}{\emph{The Journal of the Acoustical Society of
  America}} \bibinfo{volume}{81}, \bibinfo{number}{S1} (\bibinfo{year}{1987}),
  \bibinfo{pages}{S60--S60}.
\newblock


\bibitem[\protect\citeauthoryear{Correa, Ficheman, Nascimento, and
  Lopes}{Correa et~al\mbox{.}}{2009}]%
        {correa2009computer}
\bibfield{author}{\bibinfo{person}{Ana Grasielle~Dion\'{\i}sio Correa},
  \bibinfo{person}{Irene~Karaguilla Ficheman}, \bibinfo{person}{Marilena~do
  Nascimento}, {and} \bibinfo{person}{Roseli de~Deus Lopes}.}
  \bibinfo{year}{2009}\natexlab{}.
\newblock \showarticletitle{Computer Assisted Music Therapy: A Case Study of an
  Augmented Reality Musical System for Children with Cerebral Palsy
  Rehabilitation}. In \bibinfo{booktitle}{\emph{Proceedings of the 2009 Ninth
  IEEE International Conference on Advanced Learning Technologies}}
  \emph{(\bibinfo{series}{ICALT '09})}. \bibinfo{publisher}{IEEE Computer
  Society}, \bibinfo{address}{USA}, \bibinfo{pages}{218–220}.
\newblock
\showISBNx{9780769537115}
\urldef\tempurl%
\url{https://doi.org/10.1109/ICALT.2009.111}
\showDOI{\tempurl}


\bibitem[\protect\citeauthoryear{Dahlstedt}{Dahlstedt}{2015}]%
        {dahlstedt2015mapping}
\bibfield{author}{\bibinfo{person}{Palle Dahlstedt}.}
  \bibinfo{year}{2015}\natexlab{}.
\newblock \showarticletitle{Mapping Strategies and Sound Engine Design for an
  Augmented Hybrid Piano}. In \bibinfo{booktitle}{\emph{Proceedings of the
  International Conference on New Interfaces for Musical Expression}} (Baton
  Rouge, Louisiana, USA) \emph{(\bibinfo{series}{NIME 2015})}.
  \bibinfo{publisher}{The School of Music and the Center for Computation and
  Technology (CCT), Louisiana State University}, \bibinfo{address}{Baton Rouge,
  Louisiana, USA}, \bibinfo{pages}{271–276}.
\newblock
\showISBNx{9780692495476}


\bibitem[\protect\citeauthoryear{Das, Glickman, Hsiao, and Lee}{Das
  et~al\mbox{.}}{2017}]%
        {das2017music}
\bibfield{author}{\bibinfo{person}{Shantanu Das}, \bibinfo{person}{Seth
  Glickman}, \bibinfo{person}{Fu~Yen Hsiao}, {and} \bibinfo{person}{Byunghwan
  Lee}.} \bibinfo{year}{2017}\natexlab{}.
\newblock \showarticletitle{Music Everywhere--Augmented Reality Piano
  Improvisation Learning System}. In \bibinfo{booktitle}{\emph{Proceedings of
  the International Conference on New Interfaces for Musical Expression}}
  (Aalborg University Copenhagen, Copenhagen, Denmark)
  \emph{(\bibinfo{series}{NIME '17})}. \bibinfo{publisher}{PubPub},
  \bibinfo{address}{Cambridge, MA, USA}, \bibinfo{pages}{511--512}.
\newblock


\bibitem[\protect\citeauthoryear{Davis}{Davis}{1993}]%
        {davis1993user}
\bibfield{author}{\bibinfo{person}{Fred~D Davis}.}
  \bibinfo{year}{1993}\natexlab{}.
\newblock \showarticletitle{User acceptance of information technology: system
  characteristics, user perceptions and behavioral impacts}.
\newblock \bibinfo{journal}{\emph{International journal of man-machine
  studies}} \bibinfo{volume}{38}, \bibinfo{number}{3} (\bibinfo{year}{1993}),
  \bibinfo{pages}{475--487}.
\newblock


\bibitem[\protect\citeauthoryear{De~Pra, Fontana, and Tao}{De~Pra
  et~al\mbox{.}}{2014}]%
        {de2014infrared}
\bibfield{author}{\bibinfo{person}{Yuri De~Pra}, \bibinfo{person}{Federico
  Fontana}, {and} \bibinfo{person}{Linmi Tao}.}
  \bibinfo{year}{2014}\natexlab{}.
\newblock \showarticletitle{Infrared vs. Ultrasonic Finger Detection on a
  Virtual Piano Keyboard}. In \bibinfo{booktitle}{\emph{Proceedings of the 2014
  International Computer Music Conference}} \emph{(\bibinfo{series}{ICMC
  '14})}. \bibinfo{publisher}{Michigan Publishing}, \bibinfo{address}{Ann
  Arbor, MI, USA}, \bibinfo{pages}{654--658}.
\newblock


\bibitem[\protect\citeauthoryear{Dede}{Dede}{1996}]%
        {dede1996evolution}
\bibfield{author}{\bibinfo{person}{Chris Dede}.}
  \bibinfo{year}{1996}\natexlab{}.
\newblock \showarticletitle{The evolution of distance education: Emerging
  technologies and distributed learning}.
\newblock \bibinfo{journal}{\emph{American Journal of Distance Education}}
  \bibinfo{volume}{10}, \bibinfo{number}{2} (\bibinfo{year}{1996}),
  \bibinfo{pages}{4--36}.
\newblock
\urldef\tempurl%
\url{https://doi.org/10.1080/08923649609526919}
\showDOI{\tempurl}


\bibitem[\protect\citeauthoryear{Deja}{Deja}{2021a}]%
        {deja2021adaptive}
\bibfield{author}{\bibinfo{person}{Jordan~Aiko Deja}.}
  \bibinfo{year}{2021}\natexlab{a}.
\newblock \showarticletitle{Adaptive Visualisations Using Spatiotemporal and
  Heuristic Models to Support Piano Learning}. In
  \bibinfo{booktitle}{\emph{Proceedings of the 29th ACM Conference on User
  Modeling, Adaptation and Personalization}} (Utrecht, Netherlands)
  \emph{(\bibinfo{series}{UMAP '21})}. \bibinfo{publisher}{Association for
  Computing Machinery}, \bibinfo{address}{New York, NY, USA},
  \bibinfo{pages}{286–290}.
\newblock
\showISBNx{9781450383660}
\urldef\tempurl%
\url{https://doi.org/10.1145/3450613.3459656}
\showDOI{\tempurl}


\bibitem[\protect\citeauthoryear{Deja}{Deja}{2021b}]%
        {deja2021encouraging}
\bibfield{author}{\bibinfo{person}{Jordan~Aiko Deja}.}
  \bibinfo{year}{2021}\natexlab{b}.
\newblock \showarticletitle{Encouraging Improvisation in Piano Learning Using
  Adaptive Visualisations and Spatiotemporal Models}. In
  \bibinfo{booktitle}{\emph{Adjunct Publication of the 23rd International
  Conference on Mobile Human-Computer Interaction}}. \bibinfo{publisher}{ACM
  Digital Library}, \bibinfo{address}{Toulousse, France},
  \bibinfo{pages}{1--4}.
\newblock
\urldef\tempurl%
\url{https://doi.org/10.1145/3447527.3474865}
\showDOI{\tempurl}


\bibitem[\protect\citeauthoryear{Deja and Cabredo}{Deja and Cabredo}{2016}]%
        {deja2016discovering}
\bibfield{author}{\bibinfo{person}{Jordan~Aiko Deja} {and}
  \bibinfo{person}{Rafael Cabredo}.} \bibinfo{year}{2016}\natexlab{}.
\newblock \showarticletitle{Discovering Policies using Activity Models of Self
  Regulated Learners}. In \bibinfo{booktitle}{\emph{Proceedings of the 16th
  Philippine Computing Science Congress}}. \bibinfo{publisher}{Computing
  Society of the Philippines}, \bibinfo{address}{Diliman, Quezon City},
  \bibinfo{pages}{10}.
\newblock


\bibitem[\protect\citeauthoryear{Deja, Mayer, Pucihar, and Kljun}{Deja
  et~al\mbox{.}}{2022}]%
        {deja2022vision}
\bibfield{author}{\bibinfo{person}{Jordan~Aiko Deja}, \bibinfo{person}{Sven
  Mayer}, \bibinfo{person}{Klen~Čopič Pucihar}, {and}
  \bibinfo{person}{Matjaž Kljun}.} \bibinfo{year}{2022}\natexlab{}.
\newblock \bibinfo{title}{The Vision of a Human-Centered Piano}.
\newblock
\newblock
\urldef\tempurl%
\url{https://doi.org/10.48550/ARXIV.2204.06945}
\showDOI{\tempurl}


\bibitem[\protect\citeauthoryear{Delgado, Fajardo, and Molina-Solana}{Delgado
  et~al\mbox{.}}{2011}]%
        {delgado2011state}
\bibfield{author}{\bibinfo{person}{Miguel Delgado}, \bibinfo{person}{Waldo
  Fajardo}, {and} \bibinfo{person}{Miguel Molina-Solana}.}
  \bibinfo{year}{2011}\natexlab{}.
\newblock \showarticletitle{A state of the art on computational music
  performance}.
\newblock \bibinfo{journal}{\emph{Expert Systems with Applications}}
  \bibinfo{volume}{38}, \bibinfo{number}{1} (\bibinfo{year}{2011}),
  \bibinfo{pages}{155--160}.
\newblock
\showISSN{0957-4174}
\urldef\tempurl%
\url{https://doi.org/10.1016/j.eswa.2010.06.033}
\showDOI{\tempurl}


\bibitem[\protect\citeauthoryear{Delzell and Leppla}{Delzell and
  Leppla}{1992}]%
        {delzell1992gender}
\bibfield{author}{\bibinfo{person}{Judith~K. Delzell} {and}
  \bibinfo{person}{David~A. Leppla}.} \bibinfo{year}{1992}\natexlab{}.
\newblock \showarticletitle{Gender Association of Musical Instruments and
  Preferences of Fourth-Grade Students for Selected Instruments}.
\newblock \bibinfo{journal}{\emph{Journal of Research in Music Education}}
  \bibinfo{volume}{40}, \bibinfo{number}{2} (\bibinfo{year}{1992}),
  \bibinfo{pages}{93--103}.
\newblock
\urldef\tempurl%
\url{https://doi.org/10.2307/3345559}
\showDOI{\tempurl}


\bibitem[\protect\citeauthoryear{Deterding, Dixon, Khaled, and Nacke}{Deterding
  et~al\mbox{.}}{2011}]%
        {deterding2011game}
\bibfield{author}{\bibinfo{person}{Sebastian Deterding}, \bibinfo{person}{Dan
  Dixon}, \bibinfo{person}{Rilla Khaled}, {and} \bibinfo{person}{Lennart
  Nacke}.} \bibinfo{year}{2011}\natexlab{}.
\newblock \showarticletitle{From Game Design Elements to Gamefulness: Defining
  "Gamification"}. In \bibinfo{booktitle}{\emph{Proceedings of the 15th
  International Academic MindTrek Conference: Envisioning Future Media
  Environments}} (Tampere, Finland) \emph{(\bibinfo{series}{MindTrek '11})}.
  \bibinfo{publisher}{Association for Computing Machinery},
  \bibinfo{address}{New York, NY, USA}, \bibinfo{pages}{9–15}.
\newblock
\showISBNx{9781450308168}
\urldef\tempurl%
\url{https://doi.org/10.1145/2181037.2181040}
\showDOI{\tempurl}


\bibitem[\protect\citeauthoryear{Dimitracopoulou}{Dimitracopoulou}{2007}]%
        {dimitracopoulou2007computer}
\bibfield{author}{\bibinfo{person}{Angelique Dimitracopoulou}.}
  \bibinfo{year}{2007}\natexlab{}.
\newblock \showarticletitle{Computer based interaction analysis supporting
  self-regulation: Achievements and prospects of an emerging research field}.
  In \bibinfo{booktitle}{\emph{Proceedings of the IADIS International Congress
  on Cognition and Exploratory Learning in Digital Age, Algavre, Portugal}}.
  \bibinfo{publisher}{LTEE}, \bibinfo{address}{Greece}, \bibinfo{pages}{93}.
\newblock


\bibitem[\protect\citeauthoryear{Donahue, Simon, and Dieleman}{Donahue
  et~al\mbox{.}}{2019}]%
        {donahue2019piano}
\bibfield{author}{\bibinfo{person}{Chris Donahue}, \bibinfo{person}{Ian Simon},
  {and} \bibinfo{person}{Sander Dieleman}.} \bibinfo{year}{2019}\natexlab{}.
\newblock \showarticletitle{Piano Genie}. In
  \bibinfo{booktitle}{\emph{Proceedings of the 24th International Conference on
  Intelligent User Interfaces}} (Marina del Ray, California)
  \emph{(\bibinfo{series}{IUI '19})}. \bibinfo{publisher}{Association for
  Computing Machinery}, \bibinfo{address}{New York, NY, USA},
  \bibinfo{pages}{160–164}.
\newblock
\showISBNx{9781450362726}
\urldef\tempurl%
\url{https://doi.org/10.1145/3301275.3302288}
\showDOI{\tempurl}


\bibitem[\protect\citeauthoryear{Dunne, Walsh, Smyth, and Caulfield}{Dunne
  et~al\mbox{.}}{2007}]%
        {dunne2007system}
\bibfield{author}{\bibinfo{person}{Lucy Dunne}, \bibinfo{person}{P. Walsh},
  \bibinfo{person}{B. Smyth}, {and} \bibinfo{person}{B. Caulfield}.}
  \bibinfo{year}{2007}\natexlab{}.
\newblock \showarticletitle{A System for Wearable Monitoring of Seated Posture
  in Computer Users}. In \bibinfo{booktitle}{\emph{4th International Workshop
  on Wearable and Implantable Body Sensor Networks}}
  \emph{(\bibinfo{series}{BSN '07})}. \bibinfo{publisher}{Springer Berlin
  Heidelberg}, \bibinfo{address}{Berlin, Heidelberg},
  \bibinfo{pages}{203--207}.
\newblock
\showISBNx{978-3-540-70994-7}


\bibitem[\protect\citeauthoryear{Edwards, Gandini, and Forman}{Edwards
  et~al\mbox{.}}{1998}]%
        {edwards1998introduction}
\bibfield{author}{\bibinfo{person}{C.P. Edwards}, \bibinfo{person}{L. Gandini},
  {and} \bibinfo{person}{G.E. Forman}.} \bibinfo{year}{1998}\natexlab{}.
\newblock \bibinfo{booktitle}{\emph{The Hundred Languages of Children: The
  Reggio Emilia Approach--advanced Reflections}}.
\newblock \bibinfo{publisher}{Ablex Publishing Corporation},
  \bibinfo{address}{New York}.
\newblock
\showISBNx{9781567503111}
\showLCCN{97022318}


\bibitem[\protect\citeauthoryear{Ekstrom, Dermen, and Harman}{Ekstrom
  et~al\mbox{.}}{1976}]%
        {ekstrom1976manual}
\bibfield{author}{\bibinfo{person}{Ruth~B Ekstrom}, \bibinfo{person}{Diran
  Dermen}, {and} \bibinfo{person}{Harry~Horace Harman}.}
  \bibinfo{year}{1976}\natexlab{}.
\newblock \bibinfo{booktitle}{\emph{Manual for kit of factor-referenced
  cognitive tests}}. Vol.~\bibinfo{volume}{102}.
\newblock \bibinfo{publisher}{Educational testing service},
  \bibinfo{address}{Princeton, NJ, USA}.
\newblock


\bibitem[\protect\citeauthoryear{Fernandez, Paliyawan, Yin, and
  Thawonmas}{Fernandez et~al\mbox{.}}{2016}]%
        {fernandez2016piano}
\bibfield{author}{\bibinfo{person}{Carlos A~Torres Fernandez},
  \bibinfo{person}{Pujana Paliyawan}, \bibinfo{person}{Chu~Chun Yin}, {and}
  \bibinfo{person}{Ruck Thawonmas}.} \bibinfo{year}{2016}\natexlab{}.
\newblock \showarticletitle{Piano learning application with feedback provided
  by an ar virtual character}. In \bibinfo{booktitle}{\emph{5th Global
  Conference on Consumer Electronics}}. \bibinfo{publisher}{IEEE},
  \bibinfo{address}{New York, NY, USA}, \bibinfo{pages}{1--2}.
\newblock
\urldef\tempurl%
\url{https://doi.org/10.1109/GCCE.2016.7800380}
\showDOI{\tempurl}


\bibitem[\protect\citeauthoryear{Gerry, Dahl, and Serafin}{Gerry
  et~al\mbox{.}}{2019}]%
        {gerry2019adept}
\bibfield{author}{\bibinfo{person}{Lynda Gerry}, \bibinfo{person}{Sofia Dahl},
  {and} \bibinfo{person}{Stefania Serafin}.} \bibinfo{year}{2019}\natexlab{}.
\newblock \showarticletitle{ADEPT: exploring the design, pedagogy, and analysis
  of a mixed reality application for piano training}. In
  \bibinfo{booktitle}{\emph{Proceedings of 16th Sound \& Music Computing
  Conference}}. \bibinfo{publisher}{Zenodo}, \bibinfo{address}{Málaga, Spain},
  \bibinfo{pages}{2891--2892}.
\newblock
\urldef\tempurl%
\url{https://doi.org/10.5281/zenodo.3249333}
\showDOI{\tempurl}


\bibitem[\protect\citeauthoryear{Goebl and Palmer}{Goebl and Palmer}{2009}]%
        {goebl2009synchronization}
\bibfield{author}{\bibinfo{person}{Werner Goebl} {and}
  \bibinfo{person}{Caroline Palmer}.} \bibinfo{year}{2009}\natexlab{}.
\newblock \showarticletitle{Synchronization of timing and motion among
  performing musicians}.
\newblock \bibinfo{journal}{\emph{Music Perception}} \bibinfo{volume}{26},
  \bibinfo{number}{5} (\bibinfo{year}{2009}), \bibinfo{pages}{427--438}.
\newblock


\bibitem[\protect\citeauthoryear{Goguey, Gutwin, Chen, Suwanaposee, and
  Cockburn}{Goguey et~al\mbox{.}}{2021}]%
        {goguey2021interaction}
\bibfield{author}{\bibinfo{person}{Alix Goguey}, \bibinfo{person}{Carl Gutwin},
  \bibinfo{person}{Zhe Chen}, \bibinfo{person}{Pang Suwanaposee}, {and}
  \bibinfo{person}{Andy Cockburn}.} \bibinfo{year}{2021}\natexlab{}.
\newblock \showarticletitle{Interaction Pace and User Preferences}. In
  \bibinfo{booktitle}{\emph{Proceedings of the 2021 CHI Conference on Human
  Factors in Computing Systems}} (Yokohama, Japan) \emph{(\bibinfo{series}{CHI
  '21})}. \bibinfo{publisher}{Association for Computing Machinery},
  \bibinfo{address}{New York, NY, USA}, Article \bibinfo{articleno}{195},
  \bibinfo{numpages}{14}~pages.
\newblock
\showISBNx{9781450380966}
\urldef\tempurl%
\url{https://doi.org/10.1145/3411764.3445772}
\showDOI{\tempurl}


\bibitem[\protect\citeauthoryear{Goodwin and Green}{Goodwin and Green}{2013}]%
        {goodwin2013key}
\bibfield{author}{\bibinfo{person}{Adam Goodwin} {and} \bibinfo{person}{Richard
  Green}.} \bibinfo{year}{2013}\natexlab{}.
\newblock \showarticletitle{Key detection for a virtual piano teacher}. In
  \bibinfo{booktitle}{\emph{28th International Conference on Image and Vision
  Computing New Zealand}} \emph{(\bibinfo{series}{IVCNZ '13})}.
  \bibinfo{publisher}{IEEE}, \bibinfo{address}{New York, NY, USA},
  \bibinfo{pages}{282--287}.
\newblock
\urldef\tempurl%
\url{https://doi.org/10.1109/IVCNZ.2013.6727030}
\showDOI{\tempurl}


\bibitem[\protect\citeauthoryear{Gordon}{Gordon}{2003}]%
        {gordon2003music}
\bibfield{author}{\bibinfo{person}{Edwin~E. Gordon}.}
  \bibinfo{year}{2003}\natexlab{}.
\newblock \bibinfo{booktitle}{\emph{A Music Learning Theory for Newborn and
  Young Children}}.
\newblock \bibinfo{publisher}{GIA}, \bibinfo{address}{Chicago}.
\newblock
\showISBNx{9781579992590}


\bibitem[\protect\citeauthoryear{Gordon}{Gordon}{2011}]%
        {gordon2011roots}
\bibfield{author}{\bibinfo{person}{Edwin~E. Gordon}.}
  \bibinfo{year}{2011}\natexlab{}.
\newblock \emph{\bibinfo{title}{Roots of music learning theory and audiation}}.
\newblock \bibinfo{thesistype}{Ph.D. Dissertation}. \bibinfo{school}{University
  of South Carolina}, \bibinfo{address}{Chicago}.
\newblock


\bibitem[\protect\citeauthoryear{Granieri and Dooley}{Granieri and
  Dooley}{201}]%
        {granieri2019reach}
\bibfield{author}{\bibinfo{person}{Niccol{\`o} Granieri} {and}
  \bibinfo{person}{James Dooley}.} \bibinfo{year}{201}\natexlab{}.
\newblock \showarticletitle{Reach, a keyboard-based gesture recognition system
  for live piano sound modulation}. In \bibinfo{booktitle}{\emph{Proceedings of
  the International Conference on New Interfaces for Musical Expression}}
  (Porto Alegre, Brazil) \emph{(\bibinfo{series}{NIME '19})}.
  \bibinfo{publisher}{NIME Proceedings}, \bibinfo{address}{Brazil},
  \bibinfo{pages}{2}.
\newblock


\bibitem[\protect\citeauthoryear{Guest, MacQueen, and Namey}{Guest
  et~al\mbox{.}}{2011}]%
        {guest2011applied}
\bibfield{author}{\bibinfo{person}{Greg Guest}, \bibinfo{person}{Kathleen~M
  MacQueen}, {and} \bibinfo{person}{Emily~E Namey}.}
  \bibinfo{year}{2011}\natexlab{}.
\newblock \bibinfo{booktitle}{\emph{Applied thematic analysis}}.
\newblock \bibinfo{publisher}{sage publications}, \bibinfo{address}{Thousand
  Oaks}.
\newblock


\bibitem[\protect\citeauthoryear{Guo, Cui, Zhao, Li, and Hao}{Guo
  et~al\mbox{.}}{2021}]%
        {guo2021hand}
\bibfield{author}{\bibinfo{person}{Ruoxi Guo}, \bibinfo{person}{Jiahao Cui},
  \bibinfo{person}{Wanru Zhao}, \bibinfo{person}{Shuai Li}, {and}
  \bibinfo{person}{Aimin Hao}.} \bibinfo{year}{2021}\natexlab{}.
\newblock \showarticletitle{Hand-by-Hand Mentor: An AR based Training System
  for Piano Performance}. In \bibinfo{booktitle}{\emph{2021 IEEE Conference on
  Virtual Reality and 3D User Interfaces Abstracts and Workshops (VRW)}}.
  \bibinfo{publisher}{IEEE}, \bibinfo{address}{New York, NY, USA},
  \bibinfo{pages}{436--437}.
\newblock


\bibitem[\protect\citeauthoryear{Hackl and Anthes}{Hackl and Anthes}{2017}]%
        {hackl2017holokeys}
\bibfield{author}{\bibinfo{person}{Dominik Hackl} {and}
  \bibinfo{person}{Christoph Anthes}.} \bibinfo{year}{2017}\natexlab{}.
\newblock \showarticletitle{HoloKeys-An Augmented Reality Application for
  Learning the Piano}. In \bibinfo{booktitle}{\emph{Forum Media Technology}}.
  \bibinfo{publisher}{http://ceur-ws.org/}, \bibinfo{address}{Austria},
  \bibinfo{pages}{140--144}.
\newblock
\newblock
\shownote{http://ceur-ws.org/Vol-2009/fmt-proceedings-2017-paper19.pdf.}


\bibitem[\protect\citeauthoryear{Hadjakos}{Hadjakos}{2012}]%
        {hadjakos2012pianist}
\bibfield{author}{\bibinfo{person}{Aristotelis Hadjakos}.}
  \bibinfo{year}{2012}\natexlab{}.
\newblock \showarticletitle{Pianist motion capture with the Kinect depth
  camera}. In \bibinfo{booktitle}{\emph{Proceedings of the Sound and Music
  Computing Conference}}. \bibinfo{publisher}{Citeseer},
  \bibinfo{address}{Darmstadt}, \bibinfo{pages}{303--310}.
\newblock


\bibitem[\protect\citeauthoryear{Hamari, Shernoff, Rowe, Coller, Asbell-Clarke,
  and Edwards}{Hamari et~al\mbox{.}}{2016}]%
        {hamari2016challenging}
\bibfield{author}{\bibinfo{person}{Juho Hamari}, \bibinfo{person}{David~J
  Shernoff}, \bibinfo{person}{Elizabeth Rowe}, \bibinfo{person}{Brianno
  Coller}, \bibinfo{person}{Jodi Asbell-Clarke}, {and} \bibinfo{person}{Teon
  Edwards}.} \bibinfo{year}{2016}\natexlab{}.
\newblock \showarticletitle{Challenging games help students learn: An empirical
  study on engagement, flow and immersion in game-based learning}.
\newblock \bibinfo{journal}{\emph{Computers in human behavior}}
  \bibinfo{volume}{54} (\bibinfo{year}{2016}), \bibinfo{pages}{170--179}.
\newblock


\bibitem[\protect\citeauthoryear{Hanna}{Hanna}{2007}]%
        {hanna2007new}
\bibfield{author}{\bibinfo{person}{Wendell Hanna}.}
  \bibinfo{year}{2007}\natexlab{}.
\newblock \showarticletitle{The New Bloom's Taxonomy: Implications for Music
  Education}.
\newblock \bibinfo{journal}{\emph{Arts Education Policy Review}}
  \bibinfo{volume}{108}, \bibinfo{number}{4} (\bibinfo{year}{2007}),
  \bibinfo{pages}{7--16}.
\newblock
\urldef\tempurl%
\url{https://doi.org/10.3200/AEPR.108.4.7-16}
\showDOI{\tempurl}


\bibitem[\protect\citeauthoryear{Hart}{Hart}{1988}]%
        {hart1988development}
\bibfield{author}{\bibinfo{person}{Lowell~E. Hart, Sandra G.and~Staveland}.}
  \bibinfo{year}{1988}\natexlab{}.
\newblock \showarticletitle{Development of NASA-TLX (Task Load Index): Results
  of Empirical and Theoretical Research}.
\newblock \bibinfo{journal}{\emph{Advances in Psychology}}
  \bibinfo{volume}{52} (\bibinfo{year}{1988}), \bibinfo{pages}{139--183}.
\newblock
\showISSN{0166-4115}
\urldef\tempurl%
\url{https://doi.org/10.1016/S0166-4115(08)62386-9}
\showDOI{\tempurl}


\bibitem[\protect\citeauthoryear{Hassenzahl, Burmester, and Koller}{Hassenzahl
  et~al\mbox{.}}{2003}]%
        {hassenzahl2003attrakdiff}
\bibfield{author}{\bibinfo{person}{Marc Hassenzahl}, \bibinfo{person}{Michael
  Burmester}, {and} \bibinfo{person}{Franz Koller}.}
  \bibinfo{year}{2003}\natexlab{}.
\newblock \bibinfo{booktitle}{\emph{AttrakDiff: Ein Fragebogen zur Messung
  wahrgenommener hedonischer und pragmatischer Qualit{\"a}t}}.
\newblock \bibinfo{publisher}{Vieweg+Teubner Verlag},
  \bibinfo{address}{Wiesbaden}, \bibinfo{pages}{187--196}.
\newblock
\showISBNx{978-3-322-80058-9}
\urldef\tempurl%
\url{https://doi.org/10.1007/978-3-322-80058-9_19}
\showDOI{\tempurl}


\bibitem[\protect\citeauthoryear{Highben and Palmer}{Highben and
  Palmer}{2004}]%
        {highben2004effects}
\bibfield{author}{\bibinfo{person}{Zebulon Highben} {and}
  \bibinfo{person}{Caroline Palmer}.} \bibinfo{year}{2004}\natexlab{}.
\newblock \showarticletitle{Effects of Auditory and Motor Mental Practice in
  Memorized Piano Performance}.
\newblock \bibinfo{journal}{\emph{Bulletin of the Council for Research in Music
  Education}}  \bibinfo{volume}{159} (\bibinfo{year}{2004}),
  \bibinfo{pages}{58--65}.
\newblock
\showISSN{00109894}
\urldef\tempurl%
\url{http://www.jstor.org/stable/40319208}
\showURL{%
\tempurl}


\bibitem[\protect\citeauthoryear{Huang, Zhou, Yu, Wang, and Du}{Huang
  et~al\mbox{.}}{2011}]%
        {huang2011piano}
\bibfield{author}{\bibinfo{person}{Feng Huang}, \bibinfo{person}{Yu Zhou},
  \bibinfo{person}{Yao Yu}, \bibinfo{person}{Ziqiang Wang}, {and}
  \bibinfo{person}{Sidan Du}.} \bibinfo{year}{2011}\natexlab{}.
\newblock \showarticletitle{Piano AR: A Markerless Augmented Reality Based
  Piano Teaching System}. In \bibinfo{booktitle}{\emph{Proceedings of the 2011
  Third International Conference on Intelligent Human-Machine Systems and
  Cybernetics - Volume 02}} \emph{(\bibinfo{series}{IHMSC '11})}.
  \bibinfo{publisher}{IEEE Computer Society}, \bibinfo{address}{USA},
  \bibinfo{pages}{47–52}.
\newblock
\showISBNx{9781457706769}
\urldef\tempurl%
\url{https://doi.org/10.1109/IHMSC.2011.82}
\showDOI{\tempurl}


\bibitem[\protect\citeauthoryear{Huang and Lee}{Huang and Lee}{2019}]%
        {huang2019modeling}
\bibfield{author}{\bibinfo{person}{Jin Huang} {and} \bibinfo{person}{Byungjoo
  Lee}.} \bibinfo{year}{2019}\natexlab{}.
\newblock \showarticletitle{Modeling Error Rates in Spatiotemporal Moving
  Target Selection}. In \bibinfo{booktitle}{\emph{Extended Abstracts of the
  2019 CHI Conference on Human Factors in Computing Systems}} (Glasgow,
  Scotland Uk) \emph{(\bibinfo{series}{CHI EA '19})}.
  \bibinfo{publisher}{Association for Computing Machinery},
  \bibinfo{address}{New York, NY, USA}, \bibinfo{pages}{1–6}.
\newblock
\showISBNx{9781450359719}
\urldef\tempurl%
\url{https://doi.org/10.1145/3290607.3313077}
\showDOI{\tempurl}


\bibitem[\protect\citeauthoryear{Huta and Ryan}{Huta and Ryan}{2010}]%
        {huta2010pursuing}
\bibfield{author}{\bibinfo{person}{Veronika Huta} {and}
  \bibinfo{person}{Richard~M Ryan}.} \bibinfo{year}{2010}\natexlab{}.
\newblock \showarticletitle{Pursuing pleasure or virtue: The differential and
  overlapping well-being benefits of hedonic and eudaimonic motives}.
\newblock \bibinfo{journal}{\emph{Journal of happiness studies}}
  \bibinfo{volume}{11}, \bibinfo{number}{6} (\bibinfo{year}{2010}),
  \bibinfo{pages}{735--762}.
\newblock
\urldef\tempurl%
\url{https://doi.org/10.1007/s10902-009-9171-4}
\showDOI{\tempurl}


\bibitem[\protect\citeauthoryear{Karolus, Kilian, Kosch, Schmidt, and
  Wozniak}{Karolus et~al\mbox{.}}{2020}]%
        {karolus2020hit}
\bibfield{author}{\bibinfo{person}{Jakob Karolus}, \bibinfo{person}{Annika
  Kilian}, \bibinfo{person}{Thomas Kosch}, \bibinfo{person}{Albrecht Schmidt},
  {and} \bibinfo{person}{Pawe\l{}~W. Wozniak}.}
  \bibinfo{year}{2020}\natexlab{}.
\newblock \showarticletitle{Hit the Thumb Jack! Using Electromyography to
  Augment the Piano Keyboard}. In \bibinfo{booktitle}{\emph{Proceedings of the
  2020 ACM Designing Interactive Systems Conference}}.
  \bibinfo{publisher}{Association for Computing Machinery},
  \bibinfo{address}{New York, NY, USA}, \bibinfo{pages}{429–440}.
\newblock
\showISBNx{9781450369749}
\urldef\tempurl%
\url{https://doi.org/10.1145/3357236.3395500}
\showURL{%
\tempurl}


\bibitem[\protect\citeauthoryear{Karolus and Schmidt}{Karolus and
  Schmidt}{2018}]%
        {karolus2018proficiency}
\bibfield{author}{\bibinfo{person}{Jakob Karolus} {and}
  \bibinfo{person}{Albrecht Schmidt}.} \bibinfo{year}{2018}\natexlab{}.
\newblock \showarticletitle{Proficiency-Aware Systems: Adapting to the User's
  Skills and Expertise}. In \bibinfo{booktitle}{\emph{Proceedings of the 7th
  ACM International Symposium on Pervasive Displays}} (Munich, Germany)
  \emph{(\bibinfo{series}{PerDis '18})}. \bibinfo{publisher}{Association for
  Computing Machinery}, \bibinfo{address}{New York, NY, USA}, Article
  \bibinfo{articleno}{33}, \bibinfo{numpages}{2}~pages.
\newblock
\showISBNx{9781450357654}
\urldef\tempurl%
\url{https://doi.org/10.1145/3205873.3210708}
\showDOI{\tempurl}


\bibitem[\protect\citeauthoryear{Karolus and Wo{\'z}niak}{Karolus and
  Wo{\'z}niak}{2021}]%
        {karolus2021proficiency}
\bibfield{author}{\bibinfo{person}{Jakob Karolus} {and}
  \bibinfo{person}{Pawe{\l}~W. Wo{\'z}niak}.} \bibinfo{year}{2021}\natexlab{}.
\newblock \showarticletitle{Proficiency-aware systems: Designing for user
  reflection in context-aware systems}.
\newblock \bibinfo{journal}{\emph{it-Information Technology}}
  \bibinfo{volume}{63}, \bibinfo{number}{3} (\bibinfo{year}{2021}),
  \bibinfo{pages}{167--175}.
\newblock


\bibitem[\protect\citeauthoryear{Kastner}{Kastner}{2014}]%
        {kastner2014exploring}
\bibfield{author}{\bibinfo{person}{Julie~Derges Kastner}.}
  \bibinfo{year}{2014}\natexlab{}.
\newblock \showarticletitle{Exploring Informal Music Learning in a Professional
  Development Community of Music Teachers}.
\newblock \bibinfo{journal}{\emph{Bulletin of the Council for Research in Music
  Education}}  \bibinfo{volume}{202} (\bibinfo{year}{2014}),
  \bibinfo{pages}{71--89}.
\newblock
\showISSN{00109894}
\urldef\tempurl%
\url{https://doi.org/10.5406/bulcouresmusedu.202.0071}
\showDOI{\tempurl}


\bibitem[\protect\citeauthoryear{Kilian, Karolus, Kosch, Schmidt, and
  Woniak}{Kilian et~al\mbox{.}}{2021}]%
        {kilian2021em}
\bibfield{author}{\bibinfo{person}{Annika Kilian}, \bibinfo{person}{Jakob
  Karolus}, \bibinfo{person}{Thomas Kosch}, \bibinfo{person}{Albrecht Schmidt},
  {and} \bibinfo{person}{Pawe\l{}~W. Woniak}.} \bibinfo{year}{2021}\natexlab{}.
\newblock \showarticletitle{EMPiano: Electromyographic Pitch Control on the
  Piano Keyboard}. In \bibinfo{booktitle}{\emph{Extended Abstracts of the 2021
  CHI Conference on Human Factors in Computing Systems}}.
  \bibinfo{publisher}{Association for Computing Machinery},
  \bibinfo{address}{New York, NY, USA}, Article \bibinfo{articleno}{196},
  \bibinfo{numpages}{4}~pages.
\newblock
\showISBNx{9781450380959}
\urldef\tempurl%
\url{https://doi.org/10.1145/3411763.3451556}
\showURL{%
\tempurl}


\bibitem[\protect\citeauthoryear{Klepsch and Seufert}{Klepsch and
  Seufert}{2012}]%
        {klepsch2012subjective}
\bibfield{author}{\bibinfo{person}{M Klepsch} {and} \bibinfo{person}{T
  Seufert}.} \bibinfo{year}{2012}\natexlab{}.
\newblock \showarticletitle{Subjective differentiated measurement of cognitive
  load}. In \bibinfo{booktitle}{\emph{Proc. 5th Intl. Cognitive Load Theory
  Conf}}. \bibinfo{publisher}{IEE}, \bibinfo{address}{USA}.
\newblock


\bibitem[\protect\citeauthoryear{Kljun, Mariani, and Dix}{Kljun
  et~al\mbox{.}}{2015}]%
        {kljun2015transference}
\bibfield{author}{\bibinfo{person}{Matja{\v{z}} Kljun}, \bibinfo{person}{John
  Mariani}, {and} \bibinfo{person}{Alan Dix}.} \bibinfo{year}{2015}\natexlab{}.
\newblock \showarticletitle{Transference of PIM research prototype concepts to
  the mainstream: successes or failures}.
\newblock \bibinfo{journal}{\emph{Interacting with Computers}}
  \bibinfo{volume}{27}, \bibinfo{number}{2} (\bibinfo{year}{2015}),
  \bibinfo{pages}{73--98}.
\newblock
\urldef\tempurl%
\url{https://doi.org/10.1093/iwc/iwt059}
\showDOI{\tempurl}


\bibitem[\protect\citeauthoryear{Kolb}{Kolb}{2014}]%
        {kolb2014experiential}
\bibfield{author}{\bibinfo{person}{David~A Kolb}.}
  \bibinfo{year}{2014}\natexlab{}.
\newblock \bibinfo{booktitle}{\emph{Experiential learning: Experience as the
  source of learning and development}}.
\newblock \bibinfo{publisher}{FT press}, \bibinfo{address}{New Jersey}.
\newblock


\bibitem[\protect\citeauthoryear{Kosch, Funk, Schmidt, and Chuang}{Kosch
  et~al\mbox{.}}{2018a}]%
        {kosch2018identifying}
\bibfield{author}{\bibinfo{person}{Thomas Kosch}, \bibinfo{person}{Markus
  Funk}, \bibinfo{person}{Albrecht Schmidt}, {and} \bibinfo{person}{Lewis~L.
  Chuang}.} \bibinfo{year}{2018}\natexlab{a}.
\newblock \showarticletitle{Identifying Cognitive Assistance with Mobile
  Electroencephalography: A Case Study with In-Situ Projections for Manual
  Assembly}.
\newblock \bibinfo{journal}{\emph{Proc. ACM Hum.-Comput. Interact.}}
  \bibinfo{volume}{2}, \bibinfo{number}{EICS}, Article \bibinfo{articleno}{11}
  (\bibinfo{date}{jun} \bibinfo{year}{2018}), \bibinfo{numpages}{20}~pages.
\newblock
\urldef\tempurl%
\url{https://doi.org/10.1145/3229093}
\showDOI{\tempurl}


\bibitem[\protect\citeauthoryear{Kosch, Hassib, Buschek, and Schmidt}{Kosch
  et~al\mbox{.}}{2018b}]%
        {kosch2018look}
\bibfield{author}{\bibinfo{person}{Thomas Kosch}, \bibinfo{person}{Mariam
  Hassib}, \bibinfo{person}{Daniel Buschek}, {and} \bibinfo{person}{Albrecht
  Schmidt}.} \bibinfo{year}{2018}\natexlab{b}.
\newblock \showarticletitle{Look into My Eyes: Using Pupil Dilation to Estimate
  Mental Workload for Task Complexity Adaptation}. In
  \bibinfo{booktitle}{\emph{Extended Abstracts of the 2018 CHI Conference on
  Human Factors in Computing Systems}} (Montreal QC, Canada)
  \emph{(\bibinfo{series}{CHI EA '18})}. \bibinfo{publisher}{Association for
  Computing Machinery}, \bibinfo{address}{New York, NY, USA},
  \bibinfo{pages}{1–6}.
\newblock
\showISBNx{9781450356213}
\urldef\tempurl%
\url{https://doi.org/10.1145/3170427.3188643}
\showDOI{\tempurl}


\bibitem[\protect\citeauthoryear{Lam, Gutwin, Klarkowski, and Cockburn}{Lam
  et~al\mbox{.}}{2021}]%
        {lam2021effects}
\bibfield{author}{\bibinfo{person}{Kevin~C. Lam}, \bibinfo{person}{Carl
  Gutwin}, \bibinfo{person}{Madison Klarkowski}, {and} \bibinfo{person}{Andy
  Cockburn}.} \bibinfo{year}{2021}\natexlab{}.
\newblock \showarticletitle{The Effects of System Interpretation Errors on
  Learning New Input Mechanisms}. In \bibinfo{booktitle}{\emph{Proceedings of
  the 2021 CHI Conference on Human Factors in Computing Systems}} (Yokohama,
  Japan) \emph{(\bibinfo{series}{CHI '21})}. \bibinfo{publisher}{Association
  for Computing Machinery}, \bibinfo{address}{New York, NY, USA}, Article
  \bibinfo{articleno}{713}, \bibinfo{numpages}{13}~pages.
\newblock
\showISBNx{9781450380966}
\urldef\tempurl%
\url{https://doi.org/10.1145/3411764.3445366}
\showDOI{\tempurl}


\bibitem[\protect\citeauthoryear{Laroche and Kaddouch}{Laroche and
  Kaddouch}{2014}]%
        {laroche2014enacting}
\bibfield{author}{\bibinfo{person}{Julien Laroche} {and} \bibinfo{person}{Ilan
  Kaddouch}.} \bibinfo{year}{2014}\natexlab{}.
\newblock \showarticletitle{Enacting teaching and learning in the interaction
  process:“Keys” for developing skills in piano lessons through four-hand
  improvisations}.
\newblock \bibinfo{journal}{\emph{Journal of Pedagogy}} \bibinfo{volume}{5},
  \bibinfo{number}{1} (\bibinfo{year}{2014}), \bibinfo{pages}{24--47}.
\newblock
\urldef\tempurl%
\url{https://doi.org/10.2478/jped-2014-0002}
\showDOI{\tempurl}


\bibitem[\protect\citeauthoryear{Lee and Oulasvirta}{Lee and
  Oulasvirta}{2016}]%
        {lee2016modelling}
\bibfield{author}{\bibinfo{person}{Byungjoo Lee} {and} \bibinfo{person}{Antti
  Oulasvirta}.} \bibinfo{year}{2016}\natexlab{}.
\newblock \showarticletitle{Modelling Error Rates in Temporal Pointing}. In
  \bibinfo{booktitle}{\emph{Proceedings of the 2016 CHI Conference on Human
  Factors in Computing Systems}} (San Jose, California, USA)
  \emph{(\bibinfo{series}{CHI '16})}. \bibinfo{publisher}{Association for
  Computing Machinery}, \bibinfo{address}{New York, NY, USA},
  \bibinfo{pages}{1857–1868}.
\newblock
\showISBNx{9781450333627}
\urldef\tempurl%
\url{https://doi.org/10.1145/2858036.2858143}
\showDOI{\tempurl}


\bibitem[\protect\citeauthoryear{Lewis and Sauro}{Lewis and Sauro}{2009}]%
        {lewis2009factor}
\bibfield{author}{\bibinfo{person}{James~R. Lewis} {and} \bibinfo{person}{Jeff
  Sauro}.} \bibinfo{year}{2009}\natexlab{}.
\newblock \showarticletitle{The Factor Structure of the System Usability
  Scale}. In \bibinfo{booktitle}{\emph{Human Centered Design}},
  \bibfield{editor}{\bibinfo{person}{Masaaki Kurosu}} (Ed.).
  \bibinfo{publisher}{Springer Berlin Heidelberg}, \bibinfo{address}{Berlin,
  Heidelberg}, \bibinfo{pages}{94--103}.
\newblock
\showISBNx{978-3-642-02806-9}
\urldef\tempurl%
\url{https://doi.org/10.1007/978-3-642-02806-9_12}
\showDOI{\tempurl}


\bibitem[\protect\citeauthoryear{Liang, Fazekas, McPherson, and Sandler}{Liang
  et~al\mbox{.}}{2017}]%
        {liang2017piano}
\bibfield{author}{\bibinfo{person}{Beici Liang}, \bibinfo{person}{Gy{\"o}rgy
  Fazekas}, \bibinfo{person}{Andrew McPherson}, {and} \bibinfo{person}{Mark
  Sandler}.} \bibinfo{year}{2017}\natexlab{}.
\newblock \showarticletitle{Piano pedaller: a measurement system for
  classification and visualisation of piano pedalling techniques}. In
  \bibinfo{booktitle}{\emph{Proceedings of NIME New Interfaces for Musical
  Expression 2017}}. \bibinfo{publisher}{NIME}, \bibinfo{address}{Denmark},
  \bibinfo{pages}{325--359}.
\newblock


\bibitem[\protect\citeauthoryear{Liang, Wang, Sun, Liu, Yuan, Luo, and
  He}{Liang et~al\mbox{.}}{2016}]%
        {liang2016barehanded}
\bibfield{author}{\bibinfo{person}{Hui Liang}, \bibinfo{person}{Jin Wang},
  \bibinfo{person}{Qian Sun}, \bibinfo{person}{Yong-Jin Liu},
  \bibinfo{person}{Junsong Yuan}, \bibinfo{person}{Jun Luo}, {and}
  \bibinfo{person}{Ying He}.} \bibinfo{year}{2016}\natexlab{}.
\newblock \showarticletitle{Barehanded Music: Real-Time Hand Interaction for
  Virtual Piano}. In \bibinfo{booktitle}{\emph{Proceedings of the 20th ACM
  SIGGRAPH Symposium on Interactive 3D Graphics and Games}} (Redmond,
  Washington) \emph{(\bibinfo{series}{I3D '16})}.
  \bibinfo{publisher}{Association for Computing Machinery},
  \bibinfo{address}{New York, NY, USA}, \bibinfo{pages}{87–94}.
\newblock
\showISBNx{9781450340434}
\urldef\tempurl%
\url{https://doi.org/10.1145/2856400.2856411}
\showDOI{\tempurl}


\bibitem[\protect\citeauthoryear{Lyons and Zelazo}{Lyons and Zelazo}{2011}]%
        {lyons2011monitoring}
\bibfield{author}{\bibinfo{person}{Kristen~E. Lyons} {and}
  \bibinfo{person}{Philip~David Zelazo}.} \bibinfo{year}{2011}\natexlab{}.
\newblock \showarticletitle{Chapter 10 - Monitoring, metacognition, and
  executive function: Elucidating the role of self-reflection in the
  development of self-regulation}.
\newblock \bibinfo{journal}{\emph{Advances in Child Development and Behavior}}
  \bibinfo{volume}{40} (\bibinfo{year}{2011}), \bibinfo{pages}{379--412}.
\newblock
\showISSN{0065-2407}
\urldef\tempurl%
\url{https://doi.org/10.1016/B978-0-12-386491-8.00010-4}
\showDOI{\tempurl}


\bibitem[\protect\citeauthoryear{Magnusson and Mendieta}{Magnusson and
  Mendieta}{2007}]%
        {magnusson2007acoustic}
\bibfield{author}{\bibinfo{person}{Thor Magnusson} {and}
  \bibinfo{person}{Enrike~Hurtado Mendieta}.} \bibinfo{year}{2007}\natexlab{}.
\newblock \showarticletitle{The Acoustic, the Digital and the Body: A Survey on
  Musical Instruments}. In \bibinfo{booktitle}{\emph{Proceedings of the 7th
  International Conference on New Interfaces for Musical Expression}} (New
  York, New York) \emph{(\bibinfo{series}{NIME '07})}.
  \bibinfo{publisher}{Association for Computing Machinery},
  \bibinfo{address}{New York, NY, USA}, \bibinfo{pages}{94–99}.
\newblock
\showISBNx{9781450378376}
\urldef\tempurl%
\url{https://doi.org/10.1145/1279740.1279757}
\showDOI{\tempurl}


\bibitem[\protect\citeauthoryear{Maisto, Carey, and Bradizza}{Maisto
  et~al\mbox{.}}{1999}]%
        {maisto1999social}
\bibfield{author}{\bibinfo{person}{Stephen~A Maisto}, \bibinfo{person}{Kate~B
  Carey}, {and} \bibinfo{person}{Clara~M Bradizza}.}
  \bibinfo{year}{1999}\natexlab{}.
\newblock \bibinfo{booktitle}{\emph{Social learning theory.}}
\newblock \bibinfo{publisher}{The Guilford Press},
  \bibinfo{address}{Washington}.
\newblock


\bibitem[\protect\citeauthoryear{Mattmann and Troster}{Mattmann and
  Troster}{2006}]%
        {mattmann2006design}
\bibfield{author}{\bibinfo{person}{Corinne Mattmann} {and}
  \bibinfo{person}{Gerhard Troster}.} \bibinfo{year}{2006}\natexlab{}.
\newblock \showarticletitle{Design Concept of Clothing Recognizing Back
  Postures}. In \bibinfo{booktitle}{\emph{2006 3rd IEEE/EMBS International
  Summer School on Medical Devices and Biosensors}}. \bibinfo{publisher}{IEEE},
  \bibinfo{address}{New York, NY, USA}, \bibinfo{pages}{24--27}.
\newblock
\urldef\tempurl%
\url{https://doi.org/10.1109/ISSMDBS.2006.360088}
\showDOI{\tempurl}


\bibitem[\protect\citeauthoryear{McCarthy and Goble}{McCarthy and
  Goble}{2002}]%
        {mccarthy2002music}
\bibfield{author}{\bibinfo{person}{Marie McCarthy} {and}
  \bibinfo{person}{J~Scott Goble}.} \bibinfo{year}{2002}\natexlab{}.
\newblock \showarticletitle{Music education philosophy: Changing times}.
\newblock \bibinfo{journal}{\emph{Music Educators Journal}}
  \bibinfo{volume}{89}, \bibinfo{number}{1} (\bibinfo{year}{2002}),
  \bibinfo{pages}{19--26}.
\newblock
\urldef\tempurl%
\url{https://doi.org/10.2307/3399880}
\showDOI{\tempurl}


\bibitem[\protect\citeauthoryear{McPherson}{McPherson}{2015}]%
        {mcpherson2015buttons}
\bibfield{author}{\bibinfo{person}{Andrew McPherson}.}
  \bibinfo{year}{2015}\natexlab{}.
\newblock \showarticletitle{Buttons, Handles, and Keys: Advances in
  Continuous-Control Keyboard Instruments}.
\newblock \bibinfo{journal}{\emph{Comput. Music J.}} \bibinfo{volume}{39},
  \bibinfo{number}{2} (\bibinfo{date}{June} \bibinfo{year}{2015}),
  \bibinfo{pages}{28–46}.
\newblock
\showISSN{0148-9267}
\urldef\tempurl%
\url{https://doi.org/10.1162/COMJ_a_00297}
\showDOI{\tempurl}


\bibitem[\protect\citeauthoryear{McPherson and Kim}{McPherson and Kim}{2010a}]%
        {mcpherson2010toward}
\bibfield{author}{\bibinfo{person}{Andrew McPherson} {and}
  \bibinfo{person}{Youngmoo Kim}.} \bibinfo{year}{2010}\natexlab{a}.
\newblock \showarticletitle{Toward a Computationally-Enhanced Acoustic Grand
  Piano}. In \bibinfo{booktitle}{\emph{CHI '10 Extended Abstracts on Human
  Factors in Computing Systems}} (Atlanta, Georgia, USA)
  \emph{(\bibinfo{series}{CHI EA '10})}. \bibinfo{publisher}{Association for
  Computing Machinery}, \bibinfo{address}{New York, NY, USA},
  \bibinfo{pages}{4141–4146}.
\newblock
\showISBNx{9781605589305}
\urldef\tempurl%
\url{https://doi.org/10.1145/1753846.1754116}
\showDOI{\tempurl}


\bibitem[\protect\citeauthoryear{McPherson}{McPherson}{2013}]%
        {mcpherson2013portable}
\bibfield{author}{\bibinfo{person}{Andrew~P McPherson}.}
  \bibinfo{year}{2013}\natexlab{}.
\newblock \showarticletitle{Portable Measurement and Mapping of Continuous
  Piano Gesture}. In \bibinfo{booktitle}{\emph{International Conference on New
  Interfaces for Musical Expression}} \emph{(\bibinfo{series}{NIME '13})}.
  \bibinfo{publisher}{PubPub}, \bibinfo{address}{Cambridge, MA, USA},
  \bibinfo{pages}{152--157}.
\newblock


\bibitem[\protect\citeauthoryear{McPherson, Gierakowski, and Stark}{McPherson
  et~al\mbox{.}}{2013}]%
        {mcpherson2013space}
\bibfield{author}{\bibinfo{person}{Andrew~P. McPherson},
  \bibinfo{person}{Adrian Gierakowski}, {and} \bibinfo{person}{Adam~M. Stark}.}
  \bibinfo{year}{2013}\natexlab{}.
\newblock \showarticletitle{The Space between the Notes: Adding Expressive
  Pitch Control to the Piano Keyboard}. In
  \bibinfo{booktitle}{\emph{Proceedings of the SIGCHI Conference on Human
  Factors in Computing Systems}} (Paris, France) \emph{(\bibinfo{series}{CHI
  '13})}. \bibinfo{publisher}{Association for Computing Machinery},
  \bibinfo{address}{New York, NY, USA}, \bibinfo{pages}{2195–2204}.
\newblock
\showISBNx{9781450318990}
\urldef\tempurl%
\url{https://doi.org/10.1145/2470654.2481302}
\showDOI{\tempurl}


\bibitem[\protect\citeauthoryear{McPherson and Kim}{McPherson and Kim}{2010b}]%
        {mcpherson2010augmenting}
\bibfield{author}{\bibinfo{person}{Andrew~P McPherson} {and}
  \bibinfo{person}{Youngmoo~E Kim}.} \bibinfo{year}{2010}\natexlab{b}.
\newblock \showarticletitle{Augmenting the Acoustic Piano with Electromagnetic
  String Actuation and Continuous Key Position Sensing.}. In
  \bibinfo{booktitle}{\emph{International Conference on New Interfaces for
  Musical Expression}} \emph{(\bibinfo{series}{NIME '10})}.
  \bibinfo{publisher}{PubPub}, \bibinfo{address}{Cambridge, MA, USA},
  \bibinfo{pages}{217--222}.
\newblock


\bibitem[\protect\citeauthoryear{McPherson and Kim}{McPherson and Kim}{2011}]%
        {mcpherson2011multidimensional}
\bibfield{author}{\bibinfo{person}{Andrew~P. McPherson} {and}
  \bibinfo{person}{Youngmoo~E. Kim}.} \bibinfo{year}{2011}\natexlab{}.
\newblock \showarticletitle{Multidimensional Gesture Sensing at the Piano
  Keyboard}. In \bibinfo{booktitle}{\emph{Proceedings of the SIGCHI Conference
  on Human Factors in Computing Systems}} (Vancouver, BC, Canada)
  \emph{(\bibinfo{series}{CHI '11})}. \bibinfo{publisher}{Association for
  Computing Machinery}, \bibinfo{address}{New York, NY, USA},
  \bibinfo{pages}{2789–2798}.
\newblock
\showISBNx{9781450302289}
\urldef\tempurl%
\url{https://doi.org/10.1145/1978942.1979355}
\showDOI{\tempurl}


\bibitem[\protect\citeauthoryear{McPherson and Kim}{McPherson and Kim}{2012}]%
        {p2012problem}
\bibfield{author}{\bibinfo{person}{Andrew~P. McPherson} {and}
  \bibinfo{person}{Youngmoo~E. Kim}.} \bibinfo{year}{2012}\natexlab{}.
\newblock \showarticletitle{The problem of the second performer: Building a
  community around an augmented piano}.
\newblock \bibinfo{journal}{\emph{Computer Music Journal}}
  \bibinfo{volume}{36}, \bibinfo{number}{4} (\bibinfo{year}{2012}),
  \bibinfo{pages}{10--27}.
\newblock
\urldef\tempurl%
\url{https://doi.org/10.1162/COMJ\_a\_00149}
\showDOI{\tempurl}


\bibitem[\protect\citeauthoryear{Mead}{Mead}{1994}]%
        {mead1994dalcroze}
\bibfield{author}{\bibinfo{person}{Virginia~Hoge Mead}.}
  \bibinfo{year}{1994}\natexlab{}.
\newblock \bibinfo{booktitle}{\emph{Dalcroze Eurhythmics in Today's Music
  Classroom}}.
\newblock \bibinfo{publisher}{Schott}, \bibinfo{address}{New York}.
\newblock
\showISBNx{9780930448516}
\showLCCN{gb94067410}


\bibitem[\protect\citeauthoryear{Mecke, Buschek, Kiermeier, Prange, and
  Alt}{Mecke et~al\mbox{.}}{2019}]%
        {mecke2019exploring}
\bibfield{author}{\bibinfo{person}{Lukas Mecke}, \bibinfo{person}{Daniel
  Buschek}, \bibinfo{person}{Mathias Kiermeier}, \bibinfo{person}{Sarah
  Prange}, {and} \bibinfo{person}{Florian Alt}.}
  \bibinfo{year}{2019}\natexlab{}.
\newblock \showarticletitle{Exploring Intentional Behaviour Modifications for
  Password Typing on Mobile Touchscreen Devices}. In
  \bibinfo{booktitle}{\emph{Proceedings of the Fifteenth USENIX Conference on
  Usable Privacy and Security}} (Santa Clara, CA, USA)
  \emph{(\bibinfo{series}{SOUPS'19})}. \bibinfo{publisher}{USENIX Association},
  \bibinfo{address}{USA}, \bibinfo{pages}{303–318}.
\newblock
\showISBNx{9781939133052}


\bibitem[\protect\citeauthoryear{Michael and Modell}{Michael and
  Modell}{2003}]%
        {michael2003active}
\bibfield{author}{\bibinfo{person}{Joel Michael} {and}
  \bibinfo{person}{Harold~I Modell}.} \bibinfo{year}{2003}\natexlab{}.
\newblock \bibinfo{booktitle}{\emph{Active learning in secondary and college
  science classrooms: A working model for helping the learner to learn}}.
\newblock \bibinfo{publisher}{Routledge}, \bibinfo{address}{New York}.
\newblock
\urldef\tempurl%
\url{https://doi.org/10.4324/9781410609212}
\showDOI{\tempurl}


\bibitem[\protect\citeauthoryear{Moher, Liberati, Tetzlaff, Altman, and
  Group*}{Moher et~al\mbox{.}}{2009}]%
        {moher2009preferred}
\bibfield{author}{\bibinfo{person}{David Moher}, \bibinfo{person}{Alessandro
  Liberati}, \bibinfo{person}{Jennifer Tetzlaff}, \bibinfo{person}{Douglas~G
  Altman}, {and} \bibinfo{person}{PRISMA Group*}.}
  \bibinfo{year}{2009}\natexlab{}.
\newblock \showarticletitle{Preferred Reporting Items for Systematic Reviews
  and Meta-Analyses: The PRISMA Statement}.
\newblock \bibinfo{journal}{\emph{Annals of Internal Medicine}}
  \bibinfo{volume}{151}, \bibinfo{number}{4} (\bibinfo{year}{2009}),
  \bibinfo{pages}{264--269}.
\newblock
\urldef\tempurl%
\url{https://doi.org/10.7326/0003-4819-151-4-200908180-00135}
\showDOI{\tempurl}
\showeprint{https://www.acpjournals.org/doi/pdf/10.7326/0003-4819-151-4-200908180-00135}
\newblock
\shownote{PMID: 19622511.}


\bibitem[\protect\citeauthoryear{Molero, Schez-Sobrino, Vallejo, Glez-Morcillo,
  and Albusac}{Molero et~al\mbox{.}}{2021}]%
        {molero2021novel}
\bibfield{author}{\bibinfo{person}{Diego Molero}, \bibinfo{person}{Santiago
  Schez-Sobrino}, \bibinfo{person}{David Vallejo}, \bibinfo{person}{Carlos
  Glez-Morcillo}, {and} \bibinfo{person}{Javier Albusac}.}
  \bibinfo{year}{2021}\natexlab{}.
\newblock \showarticletitle{A novel approach to learning music and piano based
  on mixed reality and gamification}.
\newblock \bibinfo{journal}{\emph{Multimedia Tools and Applications}}
  \bibinfo{volume}{80}, \bibinfo{number}{1} (\bibinfo{year}{2021}),
  \bibinfo{pages}{165--186}.
\newblock
\urldef\tempurl%
\url{https://doi.org/10.1007/s11042-020-09678-9}
\showDOI{\tempurl}


\bibitem[\protect\citeauthoryear{Molloy, Huang, and W{\"u}nsche}{Molloy
  et~al\mbox{.}}{2019}]%
        {molloy2019mixed}
\bibfield{author}{\bibinfo{person}{Will Molloy}, \bibinfo{person}{Edward
  Huang}, {and} \bibinfo{person}{Burkhard~C W{\"u}nsche}.}
  \bibinfo{year}{2019}\natexlab{}.
\newblock \showarticletitle{Mixed Reality Piano Tutor: A Gamified Piano
  Practice Environment}. In \bibinfo{booktitle}{\emph{2019 International
  Conference on Electronics, Information, and Communication}}
  \emph{(\bibinfo{series}{ICEIC '19})}. \bibinfo{publisher}{IEEE},
  \bibinfo{address}{New York, NY, USA}, \bibinfo{pages}{1--7}.
\newblock
\urldef\tempurl%
\url{https://doi.org/10.23919/ELINFOCOM.2019.8706474}
\showDOI{\tempurl}


\bibitem[\protect\citeauthoryear{Montano}{Montano}{1983}]%
        {montano1984effect}
\bibfield{author}{\bibinfo{person}{David~Ricardo Montano}.}
  \bibinfo{year}{1983}\natexlab{}.
\newblock \emph{\bibinfo{title}{The effect of improvisation in given rhythms on
  rhythmic accuracy in sight reading achievement by college elementary group
  piano students}}.
\newblock \bibinfo{thesistype}{Ph.D. Dissertation}. \bibinfo{school}{University
  of Missouri}.
\newblock


\bibitem[\protect\citeauthoryear{Moro and McPherson}{Moro and
  McPherson}{2020}]%
        {moro2020performer}
\bibfield{author}{\bibinfo{person}{Giulio Moro} {and} \bibinfo{person}{Andrew~P
  McPherson}.} \bibinfo{year}{2020}\natexlab{}.
\newblock \showarticletitle{Performer experience on a continuous keyboard
  instrument}.
\newblock \bibinfo{journal}{\emph{Computer Music Journal}}
  \bibinfo{volume}{44}, \bibinfo{number}{2-3} (\bibinfo{year}{2020}),
  \bibinfo{pages}{69--91}.
\newblock


\bibitem[\protect\citeauthoryear{Mu, Li, and Wu}{Mu et~al\mbox{.}}{2010}]%
        {mu2010sitting}
\bibfield{author}{\bibinfo{person}{Lan Mu}, \bibinfo{person}{Ke Li}, {and}
  \bibinfo{person}{Chunhong Wu}.} \bibinfo{year}{2010}\natexlab{}.
\newblock \showarticletitle{A sitting posture surveillance system based on
  image processing technology}. In \bibinfo{booktitle}{\emph{2010 2nd
  International Conference on Computer Engineering and Technology}},
  Vol.~\bibinfo{volume}{1}. \bibinfo{publisher}{IEEE}, \bibinfo{address}{New
  York, NY, USA}, \bibinfo{pages}{V1--692}.
\newblock
\urldef\tempurl%
\url{https://doi.org/10.1109/ICCET.2010.5485381}
\showDOI{\tempurl}


\bibitem[\protect\citeauthoryear{Newton and Marshall}{Newton and
  Marshall}{2011}]%
        {newton2011examining}
\bibfield{author}{\bibinfo{person}{Dan Newton} {and} \bibinfo{person}{Mark~T
  Marshall}.} \bibinfo{year}{2011}\natexlab{}.
\newblock \showarticletitle{Examining How Musicians Create Augmented Musical
  Instruments.}. In \bibinfo{booktitle}{\emph{International Conference on New
  Interfaces for Musical Expression}} \emph{(\bibinfo{series}{NIME '11})}.
  \bibinfo{publisher}{Proceedings of NIME 2011}, \bibinfo{address}{Oslo,
  Norway}, \bibinfo{pages}{155--160}.
\newblock


\bibitem[\protect\citeauthoryear{Nicolls and Gillian}{Nicolls and
  Gillian}{2012}]%
        {nicolls2012gesturally}
\bibfield{author}{\bibinfo{person}{S Nicolls} {and} \bibinfo{person}{N
  Gillian}.} \bibinfo{year}{2012}\natexlab{}.
\newblock \showarticletitle{A gesturally controlled improvisation system for
  piano}. In \bibinfo{booktitle}{\emph{Proceedings of Live Interfaces:
  Performance Art Music}}. \bibinfo{publisher}{Open Conference Systems},
  \bibinfo{address}{Leeds, UK}, \bibinfo{pages}{1--7}.
\newblock


\bibitem[\protect\citeauthoryear{Ogata and Goto}{Ogata and Goto}{2017}]%
        {ogata2017keyboard}
\bibfield{author}{\bibinfo{person}{Masa Ogata} {and} \bibinfo{person}{Masataka
  Goto}.} \bibinfo{year}{2017}\natexlab{}.
\newblock \showarticletitle{Keyboard Interface with Shape-Distortion Expression
  for Interactive Performance}. In \bibinfo{booktitle}{\emph{Proceedings of the
  2017 International Computer Music Conference}} \emph{(\bibinfo{series}{ICMC
  '17})}. \bibinfo{publisher}{Michigan Publishing}, \bibinfo{address}{Ann
  Arbor, MI, USA}, \bibinfo{pages}{378--383}.
\newblock
\urldef\tempurl%
\url{http://hdl.handle.net/2027/spo.bbp2372.2017.064}
\showURL{%
\tempurl}


\bibitem[\protect\citeauthoryear{Oka and Hashimoto}{Oka and Hashimoto}{2013}]%
        {oka2013marker}
\bibfield{author}{\bibinfo{person}{Akiya Oka} {and} \bibinfo{person}{Manabu
  Hashimoto}.} \bibinfo{year}{2013}\natexlab{}.
\newblock \showarticletitle{Marker-less piano fingering recognition using
  sequential depth images}. In \bibinfo{booktitle}{\emph{The 19th Korea-Japan
  Joint Workshop on Frontiers of Computer Vision}} \emph{(\bibinfo{series}{FCV
  '13})}. \bibinfo{publisher}{IEEE}, \bibinfo{address}{New York, NY, USA},
  \bibinfo{pages}{1--4}.
\newblock
\urldef\tempurl%
\url{https://doi.org/10.1109/FCV.2013.6485449}
\showDOI{\tempurl}


\bibitem[\protect\citeauthoryear{O'Neill and Boultona}{O'Neill and
  Boultona}{1996}]%
        {oneillboysgirlsinstrument}
\bibfield{author}{\bibinfo{person}{Susan~A. O'Neill} {and}
  \bibinfo{person}{Michael~J. Boultona}.} \bibinfo{year}{1996}\natexlab{}.
\newblock \showarticletitle{Boys' and Girls' Preferences for Musical
  Instruments: A Function of Gender?}
\newblock \bibinfo{journal}{\emph{Psychology of Music}} \bibinfo{volume}{24},
  \bibinfo{number}{2} (\bibinfo{year}{1996}), \bibinfo{pages}{171--183}.
\newblock
\urldef\tempurl%
\url{https://doi.org/10.1177/0305735696242009}
\showDOI{\tempurl}


\bibitem[\protect\citeauthoryear{Orji and Vassileva}{Orji and
  Vassileva}{2021}]%
        {orji2021modelling}
\bibfield{author}{\bibinfo{person}{Fidelia~A Orji} {and}
  \bibinfo{person}{Julita Vassileva}.} \bibinfo{year}{2021}\natexlab{}.
\newblock \showarticletitle{Modelling and Quantifying Learner Motivation for
  Adaptive Systems: Current Insight and Future Perspectives}. In
  \bibinfo{booktitle}{\emph{International Conference on Human-Computer
  Interaction}}. \bibinfo{publisher}{Springer}, \bibinfo{address}{Cham},
  \bibinfo{pages}{79--92}.
\newblock


\bibitem[\protect\citeauthoryear{Overholt}{Overholt}{2005}]%
        {overholt2005overtone}
\bibfield{author}{\bibinfo{person}{Dan Overholt}.}
  \bibinfo{year}{2005}\natexlab{}.
\newblock \showarticletitle{The overtone violin: A new computer music
  instrument}. In \bibinfo{booktitle}{\emph{Proceedings of the 2017
  International Computer Music Conference}} \emph{(\bibinfo{series}{ICMC
  '05})}. \bibinfo{publisher}{Michigan Publishing}, \bibinfo{address}{Ann
  Arbor, MI, USA}, \bibinfo{pages}{1--4}.
\newblock


\bibitem[\protect\citeauthoryear{Pan, He, Zeng, Zhou, and Tang}{Pan
  et~al\mbox{.}}{2018}]%
        {pan2018pilot}
\bibfield{author}{\bibinfo{person}{Honghu Pan}, \bibinfo{person}{Xingxi He},
  \bibinfo{person}{Hong Zeng}, \bibinfo{person}{Jia Zhou}, {and}
  \bibinfo{person}{Sai Tang}.} \bibinfo{year}{2018}\natexlab{}.
\newblock \showarticletitle{Pilot Study of Piano Learning with AR Smart Glasses
  Considering Both Single and Paired Play}. In
  \bibinfo{booktitle}{\emph{International Conference on Human Aspects of IT for
  the Aged Population}}. \bibinfo{publisher}{Springer International
  Publishing}, \bibinfo{address}{Cham}, \bibinfo{pages}{561--570}.
\newblock
\showISBNx{978-3-319-92037-5}
\urldef\tempurl%
\url{https://doi.org/10.1007/978-3-319-92037-5_39}
\showDOI{\tempurl}


\bibitem[\protect\citeauthoryear{Park and Lee}{Park and Lee}{2020}]%
        {park2020intermittent}
\bibfield{author}{\bibinfo{person}{Eunji Park} {and} \bibinfo{person}{Byungjoo
  Lee}.} \bibinfo{year}{2020}\natexlab{}.
\newblock \showarticletitle{An Intermittent Click Planning Model}. In
  \bibinfo{booktitle}{\emph{Proceedings of the 2020 CHI Conference on Human
  Factors in Computing Systems}}. \bibinfo{publisher}{Association for Computing
  Machinery}, \bibinfo{address}{New York, NY, USA}, \bibinfo{pages}{1–13}.
\newblock
\showISBNx{9781450367080}
\urldef\tempurl%
\url{https://doi.org/10.1145/3313831.3376725}
\showURL{%
\tempurl}


\bibitem[\protect\citeauthoryear{Peak}{Peak}{1998}]%
        {peak1998suzuki}
\bibfield{author}{\bibinfo{person}{Lois Peak}.}
  \bibinfo{year}{1998}\natexlab{}.
\newblock \bibinfo{booktitle}{\emph{The Suzuki method of music instruction}}.
\newblock \bibinfo{publisher}{Cambridge University Press},
  \bibinfo{address}{California USA}. 345--368 pages.
\newblock


\bibitem[\protect\citeauthoryear{Raymaekers, Vermeulen, Luyten, and
  Coninx}{Raymaekers et~al\mbox{.}}{2014}]%
        {raymaekers2014game}
\bibfield{author}{\bibinfo{person}{Linsey Raymaekers}, \bibinfo{person}{Jo
  Vermeulen}, \bibinfo{person}{Kris Luyten}, {and} \bibinfo{person}{Karin
  Coninx}.} \bibinfo{year}{2014}\natexlab{}.
\newblock \showarticletitle{Game of Tones: Learning to Play Songs on a Piano
  Using Projected Instructions and Games}. In \bibinfo{booktitle}{\emph{CHI '14
  Extended Abstracts on Human Factors in Computing Systems}} (Toronto, Ontario,
  Canada) \emph{(\bibinfo{series}{CHI EA '14})}.
  \bibinfo{publisher}{Association for Computing Machinery},
  \bibinfo{address}{New York, NY, USA}, \bibinfo{pages}{411–414}.
\newblock
\showISBNx{9781450324748}
\urldef\tempurl%
\url{https://doi.org/10.1145/2559206.2574799}
\showDOI{\tempurl}


\bibitem[\protect\citeauthoryear{Robson, Plangger, Kietzmann, McCarthy, and
  Pitt}{Robson et~al\mbox{.}}{2015}]%
        {robson2015all}
\bibfield{author}{\bibinfo{person}{Karen Robson}, \bibinfo{person}{Kirk
  Plangger}, \bibinfo{person}{Jan~H Kietzmann}, \bibinfo{person}{Ian McCarthy},
  {and} \bibinfo{person}{Leyland Pitt}.} \bibinfo{year}{2015}\natexlab{}.
\newblock \showarticletitle{Is it all a game? Understanding the principles of
  gamification}.
\newblock \bibinfo{journal}{\emph{Business horizons}} \bibinfo{volume}{58},
  \bibinfo{number}{4} (\bibinfo{year}{2015}), \bibinfo{pages}{411--420}.
\newblock


\bibitem[\protect\citeauthoryear{Rodrigues, Palomino, Toda, Klock, Oliveira,
  Avila-Santos, Gasparini, and Isotani}{Rodrigues et~al\mbox{.}}{2021}]%
        {rodrigues2021personalization}
\bibfield{author}{\bibinfo{person}{Luiz Rodrigues}, \bibinfo{person}{Paula~T.
  Palomino}, \bibinfo{person}{Armando~M. Toda}, \bibinfo{person}{Ana C.~T.
  Klock}, \bibinfo{person}{Wilk Oliveira}, \bibinfo{person}{Anderson~P.
  Avila-Santos}, \bibinfo{person}{Isabela Gasparini}, {and}
  \bibinfo{person}{Seiji Isotani}.} \bibinfo{year}{2021}\natexlab{}.
\newblock \showarticletitle{Personalization Improves Gamification: Evidence
  from a Mixed-Methods Study}.
\newblock \bibinfo{journal}{\emph{Proc. ACM Hum.-Comput. Interact.}}
  \bibinfo{volume}{5}, \bibinfo{number}{CHI PLAY}, Article
  \bibinfo{articleno}{287} (\bibinfo{date}{oct} \bibinfo{year}{2021}),
  \bibinfo{numpages}{25}~pages.
\newblock
\urldef\tempurl%
\url{https://doi.org/10.1145/3474714}
\showDOI{\tempurl}


\bibitem[\protect\citeauthoryear{Rogers, R\"{o}hlig, Weing, Gugenheimer,
  K\"{o}nings, Klepsch, Schaub, Rukzio, Seufert, and Weber}{Rogers
  et~al\mbox{.}}{2014}]%
        {rogers2014piano}
\bibfield{author}{\bibinfo{person}{Katja Rogers}, \bibinfo{person}{Amrei
  R\"{o}hlig}, \bibinfo{person}{Matthias Weing}, \bibinfo{person}{Jan
  Gugenheimer}, \bibinfo{person}{Bastian K\"{o}nings}, \bibinfo{person}{Melina
  Klepsch}, \bibinfo{person}{Florian Schaub}, \bibinfo{person}{Enrico Rukzio},
  \bibinfo{person}{Tina Seufert}, {and} \bibinfo{person}{Michael Weber}.}
  \bibinfo{year}{2014}\natexlab{}.
\newblock \showarticletitle{P.I.A.N.O.: Faster Piano Learning with Interactive
  Projection}. In \bibinfo{booktitle}{\emph{Proceedings of the Ninth ACM
  International Conference on Interactive Tabletops and Surfaces}} (Dresden,
  Germany) \emph{(\bibinfo{series}{ITS '14})}. \bibinfo{publisher}{Association
  for Computing Machinery}, \bibinfo{address}{New York, NY, USA},
  \bibinfo{pages}{149–158}.
\newblock
\showISBNx{9781450325875}
\urldef\tempurl%
\url{https://doi.org/10.1145/2669485.2669514}
\showDOI{\tempurl}


\bibitem[\protect\citeauthoryear{Russell-Bowie}{Russell-Bowie}{2013}]%
        {russell2013mission}
\bibfield{author}{\bibinfo{person}{Deirdre Russell-Bowie}.}
  \bibinfo{year}{2013}\natexlab{}.
\newblock \showarticletitle{Mission Impossible or Possible Mission? Changing
  Confidence and Attitudes of Primary Preservice Music Education Students Using
  Kolb's Experiential Learning Theory}.
\newblock \bibinfo{journal}{\emph{Australian Journal of Music Education}}
  \bibinfo{volume}{2} (\bibinfo{year}{2013}), \bibinfo{pages}{46--63}.
\newblock
\showISSN{0004-9484}


\bibitem[\protect\citeauthoryear{Sami~Uddin and Gutwin}{Sami~Uddin and
  Gutwin}{2021}]%
        {uddin2021image}
\bibfield{author}{\bibinfo{person}{Md. Sami~Uddin} {and} \bibinfo{person}{Carl
  Gutwin}.} \bibinfo{year}{2021}\natexlab{}.
\newblock \showarticletitle{The Image of the Interface: How People Use
  Landmarks to Develop Spatial Memory of Commands in Graphical Interfaces}. In
  \bibinfo{booktitle}{\emph{Proceedings of the 2021 CHI Conference on Human
  Factors in Computing Systems}} (Yokohama, Japan) \emph{(\bibinfo{series}{CHI
  '21})}. \bibinfo{publisher}{Association for Computing Machinery},
  \bibinfo{address}{New York, NY, USA}, Article \bibinfo{articleno}{515},
  \bibinfo{numpages}{17}~pages.
\newblock
\showISBNx{9781450380966}
\urldef\tempurl%
\url{https://doi.org/10.1145/3411764.3445050}
\showDOI{\tempurl}


\bibitem[\protect\citeauthoryear{Sandnes and Eika}{Sandnes and Eika}{2019}]%
        {sandnes2019enhanced}
\bibfield{author}{\bibinfo{person}{Frode~Eika Sandnes} {and}
  \bibinfo{person}{Evelyn Eika}.} \bibinfo{year}{2019}\natexlab{}.
\newblock \showarticletitle{Enhanced Learning of Jazz Chords with a Projector
  Based Piano Keyboard Augmentation}. In
  \bibinfo{booktitle}{\emph{International Conference on Innovative Technologies
  and Learning}}. \bibinfo{publisher}{Springer}, \bibinfo{address}{Cham},
  \bibinfo{pages}{194--203}.
\newblock


\bibitem[\protect\citeauthoryear{Santini}{Santini}{2020}]%
        {santiniaugmented}
\bibfield{author}{\bibinfo{person}{Giovanni Santini}.}
  \bibinfo{year}{2020}\natexlab{}.
\newblock \showarticletitle{Augmented Piano in Augmented Reality}. In
  \bibinfo{booktitle}{\emph{International Conference on New Interfaces for
  Musical Expression}} \emph{(\bibinfo{series}{NIME '20})}.
  \bibinfo{publisher}{PubPub}, \bibinfo{address}{Cambridge, MA, USA},
  \bibinfo{pages}{411--415}.
\newblock


\bibitem[\protect\citeauthoryear{Santos, Chen, Taketomi, Yamamoto, Miyazaki,
  and Kato}{Santos et~al\mbox{.}}{2014}]%
        {santos2013augmented}
\bibfield{author}{\bibinfo{person}{Marc Ericson~C Santos},
  \bibinfo{person}{Angie Chen}, \bibinfo{person}{Takafumi Taketomi},
  \bibinfo{person}{Goshiro Yamamoto}, \bibinfo{person}{Jun Miyazaki}, {and}
  \bibinfo{person}{Hirokazu Kato}.} \bibinfo{year}{2014}\natexlab{}.
\newblock \showarticletitle{Augmented Reality Learning Experiences: Survey of
  Prototype Design and Evaluation}.
\newblock \bibinfo{journal}{\emph{IEEE Transactions on Learning Technologies}}
  \bibinfo{volume}{7}, \bibinfo{number}{1} (\bibinfo{year}{2014}),
  \bibinfo{pages}{38--56}.
\newblock
\urldef\tempurl%
\url{https://doi.org/10.1109/TLT.2013.37}
\showDOI{\tempurl}


\bibitem[\protect\citeauthoryear{Schmalstieg and Wagner}{Schmalstieg and
  Wagner}{2007}]%
        {schmalstieg2007experiences}
\bibfield{author}{\bibinfo{person}{Dieter Schmalstieg} {and}
  \bibinfo{person}{Daniel Wagner}.} \bibinfo{year}{2007}\natexlab{}.
\newblock \showarticletitle{Experiences with Handheld Augmented Reality}. In
  \bibinfo{booktitle}{\emph{6th IEEE and ACM International Symposium on Mixed
  and Augmented Reality}} \emph{(\bibinfo{series}{ISMAR '07})}.
  \bibinfo{publisher}{IEEE}, \bibinfo{address}{New York, NY, USA},
  \bibinfo{pages}{3--18}.
\newblock
\urldef\tempurl%
\url{https://doi.org/10.1109/ISMAR.2007.4538819}
\showDOI{\tempurl}


\bibitem[\protect\citeauthoryear{Schneegass, Olsson, Mayer, and
  Van~Laerhoven}{Schneegass et~al\mbox{.}}{2016}]%
        {schneegass2016mobile}
\bibfield{author}{\bibinfo{person}{Stefan Schneegass}, \bibinfo{person}{Thomas
  Olsson}, \bibinfo{person}{Sven Mayer}, {and} \bibinfo{person}{Kristof
  Van~Laerhoven}.} \bibinfo{year}{2016}\natexlab{}.
\newblock \showarticletitle{Mobile interactions augmented by wearable
  computing: A design space and vision}.
\newblock \bibinfo{journal}{\emph{International Journal of Mobile Human
  Computer Interaction (IJMHCI)}} \bibinfo{volume}{8}, \bibinfo{number}{4}
  (\bibinfo{year}{2016}), \bibinfo{pages}{104--114}.
\newblock
\urldef\tempurl%
\url{https://doi.org/10.4018/IJMHCI.2016100106}
\showDOI{\tempurl}


\bibitem[\protect\citeauthoryear{Scott}{Scott}{2011}]%
        {scott2011contemplating}
\bibfield{author}{\bibinfo{person}{Sheila Scott}.}
  \bibinfo{year}{2011}\natexlab{}.
\newblock \showarticletitle{Contemplating a constructivist stance for active
  learning within music education}.
\newblock \bibinfo{journal}{\emph{Arts Education Policy Review}}
  \bibinfo{volume}{112}, \bibinfo{number}{4} (\bibinfo{year}{2011}),
  \bibinfo{pages}{191--198}.
\newblock
\urldef\tempurl%
\url{https://doi.org/10.1080/10632913.2011.592469}
\showDOI{\tempurl}


\bibitem[\protect\citeauthoryear{Shamrock}{Shamrock}{1997}]%
        {shamrock1997orff}
\bibfield{author}{\bibinfo{person}{Mary Shamrock}.}
  \bibinfo{year}{1997}\natexlab{}.
\newblock \showarticletitle{Orff-Schulwerk: An Integrated Foundation: This
  article on the methodologies and practices of Orff-Schulwerk was first
  published in the Music Educators Journal in February 1986}.
\newblock \bibinfo{journal}{\emph{Music Educators Journal}}
  \bibinfo{volume}{83}, \bibinfo{number}{6} (\bibinfo{year}{1997}),
  \bibinfo{pages}{41--44}.
\newblock


\bibitem[\protect\citeauthoryear{Silva, Kergomard, Vergez, and Gilbert}{Silva
  et~al\mbox{.}}{2008}]%
        {silva2008interaction}
\bibfield{author}{\bibinfo{person}{Fabrice Silva}, \bibinfo{person}{Jean
  Kergomard}, \bibinfo{person}{Christophe Vergez}, {and}
  \bibinfo{person}{Jo{\"e}l Gilbert}.} \bibinfo{year}{2008}\natexlab{}.
\newblock \showarticletitle{Interaction of reed and acoustic resonator in
  clarinetlike systems}.
\newblock \bibinfo{journal}{\emph{The Journal of the Acoustical Society of
  America}} \bibinfo{volume}{124}, \bibinfo{number}{5} (\bibinfo{year}{2008}),
  \bibinfo{pages}{3284--3295}.
\newblock


\bibitem[\protect\citeauthoryear{Simpson}{Simpson}{1966}]%
        {simpson1966classification}
\bibfield{author}{\bibinfo{person}{Elizabeth~J Simpson}.}
  \bibinfo{year}{1966}\natexlab{}.
\newblock \bibinfo{booktitle}{\emph{The classification of educational
  objectives, psychomotor domain}}.
\newblock \bibinfo{type}{{T}echnical {R}eport}. \bibinfo{institution}{Illinois
  Univ, Urbana}, \bibinfo{address}{New York, NY, USA}.
\newblock


\bibitem[\protect\citeauthoryear{Sloboda}{Sloboda}{1992}]%
        {sloboda1992transitions}
\bibfield{author}{\bibinfo{person}{Michael J.~A. Sloboda, John A.and~Howe}.}
  \bibinfo{year}{1992}\natexlab{}.
\newblock \showarticletitle{Transitions in the Early Musical Careers of Able
  Young Musicians: Choosing Instruments and Teachers}.
\newblock \bibinfo{journal}{\emph{Journal of Research in Music Education}}
  \bibinfo{volume}{40}, \bibinfo{number}{4} (\bibinfo{year}{1992}),
  \bibinfo{pages}{283--294}.
\newblock
\urldef\tempurl%
\url{https://doi.org/10.2307/3345836}
\showDOI{\tempurl}


\bibitem[\protect\citeauthoryear{Smith}{Smith}{1983}]%
        {smith1983homer}
\bibfield{author}{\bibinfo{person}{Gregory~Eugene Smith}.}
  \bibinfo{year}{1983}\natexlab{}.
\newblock \bibinfo{booktitle}{\emph{Homer, Gregory, and Bill Evans? The theory
  of formulaic composition in the context of jazz piano improvisation}}.
\newblock \bibinfo{publisher}{Harvard University}, \bibinfo{address}{Cambridge,
  MA, USA}.
\newblock


\bibitem[\protect\citeauthoryear{Solis and Nettl}{Solis and Nettl}{2009}]%
        {solis2009musical}
\bibfield{author}{\bibinfo{person}{Gabriel Solis} {and} \bibinfo{person}{Bruno
  Nettl}.} \bibinfo{year}{2009}\natexlab{}.
\newblock \bibinfo{booktitle}{\emph{Musical improvisation: Art, education, and
  society}}.
\newblock \bibinfo{publisher}{University of Illinois Press},
  \bibinfo{address}{Illinois, USA}.
\newblock


\bibitem[\protect\citeauthoryear{Speicher, Hall, and Nebeling}{Speicher
  et~al\mbox{.}}{2019}]%
        {speicher2019what}
\bibfield{author}{\bibinfo{person}{Maximilian Speicher},
  \bibinfo{person}{Brian~D. Hall}, {and} \bibinfo{person}{Michael Nebeling}.}
  \bibinfo{year}{2019}\natexlab{}.
\newblock \bibinfo{booktitle}{\emph{What is Mixed Reality?}}
\newblock \bibinfo{publisher}{Association for Computing Machinery},
  \bibinfo{address}{New York, NY, USA}, \bibinfo{pages}{1–15}.
\newblock
\showISBNx{9781450359702}
\urldef\tempurl%
\url{https://doi.org/10.1145/3290605.3300767}
\showURL{%
\tempurl}


\bibitem[\protect\citeauthoryear{Stanbury, Said, and Kang}{Stanbury
  et~al\mbox{.}}{2021}]%
        {stanbury2021holokeys}
\bibfield{author}{\bibinfo{person}{Austin~Jerald Stanbury},
  \bibinfo{person}{Ines Said}, {and} \bibinfo{person}{Hyo~Jeong Kang}.}
  \bibinfo{year}{2021}\natexlab{}.
\newblock \showarticletitle{HoloKeys: Interactive Piano Education Using
  Augmented Reality and IoT}. In \bibinfo{booktitle}{\emph{Proceedings of the
  27th ACM Symposium on Virtual Reality Software and Technology}} (Osaka,
  Japan) \emph{(\bibinfo{series}{VRST '21})}. \bibinfo{publisher}{Association
  for Computing Machinery}, \bibinfo{address}{New York, NY, USA}, Article
  \bibinfo{articleno}{76}, \bibinfo{numpages}{3}~pages.
\newblock
\showISBNx{9781450390927}
\urldef\tempurl%
\url{https://doi.org/10.1145/3489849.3489921}
\showDOI{\tempurl}


\bibitem[\protect\citeauthoryear{Stryker and Leaver}{Stryker and
  Leaver}{1997}]%
        {stryker1997content}
\bibfield{author}{\bibinfo{person}{Stephen~B Stryker} {and}
  \bibinfo{person}{Betty~L Leaver}.} \bibinfo{year}{1997}\natexlab{}.
\newblock \showarticletitle{Content-based instruction: From theory to
  practice}.
\newblock \bibinfo{journal}{\emph{Content-based instruction in foreign language
  education: Models and methods}} \bibinfo{volume}{8}, \bibinfo{number}{1}
  (\bibinfo{year}{1997}), \bibinfo{pages}{3--28}.
\newblock


\bibitem[\protect\citeauthoryear{Sun and Chiang}{Sun and Chiang}{2018}]%
        {sun2018mr}
\bibfield{author}{\bibinfo{person}{Chung-Hsuan Sun} {and}
  \bibinfo{person}{Pei-Ying Chiang}.} \bibinfo{year}{2018}\natexlab{}.
\newblock \showarticletitle{Mr. Piano: A Portable Piano Tutoring System}. In
  \bibinfo{booktitle}{\emph{2018 IEEE XXV International Conference on
  Electronics, Electrical Engineering and Computing}}
  \emph{(\bibinfo{series}{INTERCON '18})}. \bibinfo{publisher}{IEEE},
  \bibinfo{address}{New York, NY, USA}, \bibinfo{pages}{1--4}.
\newblock
\urldef\tempurl%
\url{https://doi.org/10.1109/INTERCON.2018.8526423}
\showDOI{\tempurl}


\bibitem[\protect\citeauthoryear{Takegawa, Terada, and Tsukamoto}{Takegawa
  et~al\mbox{.}}{2012}]%
        {takegawa2012piano}
\bibfield{author}{\bibinfo{person}{Yoshinari Takegawa},
  \bibinfo{person}{Tsutomu Terada}, {and} \bibinfo{person}{Masahiko
  Tsukamoto}.} \bibinfo{year}{2012}\natexlab{}.
\newblock \showarticletitle{A piano learning support system considering
  rhythm}. In \bibinfo{booktitle}{\emph{Non-Cochlear Sound: Proceedings of the
  38th International Computer Music Conference}} \emph{(\bibinfo{series}{ICMC
  '12})}. \bibinfo{publisher}{Michigan Publishing}, \bibinfo{address}{Ann
  Arbor, MI, USA}, \bibinfo{pages}{325--332}.
\newblock
\urldef\tempurl%
\url{http://hdl.handle.net/2027/spo.bbp2372.2012.061}
\showURL{%
\tempurl}


\bibitem[\protect\citeauthoryear{Tan, Slivovsky, and Pentland}{Tan
  et~al\mbox{.}}{2001}]%
        {tan2001sensing}
\bibfield{author}{\bibinfo{person}{Hong~Z Tan}, \bibinfo{person}{Lynne~A
  Slivovsky}, {and} \bibinfo{person}{Alex Pentland}.}
  \bibinfo{year}{2001}\natexlab{}.
\newblock \showarticletitle{A sensing chair using pressure distribution
  sensors}.
\newblock \bibinfo{journal}{\emph{IEEE/ASME Transactions On Mechatronics}}
  \bibinfo{volume}{6}, \bibinfo{number}{3} (\bibinfo{year}{2001}),
  \bibinfo{pages}{261--268}.
\newblock


\bibitem[\protect\citeauthoryear{Thomas}{Thomas}{1970}]%
        {thomas1970manhattanville}
\bibfield{author}{\bibinfo{person}{Ronald~B Thomas}.}
  \bibinfo{year}{1970}\natexlab{}.
\newblock \bibinfo{booktitle}{\emph{Manhattanville Music Curriculum Program.
  Final Report.}}
\newblock \bibinfo{type}{{T}echnical {R}eport}.
  \bibinfo{institution}{Manhattanville Coll., Purchase, NY.},
  \bibinfo{address}{New York, NY, USA}. \bibinfo{pages}{459} pages.
\newblock


\bibitem[\protect\citeauthoryear{Toda, Klock, Oliveira, Palomino, Rodrigues,
  Shi, Bittencourt, Gasparini, Isotani, and Cristea}{Toda
  et~al\mbox{.}}{2019a}]%
        {toda2019analysing}
\bibfield{author}{\bibinfo{person}{Armando~M Toda}, \bibinfo{person}{Ana~CT
  Klock}, \bibinfo{person}{Wilk Oliveira}, \bibinfo{person}{Paula~T Palomino},
  \bibinfo{person}{Luiz Rodrigues}, \bibinfo{person}{Lei Shi},
  \bibinfo{person}{Ig Bittencourt}, \bibinfo{person}{Isabela Gasparini},
  \bibinfo{person}{Seiji Isotani}, {and} \bibinfo{person}{Alexandra~I
  Cristea}.} \bibinfo{year}{2019}\natexlab{a}.
\newblock \showarticletitle{Analysing gamification elements in educational
  environments using an existing Gamification taxonomy}.
\newblock \bibinfo{journal}{\emph{Smart Learning Environments}}
  \bibinfo{volume}{6}, \bibinfo{number}{1} (\bibinfo{year}{2019}),
  \bibinfo{pages}{1--14}.
\newblock


\bibitem[\protect\citeauthoryear{Toda, Palomino, Oliveira, Rodrigues, Klock,
  Gasparini, Cristea, and Isotani}{Toda et~al\mbox{.}}{2019b}]%
        {toda2019gamify}
\bibfield{author}{\bibinfo{person}{Armando~M Toda}, \bibinfo{person}{Paula~T
  Palomino}, \bibinfo{person}{Wilk Oliveira}, \bibinfo{person}{Luiz Rodrigues},
  \bibinfo{person}{Ana~CT Klock}, \bibinfo{person}{Isabela Gasparini},
  \bibinfo{person}{Alexandra~I Cristea}, {and} \bibinfo{person}{Seiji
  Isotani}.} \bibinfo{year}{2019}\natexlab{b}.
\newblock \showarticletitle{How to gamify learning systems? an experience
  report using the design sprint method and a taxonomy for gamification
  elements in education}.
\newblock \bibinfo{journal}{\emph{Journal of Educational Technology \&
  Society}} \bibinfo{volume}{22}, \bibinfo{number}{3} (\bibinfo{year}{2019}),
  \bibinfo{pages}{47--60}.
\newblock


\bibitem[\protect\citeauthoryear{Trujano, Khan, and Maes}{Trujano
  et~al\mbox{.}}{2018}]%
        {trujano2018arpiano}
\bibfield{author}{\bibinfo{person}{Fernando Trujano}, \bibinfo{person}{Mina
  Khan}, {and} \bibinfo{person}{Pattie Maes}.} \bibinfo{year}{2018}\natexlab{}.
\newblock \showarticletitle{ARPiano Efficient Music Learning Using Augmented
  Reality}. In \bibinfo{booktitle}{\emph{Innovative Technologies and
  Learning}}, \bibfield{editor}{\bibinfo{person}{Ting-Ting Wu},
  \bibinfo{person}{Yueh-Min Huang}, \bibinfo{person}{Rustam Shadiev},
  \bibinfo{person}{Lin Lin}, {and} \bibinfo{person}{Andreja~Isteni{\v{c}}
  Star{\v{c}}i{\v{c}}}} (Eds.). \bibinfo{publisher}{Springer International
  Publishing}, \bibinfo{address}{Cham}, \bibinfo{pages}{3--17}.
\newblock
\showISBNx{978-3-319-99737-7}
\urldef\tempurl%
\url{https://doi.org/10.1007/978-3-319-99737-7_1}
\showDOI{\tempurl}


\bibitem[\protect\citeauthoryear{Turchet}{Turchet}{2018}]%
        {turchet2018some}
\bibfield{author}{\bibinfo{person}{Luca Turchet}.}
  \bibinfo{year}{2018}\natexlab{}.
\newblock \showarticletitle{Some Reflections on the Relation between Augmented
  and Smart Musical Instruments}.
\newblock In \bibinfo{booktitle}{\emph{Proceedings of the Audio Mostly 2018 on
  Sound in Immersion and Emotion}}. \bibinfo{publisher}{Association for
  Computing Machinery}, \bibinfo{address}{New York, NY, USA}.
\newblock
\showISBNx{9781450366090}
\urldef\tempurl%
\url{https://doi.org/10.1145/3243274.3243281}
\showDOI{\tempurl}


\bibitem[\protect\citeauthoryear{Vassileva}{Vassileva}{2012}]%
        {vassileva2012motivating}
\bibfield{author}{\bibinfo{person}{Julita Vassileva}.}
  \bibinfo{year}{2012}\natexlab{}.
\newblock \showarticletitle{Motivating participation in social computing
  applications: a user modeling perspective}.
\newblock \bibinfo{journal}{\emph{User Modeling and User-Adapted Interaction}}
  \bibinfo{volume}{22}, \bibinfo{number}{1} (\bibinfo{year}{2012}),
  \bibinfo{pages}{177--201}.
\newblock


\bibitem[\protect\citeauthoryear{Walder}{Walder}{2016}]%
        {walder2016modelling}
\bibfield{author}{\bibinfo{person}{Christian Walder}.}
  \bibinfo{year}{2016}\natexlab{}.
\newblock \showarticletitle{Modelling Symbolic Music: Beyond the Piano Roll}.
  In \bibinfo{booktitle}{\emph{Proceedings of The 8th Asian Conference on
  Machine Learning}} \emph{(\bibinfo{series}{Proceedings of Machine Learning
  Research}, Vol.~\bibinfo{volume}{63})},
  \bibfield{editor}{\bibinfo{person}{Robert~J. Durrant} {and}
  \bibinfo{person}{Kee-Eung Kim}} (Eds.). \bibinfo{publisher}{PMLR},
  \bibinfo{address}{The University of Waikato, Hamilton, New Zealand},
  \bibinfo{pages}{174--189}.
\newblock
\urldef\tempurl%
\url{http://proceedings.mlr.press/v63/walder88.html}
\showURL{%
\tempurl}


\bibitem[\protect\citeauthoryear{Waldron}{Waldron}{2009}]%
        {waldron2009exploring}
\bibfield{author}{\bibinfo{person}{Janice Waldron}.}
  \bibinfo{year}{2009}\natexlab{}.
\newblock \showarticletitle{Exploring a virtual music community of practice:
  Informal music learning on the Internet}.
\newblock \bibinfo{journal}{\emph{Journal of Music, Technology \& Education}}
  \bibinfo{volume}{2}, \bibinfo{number}{2-3} (\bibinfo{year}{2009}),
  \bibinfo{pages}{97--112}.
\newblock
\urldef\tempurl%
\url{https://doi.org/10.1386/jmte.2.2-3.97_1}
\showDOI{\tempurl}


\bibitem[\protect\citeauthoryear{Webster}{Webster}{2011}]%
        {webster2011construction}
\bibfield{author}{\bibinfo{person}{Peter~R Webster}.}
  \bibinfo{year}{2011}\natexlab{}.
\newblock \showarticletitle{Construction of music learning}.
\newblock \bibinfo{journal}{\emph{MENC handbook of research on music learning}}
   \bibinfo{volume}{1} (\bibinfo{year}{2011}), \bibinfo{pages}{35--83}.
\newblock
\urldef\tempurl%
\url{https://doi.org/10.1093/acprof:osobl/9780195386677.003.0002}
\showDOI{\tempurl}


\bibitem[\protect\citeauthoryear{Weing, R\"{o}hlig, Rogers, Gugenheimer,
  Schaub, K\"{o}nings, Rukzio, and Weber}{Weing et~al\mbox{.}}{2013}]%
        {weing2013piano}
\bibfield{author}{\bibinfo{person}{Matthias Weing}, \bibinfo{person}{Amrei
  R\"{o}hlig}, \bibinfo{person}{Katja Rogers}, \bibinfo{person}{Jan
  Gugenheimer}, \bibinfo{person}{Florian Schaub}, \bibinfo{person}{Bastian
  K\"{o}nings}, \bibinfo{person}{Enrico Rukzio}, {and} \bibinfo{person}{Michael
  Weber}.} \bibinfo{year}{2013}\natexlab{}.
\newblock \showarticletitle{P.I.A.N.O.: Enhancing Instrument Learning via
  Interactive Projected Augmentation}. In \bibinfo{booktitle}{\emph{Proceedings
  of the 2013 ACM Conference on Pervasive and Ubiquitous Computing Adjunct
  Publication}} (Zurich, Switzerland) \emph{(\bibinfo{series}{UbiComp '13
  Adjunct})}. \bibinfo{publisher}{Association for Computing Machinery},
  \bibinfo{address}{New York, NY, USA}, \bibinfo{pages}{75–78}.
\newblock
\showISBNx{9781450322157}
\urldef\tempurl%
\url{https://doi.org/10.1145/2494091.2494113}
\showDOI{\tempurl}


\bibitem[\protect\citeauthoryear{Westerlund}{Westerlund}{2003}]%
        {westerlund2003reconsidering}
\bibfield{author}{\bibinfo{person}{Heidi Westerlund}.}
  \bibinfo{year}{2003}\natexlab{}.
\newblock \showarticletitle{Reconsidering Aesthetic Experience in Praxial Music
  Education}.
\newblock \bibinfo{journal}{\emph{Philosophy of music education review}}
  \bibinfo{volume}{11}, \bibinfo{number}{1} (\bibinfo{year}{2003}),
  \bibinfo{pages}{45--62}.
\newblock
\showISSN{10635734, 15433412}
\urldef\tempurl%
\url{https://www.jstor.org/stable/40327197}
\showURL{%
\tempurl}


\bibitem[\protect\citeauthoryear{Wigram}{Wigram}{2004}]%
        {wigram2004improvisation}
\bibfield{author}{\bibinfo{person}{Tony Wigram}.}
  \bibinfo{year}{2004}\natexlab{}.
\newblock \bibinfo{booktitle}{\emph{Improvisation: Methods and techniques for
  music therapy clinicians, educators, and students}}.
\newblock \bibinfo{publisher}{Jessica Kingsley Publishers},
  \bibinfo{address}{United Kingdom}.
\newblock


\bibitem[\protect\citeauthoryear{Williams and Moser}{Williams and
  Moser}{2019}]%
        {williams2019art}
\bibfield{author}{\bibinfo{person}{Michael Williams} {and}
  \bibinfo{person}{Tami Moser}.} \bibinfo{year}{2019}\natexlab{}.
\newblock \showarticletitle{The art of coding and thematic exploration in
  qualitative research}.
\newblock \bibinfo{journal}{\emph{International Management Review}}
  \bibinfo{volume}{15}, \bibinfo{number}{1} (\bibinfo{year}{2019}),
  \bibinfo{pages}{45--55}.
\newblock


\bibitem[\protect\citeauthoryear{Wozniak, Goyal, Kucharski, Lischke, Mayer, and
  Fjeld}{Wozniak et~al\mbox{.}}{2016}]%
        {wozniak2016ramparts}
\bibfield{author}{\bibinfo{person}{Pawe\l{} Wozniak}, \bibinfo{person}{Nitesh
  Goyal}, \bibinfo{person}{Przemys\l{}aw Kucharski}, \bibinfo{person}{Lars
  Lischke}, \bibinfo{person}{Sven Mayer}, {and} \bibinfo{person}{Morten
  Fjeld}.} \bibinfo{year}{2016}\natexlab{}.
\newblock \showarticletitle{RAMPARTS: Supporting Sensemaking with
  Spatially-Aware Mobile Interactions}. In
  \bibinfo{booktitle}{\emph{Proceedings of the 2016 CHI Conference on Human
  Factors in Computing Systems}} (San Jose, California, USA)
  \emph{(\bibinfo{series}{CHI '16})}. \bibinfo{publisher}{Association for
  Computing Machinery}, \bibinfo{address}{New York, NY, USA},
  \bibinfo{pages}{2447–2460}.
\newblock
\showISBNx{9781450333627}
\urldef\tempurl%
\url{https://doi.org/10.1145/2858036.2858491}
\showDOI{\tempurl}


\bibitem[\protect\citeauthoryear{Wrigley and Emmerson}{Wrigley and
  Emmerson}{2013}]%
        {wrigley2013ecological}
\bibfield{author}{\bibinfo{person}{William~J Wrigley} {and}
  \bibinfo{person}{Stephen~B Emmerson}.} \bibinfo{year}{2013}\natexlab{}.
\newblock \showarticletitle{Ecological development and validation of a music
  performance rating scale for five instrument families}.
\newblock \bibinfo{journal}{\emph{Psychology of Music}} \bibinfo{volume}{41},
  \bibinfo{number}{1} (\bibinfo{year}{2013}), \bibinfo{pages}{97--118}.
\newblock
\urldef\tempurl%
\url{https://doi.org/10.1177/0305735611418552}
\showDOI{\tempurl}


\bibitem[\protect\citeauthoryear{Wristen}{Wristen}{2005}]%
        {wristen2005cognition}
\bibfield{author}{\bibinfo{person}{Brenda Wristen}.}
  \bibinfo{year}{2005}\natexlab{}.
\newblock \showarticletitle{Cognition and motor execution in piano
  sight-reading: A review of literature}.
\newblock \bibinfo{journal}{\emph{Update: Applications of Research in Music
  Education}} \bibinfo{volume}{24}, \bibinfo{number}{1} (\bibinfo{year}{2005}),
  \bibinfo{pages}{44--56}.
\newblock


\bibitem[\protect\citeauthoryear{Xia and Dannenberg}{Xia and
  Dannenberg}{2017}]%
        {xia2017improvised}
\bibfield{author}{\bibinfo{person}{Guangyu Xia} {and} \bibinfo{person}{Roger~B
  Dannenberg}.} \bibinfo{year}{2017}\natexlab{}.
\newblock \showarticletitle{Improvised duet interaction: learning improvisation
  techniques for automatic accompaniment.}. In
  \bibinfo{booktitle}{\emph{International Conference on New Interfaces for
  Musical Expression}} \emph{(\bibinfo{series}{NIME '17})}.
  \bibinfo{publisher}{NIME}, \bibinfo{address}{Copenhagen, Denmark},
  \bibinfo{pages}{110--114}.
\newblock


\bibitem[\protect\citeauthoryear{Xiao, Aguilera, Williams, and Ishii}{Xiao
  et~al\mbox{.}}{2013}]%
        {xiao2013mirrorfugue}
\bibfield{author}{\bibinfo{person}{Xiao Xiao}, \bibinfo{person}{Paula
  Aguilera}, \bibinfo{person}{Jonathan Williams}, {and}
  \bibinfo{person}{Hiroshi Ishii}.} \bibinfo{year}{2013}\natexlab{}.
\newblock \showarticletitle{MirrorFugue iii: Conjuring the Recorded Pianist}.
  In \bibinfo{booktitle}{\emph{CHI '13 Extended Abstracts on Human Factors in
  Computing Systems}} (Paris, France) \emph{(\bibinfo{series}{CHI EA '13})}.
  \bibinfo{publisher}{Association for Computing Machinery},
  \bibinfo{address}{New York, NY, USA}, \bibinfo{pages}{2891–2892}.
\newblock
\showISBNx{9781450319522}
\urldef\tempurl%
\url{https://doi.org/10.1145/2468356.2479564}
\showDOI{\tempurl}


\bibitem[\protect\citeauthoryear{Xiao and Ishii}{Xiao and Ishii}{2010}]%
        {xiao2010mirrorfugue}
\bibfield{author}{\bibinfo{person}{Xiao Xiao} {and} \bibinfo{person}{Hiroshi
  Ishii}.} \bibinfo{year}{2010}\natexlab{}.
\newblock \showarticletitle{MirrorFugue: Communicating Hand Gesture in Remote
  Piano Collaboration}. In \bibinfo{booktitle}{\emph{Proceedings of the Fifth
  International Conference on Tangible, Embedded, and Embodied Interaction}}
  (Funchal, Portugal) \emph{(\bibinfo{series}{TEI '11})}.
  \bibinfo{publisher}{Association for Computing Machinery},
  \bibinfo{address}{New York, NY, USA}, \bibinfo{pages}{13–20}.
\newblock
\showISBNx{9781450304788}
\urldef\tempurl%
\url{https://doi.org/10.1145/1935701.1935705}
\showDOI{\tempurl}


\bibitem[\protect\citeauthoryear{Xiao and Ishii}{Xiao and Ishii}{2011}]%
        {xiao2011duet}
\bibfield{author}{\bibinfo{person}{Xiao Xiao} {and} \bibinfo{person}{Hiroshi
  Ishii}.} \bibinfo{year}{2011}\natexlab{}.
\newblock \showarticletitle{Duet for Solo Piano: MirrorFugue for Single User
  Playing with Recorded Performances}. In \bibinfo{booktitle}{\emph{CHI '11
  Extended Abstracts on Human Factors in Computing Systems}} (Vancouver, BC,
  Canada) \emph{(\bibinfo{series}{CHI EA '11})}.
  \bibinfo{publisher}{Association for Computing Machinery},
  \bibinfo{address}{New York, NY, USA}, \bibinfo{pages}{1285–1290}.
\newblock
\showISBNx{9781450302685}
\urldef\tempurl%
\url{https://doi.org/10.1145/1979742.1979762}
\showDOI{\tempurl}


\bibitem[\protect\citeauthoryear{Xiao, Tome, and Ishii}{Xiao
  et~al\mbox{.}}{2014}]%
        {xiao2014andante}
\bibfield{author}{\bibinfo{person}{Xiao Xiao}, \bibinfo{person}{Basheer Tome},
  {and} \bibinfo{person}{Hiroshi Ishii}.} \bibinfo{year}{2014}\natexlab{}.
\newblock \showarticletitle{Andante: Walking Figures on the Piano Keyboard to
  Visualize Musical Motion}. In \bibinfo{booktitle}{\emph{International
  Conference on New Interfaces for Musical Expression}}
  \emph{(\bibinfo{series}{NIME '14})}. \bibinfo{publisher}{Zenodo},
  \bibinfo{address}{Málaga, Spain}, \bibinfo{pages}{629--632}.
\newblock
\urldef\tempurl%
\url{https://doi.org/10.5281/zenodo.1178987}
\showDOI{\tempurl}


\bibitem[\protect\citeauthoryear{Yang and Essl}{Yang and Essl}{2012}]%
        {yang2012augmented}
\bibfield{author}{\bibinfo{person}{Qi Yang} {and} \bibinfo{person}{Georg
  Essl}.} \bibinfo{year}{2012}\natexlab{}.
\newblock \showarticletitle{Augmented Piano Performance using a Depth Camera.}.
  In \bibinfo{booktitle}{\emph{International Conference on New Interfaces for
  Musical Expression}} \emph{(\bibinfo{series}{NIME '12})}.
  \bibinfo{publisher}{NIME}, \bibinfo{address}{Michigan USA},
  \bibinfo{pages}{1--2}.
\newblock


\bibitem[\protect\citeauthoryear{Yang and Essl}{Yang and Essl}{2013}]%
        {yang2013visual}
\bibfield{author}{\bibinfo{person}{Qi Yang} {and} \bibinfo{person}{Georg
  Essl}.} \bibinfo{year}{2013}\natexlab{}.
\newblock \showarticletitle{Visual Associations in Augmented Keyboard
  Performance.}. In \bibinfo{booktitle}{\emph{International Conference on New
  Interfaces for Musical Expression}} \emph{(\bibinfo{series}{NIME '13})}.
  \bibinfo{publisher}{NIME}, \bibinfo{address}{Daejeon, Korea},
  \bibinfo{pages}{252--255}.
\newblock


\bibitem[\protect\citeauthoryear{Yang}{Yang}{2012}]%
        {yang2012building}
\bibfield{author}{\bibinfo{person}{Ya-Ting~Carolyn Yang}.}
  \bibinfo{year}{2012}\natexlab{}.
\newblock \showarticletitle{Building virtual cities, inspiring intelligent
  citizens: Digital games for developing students’ problem solving and
  learning motivation}.
\newblock \bibinfo{journal}{\emph{Computers \& Education}}
  \bibinfo{volume}{59}, \bibinfo{number}{2} (\bibinfo{year}{2012}),
  \bibinfo{pages}{365--377}.
\newblock
\urldef\tempurl%
\url{https://doi.org/10.1016/j.compedu.2012.01.012}
\showDOI{\tempurl}


\bibitem[\protect\citeauthoryear{Yuksel, Oleson, Harrison, Peck, Afergan,
  Chang, and Jacob}{Yuksel et~al\mbox{.}}{2016}]%
        {yuksel2016learn}
\bibfield{author}{\bibinfo{person}{Beste~F. Yuksel}, \bibinfo{person}{Kurt~B.
  Oleson}, \bibinfo{person}{Lane Harrison}, \bibinfo{person}{Evan~M. Peck},
  \bibinfo{person}{Daniel Afergan}, \bibinfo{person}{Remco Chang}, {and}
  \bibinfo{person}{Robert~JK Jacob}.} \bibinfo{year}{2016}\natexlab{}.
\newblock \showarticletitle{Learn Piano with BACh: An Adaptive Learning
  Interface That Adjusts Task Difficulty Based on Brain State}. In
  \bibinfo{booktitle}{\emph{Proceedings of the 2016 CHI Conference on Human
  Factors in Computing Systems}} (San Jose, California, USA)
  \emph{(\bibinfo{series}{CHI '16})}. \bibinfo{publisher}{Association for
  Computing Machinery}, \bibinfo{address}{New York, NY, USA},
  \bibinfo{pages}{5372–5384}.
\newblock
\showISBNx{9781450333627}
\urldef\tempurl%
\url{https://doi.org/10.1145/2858036.2858388}
\showDOI{\tempurl}


\bibitem[\protect\citeauthoryear{Zandt-Escobar, Caramiaux, and
  Tanaka}{Zandt-Escobar et~al\mbox{.}}{2014}]%
        {zandt2014piaf}
\bibfield{author}{\bibinfo{person}{Van Zandt-Escobar},
  \bibinfo{person}{Baptiste Caramiaux}, {and} \bibinfo{person}{Atau Tanaka}.}
  \bibinfo{year}{2014}\natexlab{}.
\newblock \showarticletitle{Piaf: A tool for augmented piano performance using
  gesture variation following}. In \bibinfo{booktitle}{\emph{Proceedings of the
  International Conference on New Interfaces for Musical Expression}}
  \emph{(\bibinfo{series}{NIME '14})}. \bibinfo{publisher}{PubPub},
  \bibinfo{address}{Cambridge, MA, USA}, \bibinfo{pages}{167--170}.
\newblock


\bibitem[\protect\citeauthoryear{Zaqout, Elhissi, Jarour, and Elowini}{Zaqout
  et~al\mbox{.}}{2015}]%
        {zaqout2015augmented}
\bibfield{author}{\bibinfo{person}{Ihab Zaqout}, \bibinfo{person}{Samar
  Elhissi}, \bibinfo{person}{Aya Jarour}, {and} \bibinfo{person}{Heba
  Elowini}.} \bibinfo{year}{2015}\natexlab{}.
\newblock \showarticletitle{Augmented Piano Reality}.
\newblock \bibinfo{journal}{\emph{International Journal of Hybrid Information
  Technology}} \bibinfo{volume}{8}, \bibinfo{number}{10}
  (\bibinfo{year}{2015}), \bibinfo{pages}{141--152}.
\newblock
\urldef\tempurl%
\url{https://doi.org/10.14257/ijhit.2015.8.10.13}
\showDOI{\tempurl}


\bibitem[\protect\citeauthoryear{Zeng, He, and Pan}{Zeng et~al\mbox{.}}{2019}]%
        {zeng2019funpianoar}
\bibfield{author}{\bibinfo{person}{Hong Zeng}, \bibinfo{person}{Xingxi He},
  {and} \bibinfo{person}{Honghu Pan}.} \bibinfo{year}{2019}\natexlab{}.
\newblock \showarticletitle{FunPianoAR: a novel AR application for piano
  learning considering paired play based on multi-marker tracking}. In
  \bibinfo{booktitle}{\emph{Journal of Physics: Conference Series}},
  Vol.~\bibinfo{volume}{1229}. \bibinfo{publisher}{IOP Publishing},
  \bibinfo{address}{Bristol, United Kingdom}, \bibinfo{pages}{012072}.
\newblock
\urldef\tempurl%
\url{https://doi.org/10.1088/1742-6596/1229/1/012072}
\showDOI{\tempurl}


\bibitem[\protect\citeauthoryear{Zhang, Shen, Ong, and Nee}{Zhang
  et~al\mbox{.}}{2010}]%
        {zhang2010affordable}
\bibfield{author}{\bibinfo{person}{D Zhang}, \bibinfo{person}{Yan Shen},
  \bibinfo{person}{Soh-Khim Ong}, {and} \bibinfo{person}{Andrew~YC Nee}.}
  \bibinfo{year}{2010}\natexlab{}.
\newblock \showarticletitle{An Affordable Augmented Reality Based
  Rehabilitation System for Hand Motions}. In \bibinfo{booktitle}{\emph{2010
  International Conference on Cyberworlds}} \emph{(\bibinfo{series}{CW '10})}.
  \bibinfo{publisher}{IEEE}, \bibinfo{address}{New York, NY, USA},
  \bibinfo{pages}{346--353}.
\newblock
\urldef\tempurl%
\url{https://doi.org/10.1109/CW.2010.31}
\showDOI{\tempurl}


\bibitem[\protect\citeauthoryear{Zimmerman and Moylan}{Zimmerman and
  Moylan}{2009}]%
        {zimmerman2009self}
\bibfield{author}{\bibinfo{person}{Barry~J Zimmerman} {and}
  \bibinfo{person}{Adam~R Moylan}.} \bibinfo{year}{2009}\natexlab{}.
\newblock \showarticletitle{Self-regulation: Where metacognition and motivation
  intersect}.
\newblock In \bibinfo{booktitle}{\emph{Handbook of metacognition in
  education}}. \bibinfo{publisher}{Routledge}, \bibinfo{address}{UK},
  \bibinfo{pages}{311--328}.
\newblock


\bibitem[\protect\citeauthoryear{Zor{\v{c}}, {\v{C}}opi{\v{c}}~Pucihar, and
  Kljun}{Zor{\v{c}} et~al\mbox{.}}{2019}]%
        {zorvc2019preprivcljive}
\bibfield{author}{\bibinfo{person}{Luka Zor{\v{c}}}, \bibinfo{person}{Klen
  {\v{C}}opi{\v{c}}~Pucihar}, {and} \bibinfo{person}{Matja{\v{z}} Kljun}.}
  \bibinfo{year}{2019}\natexlab{}.
\newblock \showarticletitle{Prepri{\v{c}}ljive tehnologije za spodbujanje
  pravilne dr{\v{z}}e telesa pri sedenju -- Persuasive technologies for
  promoting correct sitting posture}. In \bibinfo{booktitle}{\emph{IS 2019 -
  Information society multi conference - Human-Computer Interaction in
  Information Society}}. \bibinfo{publisher}{Jožef Stefan Institute},
  \bibinfo{address}{Slovenia}, \bibinfo{pages}{21--24}.
\newblock


\end{thebibliography}


\received{February 2022}
\received[revised]{April 2022}
\received[accepted]{June 2022}

\end{document}